\documentclass[a4paper,12pt]{article}
\usepackage[utf8]{inputenc}
\pdfoutput=1
\usepackage{a4wide}
\usepackage{expdlist}
\usepackage{amsmath,amsfonts,amssymb}
\usepackage{graphicx}
\usepackage{enumerate}
\usepackage{hyperref}
\usepackage{latexsym}
\usepackage{hepnicenames}
\usepackage{enumitem}
\usepackage{soul}
\usepackage[normalem]{ulem}
\usepackage{wasysym}
\usepackage{makecell}
\usepackage{bbm}

\usepackage{tabularx}
\usepackage{array}
\usepackage{booktabs}
\usepackage{comment}

\font\tenrsfs=rsfs10 at 12pt
\font\sevenrsfs=rsfs7
\font\fiversfs=rsfs5
\newfam\rsfsfam
\textfont\rsfsfam=\tenrsfs
\scriptfont\rsfsfam=\sevenrsfs
\scriptscriptfont\rsfsfam=\fiversfs

\numberwithin{equation}{section}

\usepackage{mathrsfs}
\usepackage{braket}
\usepackage{titling}
\usepackage{amsmath}
\usepackage{slashed}
\usepackage{amssymb}
\usepackage{epsfig}
\usepackage{graphicx}
\usepackage{color}
\usepackage{rotating}
\usepackage{hyperref}
\usepackage[table,xcdraw,dvipsnames]{xcolor}
\usepackage[compress,numbers,sort]{natbib}
\usepackage{bbold}
\usepackage{colortbl}
\usepackage{pdflscape}
\usepackage{color}
\usepackage{mathtools}
\usepackage{colortbl}
\definecolor{Gray}{gray}{0.95}
\definecolor{RGray}{gray}{0.85}
\definecolor{CGray}{gray}{0.93}

\newcommand{\SU}{{\rm SU}}
\newcommand{\SO}{{\rm SO}}

\newcommand{\U}{{\rm U}}

\newcommand{\V}{{\cal V}}

\newcommand{\PQ}{{\rm PQ}}

\newcommand{\1}{{\textbf{1}}}
\newcommand{\3}{{\textbf{3}}}

\newcommand{\6}{\textbf{6}}

\definecolor{nicered}{rgb}{0.7,0.1,0.1}
\definecolor{nicegreen}{rgb}{0.1,0.5,0.1}
\definecolor{red}{rgb}{1.0, 0, 0}
\definecolor{niceblue}{rgb}{0,0,0.8}
\definecolor{red}{rgb}{1.0, 0, 0}
\hypersetup{colorlinks,bookmarksopen,bookmarksnumbered,
linkcolor=blus,pdfstartview=FitH,urlcolor=rossos,citecolor=verde}
\allowdisplaybreaks

\definecolor{rossos}{cmyk}{0,1,1,0.55}
\definecolor{rossoc}{cmyk}{0,1,1,0.2}
\definecolor{blu}{cmyk}{1,1,0,0.3}
\definecolor{blus}{cmyk}{1,1,0,0.6}
\definecolor{bluc}{cmyk}{1,1,0,0.1}
\definecolor{verde}{cmyk}{0.92,0,0.59,0.25}
\definecolor{verdec}{cmyk}{0.92,0,0.59,0.15}
\definecolor{verdes}{cmyk}{0.92,0,0.59,0.4}

\def\eq#1{{Eq.~(\ref{#1})}}
\def\eqs#1#2{{Eqs.~(\ref{#1})--(\ref{#2})}}
\def\fig#1{{Fig.~\ref{#1}}}

\def\Table#1{{Table~\ref{#1}}}

\def\sect#1{{Sect.~\ref{#1}}}

\def\app#1{{App.~\ref{#1}}}
\def\apps#1#2{{Apps.~\ref{#1}--\ref{#2}}}
\def\vev#1{\left\langle #1\right\rangle}
\def\abs#1{\left| #1\right|}

\def\diag{\mbox{diag}\,}
\renewcommand{\bar}{\overline}

\newcommand{\beq}{\begin{equation}}
\newcommand{\eeq}{\end{equation}}
\newcommand{\bea}{\begin{eqnarray}}
\newcommand{\eea}{\end{eqnarray}}

\renewcommand{\[}{\left[}

\renewcommand{\(}{\left(}
\renewcommand{\)}{\right)}

\newcommand{\mpl}{M_{\rm Pl}}
\DeclareMathOperator{\upp}{upp}

\newcommand{\Matrix}[1]{\left(\begin{smallmatrix}#1\end{smallmatrix}\right)}

\def\colorEW{\textcolor{blue}}
\def\colorPQ{\textcolor{red}}
\def\OUTER{\textcolor{teal}}
\newcommand{\parY}[2]{Y^{#1}_{#2}}
\newcommand{\vevv}[2]{\colorEW{v^{#1}_{\textcolor{black}{#2}}}}
\newcommand{\vevvc}[2]{\colorEW{v^{#1 *}}{}^{\textcolor{black}{ #2}}}
\newcommand{\vevV}[1]{\colorPQ{V}_{\!\textcolor{black}{#1}}}
\newcommand{\vevVc}[1]{\colorPQ{V^{\ast}}{}^{\textcolor{black}{#1}}}
\newcommand{\vevZ}[1]{\colorPQ{Z^{\textcolor{black}{#1}}}}
\newcommand{\vevZc}[1]{\colorPQ{Z^\ast_{\textcolor{black}{#1}}}}
\def\vevW{\colorPQ{W}}
\def\parA{\mathcal{A}}
\def\parB{\mathcal{B}}
\def\parBp{\mathcal{B}'}
\def\parC{\mathcal{C}}
\def\parD{\mathcal{D}}
\def\parAt{\tilde{\parA}}
\def\parBt{\tilde{\parB}}
\def\parCt{\tilde{\parC}}
\def\parDt{\tilde{\parD}}
\def\parEt{\tilde{\mathcal{E}}}
\def\parFt{\tilde{\mathcal{F}}}
\def\parGt{\tilde{\mathcal{G}}}
\def\YES{\not\to 0}
\def\NO{\to 0}
\def\LUV{\Lambda_{\rm UV}}
\def\EMPH{\emph}

\newcommand{\mymatrixs}[1]{
	\left(
	\begin{smallmatrix}
	#1
	\end{smallmatrix}
	\right)
}
\newcommand{\mymatrix}[1]{
	\begin{pmatrix}
	#1
	\end{pmatrix}
}

\makeatletter
\def\namedlabel#1#2{\begingroup
    #2%
    \def\@currentlabel{#2}%
    \phantomsection\label{#1}\endgroup
}
\makeatother

\begin{document}

\begin{center}  
%\vspace{3cm}
{\Large\bf\color{blus} High-quality Peccei-Quinn symmetry from the \\ 
\vspace{0.2cm}
interplay of vertical and horizontal gauge symmetries} \\
\vspace{1cm}

{\bf Luca Di Luzio$^{a}$, Giacomo Landini$^{b}$, Federico Mescia$^{c}$, Vasja Susič$^{c}$ }\\[7mm]

{\it $^a$Istituto Nazionale di Fisica Nucleare, Sezione di Padova, \\ 
Via F.~Marzolo 8, 35131 Padova, Italy}\\[1mm] 
{\it $^b$Instituto de Física Corpuscular  (IFIC), Universitat de Val\`{e}ncia-CSIC, \\ 
Parc Cient\'{i}fic UV, C/ Catedrático José Beltrán 2, E-46980 Paterna, Spain} \\[1mm] 
{\it $^c$Istituto Nazionale di Fisica Nucleare, Laboratori Nazionali di Frascati, \\ C.P.~13, 00044 Frascati, Italy}

\vspace{0.5cm}
\begin{abstract}
We explore a class of axion models where an accidental $\mathrm{U}(1)$ Peccei-Quinn (PQ) symmetry automatically emerges  from the interplay of vertical (grand-unified) and horizontal (flavor) gauge symmetries. 
We study a specific Pati-Salam realization in detail, and aim to generalize the conclusions. We show that our specific model offers protection from PQ-violating operators to high dimension, and demonstrate that the model can reproduce the Standard Model flavor structure. A distinctive feature of the vertical-horizontal setup is the presence of parametrically light fermions, known as anomalons, which are introduced to cancel the gauge anomalies of the flavor symmetry. We also identify a major challenge to building a fully realistic model, most notably that of Landau poles in gauge couplings before the Planck scale.
For the specific model investigated, the pre-inflationary PQ-breaking scenario predicts the axion mass window to be $m_a \in [2 \times 10^{-8}, 10^{-3}]\,\mathrm{eV}$. Conversely, a high-quality axion may be obtained instead in the post-inflationary scenario, with axion mass $m_a \gtrsim 0.01\,\mathrm{eV}$, and 
anomalon masses predicted below the $\mathrm{eV}$ scale. We elaborate on anomalons' cosmological production in the early universe, highlighting how measurements of $\Delta N_{\rm eff}$ could serve as a low-energy probe of the ultraviolet dynamics addressing the PQ quality problem.
\end{abstract}
\thispagestyle{empty}
\bigskip
\bigskip
\end{center}

\clearpage

\tableofcontents

\section{Introduction}
\label{sec:intro}
The axion solution to the strong CP problem \cite{Peccei:1977hh,Peccei:1977ur,Weinberg:1977ma,Wilczek:1977pj}, while elegant and potentially testable, raises the puzzling question 
of the origin of Peccei-Quinn (PQ) symmetry. In quantum field theories, global symmetries are not considered fundamental but are rather seen as accidental. 
While some axion models \cite{Kim:1979if,Shifman:1979if,Zhitnitsky:1980tq,Dine:1981rt} impose an effective $\U(1)_{\rm PQ}$ symmetry `by hand'', a robust PQ theory should preferably generate this symmetry in an automatic way \cite{Georgi:1981pu}.
Furthermore, the PQ symmetry needs to be of very ``high quality'', 
since even a tiny 
explicit 
breaking of the $\U(1)_{\rm PQ}$ 
from the ultraviolet (UV) completion 
would spoil the PQ solution to the strong CP problem.  
The latter constitutes the (in)famous PQ quality problem \cite{Georgi:1981pu,Dine:1986bg,Barr:1992qq,Kamionkowski:1992mf,Holman:1992us,Ghigna:1992iv}. 

Different approaches to the PQ quality problem  have been proposed  so far.
A prevalent strategy involves extending the Standard Model (SM) gauge symmetry, using either continuous or discrete local symmetries that are exact in the UV regime,  
so that the PQ symmetry arises as an accidental global symmetry, 
and is eventually broken by some higher-order effective operators
(see, for example, Refs.~\cite{Randall:1992ut,Dobrescu:1996jp,Babu:2002ic,Redi:2016esr,Fukuda:2017ylt,DiLuzio:2017tjx,Bonnefoy:2018ibr,Lillard:2018fdt,Gavela:2018paw,Lee:2018yak,Ardu:2020qmo,Yin:2020dfn,Contino:2021ayn,Yin:2024txg}). 
However, those models often display two unpleasant features: 
$i)$ the presence of extra gauge symmetries, with no relation to the SM structure and 
$ii)$ the lack of testability of the UV mechanism addressing the PQ quality problem. 
Ideally, the origin of the extended gauge symmetry,  required for the PQ symmetry to emerge accidentally,  should be more naturally linked to the SM structure. This could be achieved through either grand-unified extensions of the SM (vertical symmetries) or by gauging the SM flavor group (horizontal symmetries).

A different strategy, relying on the interplay between grand-unified and flavor gauge symmetries,  
was put forth in Ref.~\cite{DiLuzio:2020qio},\footnote{A conceptually 
similar approach, based on $\SU(5) \times \SU(3)_{\rm H}$, 
can be found in Refs.~\cite{Berezhiani:1985in,Berezhiani:1990wn} (see also \cite{Berezhiani:1983hm}). 
Note, however, that the latter model requires an extra 
symmetry in order to forbid certain PQ-breaking 
cubic operators in the scalar potential. 
The proposal in Ref.~\cite{DiLuzio:2020qio} builds instead upon \cite{Chang:1987hz} (see also \cite{Chang:1984ip}), which  
considered a \emph{global} flavor group, thus 
not addressing the PQ 
quality problem and resulting in significantly different phenomenological implications. Other approaches to the PQ quality problem that rely on grand unified theories (GUTs) have been discussed e.g.~in \cite{Vecchi:2021shj,Babu:2024udi}, while solutions based solely of 
flavor gauge symmetries have been explored in \cite{Darme:2021cxx,Darme:2022uzl}.} 
based on the gauge group $\SO(10) \times \SU(3)_f$, where $\SU(3)_f$ denotes the flavor group of SO(10), 
in which the three SM families (plus sterile neutrinos) 
are embedded into a $(16,\3)$ of $\SO(10) \times \SU(3)_f$. 
In this model, the accidental $\U(1)_{\rm PQ}$ 
arises from the interplay between the SO(10) and $\SU(3)_f$ symmetries, 
which furthermore forbid dangerous
PQ-breaking effective operators 
involving large-scale vacuum expectation values (VEVs), 
thus providing a solution to the PQ quality problem.
A remarkable feature inherent to the model is the 
presence of parametrically light fermion fields, 
also known as anomalons,  
introduced to cancel the gauge anomalies of the 
flavor symmetry. 
However, some aspects of the model, 
such as reproducing the SM flavor pattern and the unification constraints, 
as well as the phenomenology of the light anomalon fields 
were left unaddressed 
due to the  complexity of the $\SO(10)$ symmetry breaking. 

It is the purpose of the present study to fill 
some of the missing gaps mentioned above, but working with a simpler model which shares some structural analogies with the $\SO(10)$ model of Ref.~\cite{DiLuzio:2020qio}. 
The model presented here, based on the Pati-Salam \cite{Pati:1974yy} gauge group\footnote{For 
other Pati-Salam axion models, without accidental $\U(1)_{\rm PQ}$, 
see e.g.~\cite{Saad:2017pqj,DiLuzio:2020xgc}.} and the gauging of the $\SU(3)_{f_R}$ flavor group, 
turns out to be more tractable from the point of view of the symmetry breaking 
and the SM flavor pattern, and hence it can be analyzed in greater detail. 
A notable feature of this class of models 
consists in the non-trivial embedding of the 
axion field in the scalar multiplets of the extended
gauge symmetries. This establishes a connection of the 
axion decay constant with the $B-L$ and flavor breaking scales, 
thus leading to specific axion mass predictions,  
which reflects a broader strategy aimed at narrowing down axion mass ranges in GUTs (see e.g.~\cite{Ernst:2018bib,DiLuzio:2018gqe,FileviezPerez:2019ssf,FileviezPerez:2019fku}). 

An important aspect of our work consists in 
the calculation of the neutrino-anomalon spectrum 
and the investigation of the 
cosmological production of the anomalon fields in the early universe. Given that the anomalons are charged only under the
gauged flavor symmetry, they can be produced
via flavor interactions, neutrino mixing and 
effective Yukawa-like operators.  
Moreover, since the anomalons 
are expected to be lighter than the $\mathrm{eV}$ scale, 
at least for the most favorable
values of the PQ-breaking scale 
that address PQ quality, 
they contribute to the dark radiation 
of the universe. 
We hence discuss present bounds from Planck'18 \cite{Planck:2018vyg} 
and highlight how future measurements of $\Delta N_{\rm eff}$ could serve as a low-energy probe of the UV dynamics addressing the PQ quality problem.

Finally, the detailed analysis of the model also reveals shortcomings, which seem to present a broader challenge for models based on the interplay of vertical and horizontal symmetries. The proliferation of scalar degrees of freedom, due to non-trivial transformation under the flavor group, typically leads to a Landau pole at energies much lower than the Planck scale, 
which calls into question the effectiveness of the quality-protection mechanism. While the present work illustrates this issue, it ultimately remains unresolved, and presumably requires careful model building in an even more minimal setup (in terms of degrees of freedom).

The paper is structured as follows. 
In \sect{sec:PS-model} we 
present the Pati-Salam model with gauged $\SU(3)_{f_R}$ flavor group, which leads to the emergence of the 
accidental $\U(1)_{\rm PQ}$. 
This section also includes a detailed discussion 
of the axion embedding and explores the potential 
of the model for addressing the PQ quality problem.
\sect{sec:SMflavour} is devoted to show that the model 
can reproduce the SM flavor structure, 
including the neutrino sector that mixes with the 
SM-singlet anomalon fields. 
\sect{sec:axionpheno} explores the phenomenological 
consequences for axion physics, 
touching upon astrophysical limits, axion cosmology 
and direct searches. 
Next, in \sect{sec:anomalon-cosmology} we study the 
cosmological production of the anomalon fields
in the early universe and their potential detectability 
as a component of dark radiation. 
We conclude in \sect{sec:concl}, 
while more technical details about 
the derivation of 
axion couplings and 
the Pati-Salam model, the issue of Landau poles,
as well as the implications of the present study for 
the $\SO(10)$ model of Ref.~\cite{DiLuzio:2020qio}, 
are deferred to \apps{sec:loweaxion}{sec:SO10model}.

\section{A Pati-Salam model with accidental $\U(1)_{\rm PQ}$ \label{sec:PS-model}}

Our goal is to identify a concrete model, based on the interplay of vertical and horizontal gauge symmetries, 
which guarantees an accidental $\U(1)_{\rm PQ}$ that is also protected at the effective field theory (EFT) level. 
A model along those lines, based on the SO(10) group and a flavor $\SU(3)_{f}$ symmetry, 
was proposed in Ref.~\cite{DiLuzio:2020qio}. 
Some aspects of the latter model, 
such as reproducing the SM flavor pattern,  
were left unaddressed 
due to the complexity of the $\SO(10)$ symmetry breaking. 
Therefore, our objective here is to identify a relatively simpler model, which is more tractable from the point of view of the symmetry breaking 
and the SM flavor pattern. 
We will come back to the consequences 
of the present analysis on the $\SO(10)$ model of Ref.~\cite{DiLuzio:2020qio} in \app{sec:SO10model}. 

The model we investigate is based on the Pati-Salam group with gauged right-flavor, and its field content is displayed in Table \ref{tab:PSirrep}. It exhibits an accidental $\U(1)_{\rm PQ}$ protected also at the EFT level, as well as a consistent flavor breaking pattern. The model building choices that lead to its composition are elaborated upon in the points below:
\begin{itemize}[leftmargin=0.5cm,itemsep=0.0cm]
\item \underline{Vertical symmetry:}\\
    In the $\SO(10) \times \SU(3)_{f}$ model \cite{DiLuzio:2020qio}, the key to understand the protection of the $\U(1)_{\rm PQ}$ symmetry was the non-trivial action of the $\mathbb{Z}_4 \times \mathbb{Z}_3$ center of the gauge group. 
    It is hence natural to consider the possibility of replacing $\SO(10)$ with the Pati-Salam gauge group $G_{\rm PS} \equiv \SU(4)_{\rm PS} \times \SU(2)_L \times \SU(2)_R$, which also features a $\mathbb{Z}_4$ center in its first factor, although it operates in a fundamentally different way compared to $\SO(10)$.\footnote{The generator of the center for an $\SU(4)$ irreducible representation with a Dynkin label $[n_{1}n_{2}n_{3}]$ is $\exp(i 2\pi (n_{1}+2n_{2}-n_{3})/4)$, where the linear combination of $n_{i}$ corresponds to the difference in number of upper and lower fundamental indices when the representation is written as a tensor; its value $\mathrm{mod}(4)$ is known as quadrality~\cite{Slansky:1981yr}. The center of an irreducible representation (irrep) $[n_1 n_2 n_3 n_4 n_5]$ of $\SO(10)$, on the other hand, is generated by $\exp(i2\pi(2n_1+2n_3-n_4+n_5)/4)$ (using the convention of Slansky~\cite{Slansky:1981yr}), where an odd $\mathbb{Z}_{4}$ charge signifies the representation as spinorial.}
\item \underline{Horizontal symmetry:}\\
    In terms of the Pati-Salam group $G_{\rm PS}$, the SM fermion content is minimally embedded into 3 copies of $(4,2,1)$ and $(\overline{4},1,2)$, with each copy of the latter featuring also a right-handed neutrino. The global flavor group (its non-abelian part) of this setup is $\SU(3)_{f_L}\times \SU(3)_{f_R}$, corresponding to the $\SU(3)$ rotations of the SM fields contained in $Q_L$ and $Q_R$. The gauged horizontal group is then chosen as a subgroup of the global flavor group. One might be tempted into a sequential breaking of right-related parts (the vertical $\SU(2)_R$ and the flavor $\SU(3)_{f_{R}}$) at a high scale, followed by a breaking of left-related parts (the vertical $\SU(2)_L$ and a horizontal $\SU(3)_{f_{L}}$) at the electroweak (EW) scale. Such a setup is not viable for phenomenological reasons, however, since it implies unwanted EW-scale L-flavor gauge bosons. We conclude the horizontal group must be broken already at the high scale; for simplicity and in analogy with the $\SO(10)$ model \cite{DiLuzio:2020qio}, it is chosen to be $\SU(3)_{f}$, so the 
    center $\mathbb{Z}_{4}\times\mathbb{Z}_{3}$ is retained as well.
    \par
    There are multiple possibilities of embedding $\SU(3)_{f}\subset \SU(3)_{f_L}\times \SU(3)_{f_R}$, but only some may viably lead to an accidental PQ symmetry. A choice of embedding is equivalent to a choice of horizontal assignments under $\SU(3)_{f}$ for fermions in $(4,2,1)$ and $(\overline{4},1,2)$ of Pati-Salam; since we have three copies of each, $Q_L$ and $Q_{R}$ can each transform as a singlet, triplet or anti-triplet. 
    \par
    A left-right symmetric choice would be to embed the horizontal symmetry diagonally into the global flavor group, i.e.~where $Q_L$ and $Q_R$ transform as a $3$ (or equivalently both as $\overline{3}$). A Yukawa term $\overline{Q}_L Q_{R} \Phi$ would imply the scalar representation $\Phi$ that contains the SM Higgs to transform as either a singlet or an octet under the horizontal symmetry, allowing for a gauge-invariant term $\Phi^{2}$, which breaks PQ if $\Phi$ is charged under it. Without additional vector-like quarks as in the KSVZ-type axion models, $Q_{L}$ and $Q_{R}$ must be charged left-right asymetrically under PQ for it to be anomalous under QCD, and hence $\Phi$ must be charged as well. Therefore a left-right asymmetric choice of horizontal assignment is necessary for the emergence of the accidental PQ symmetry.
    \par
    Among the asymmetric horizontal assignments, we can choose either $Q_L$ and $Q_{R}$ to transform in the opposite way (one as a $3$ and one as a $\overline{3}$ of $\SU(3)_{f}$), or that only one transforms non-trivially. The first case is analogous to the $\SO(10)$ case, since $\overline{Q}_L,Q_{R}\subset \overline{16}$, but this makes the $\SU(3)_{f}^{3}$ anomaly twice as large as the second case. We choose for our model the second option and the manifestly left-right asymmetric choice $\SU(3)_{f}=\SU(3)_{f_{R}}$, and make the according horizontal assignments, cf.~Table~\ref{tab:PSirrep}. 
\item \underline{Anomalons:}\\
    The above discussion on horizontal symmetry showed the left-right symmetric choice --- the only choice free of the $\SU(3)_{f}^{3}$ gauge anomaly --- being incompatible with an accidental PQ. 
    This necessarily requires the introduction of new fermions --- dubbed anomalons and labeled $\Psi_R$ --- to facilitate the cancellation of gauge anomalies. Since $Q_{L}$ and $Q_{R}$ already by themselves cancel the anomalies associated to the PS group, it is most straightforward if the anomalons are PS singlets. 
    \par
    For our $\SU(3)_{f_{R}}$ choice of horizontal symmetry, 
    the $\SU(3)_{f_{R}}^{3}$ anomaly is most simply canceled by introducing $8$ copies of anomalons transforming as $\overline{\3}$ under flavor, which cancel the anomaly from $8$ flavor-triplets in $Q_{R}$. The only alternative that does not overcompensate the anomaly from Pati-Salam multiplets is one copy of $\overline{3}$ and one copy of $\overline{6}$.\footnote{
        For an $\SU(3)^{3}$ triangular anomaly involving the representation $R$ of fermions, the anomaly coefficient $\kappa_R$ in \hbox{$\mathrm{Tr}(\{T_{a},T_{b}\}T_{c})=\kappa_{R}\,d_{abc}$} has values \hbox{$\kappa_{\overline{3}}=-1/2$} and \hbox{$\kappa_{\overline{6}}=-7/2$}, hence $7$ anti-triplets can be replaced with one anti-sextet. This can be confirmed by using the \texttt{TriangularAnomalyValue} function in \texttt{GroupMath}~\cite{Fonseca:2020vke}, which however switches conventions in
        labeling $6$ and $\overline{6}$. The next smallest representation that can contribute with a negative value is $\overline{10}$ with $\kappa_{\overline{10}}=-27/2$, already overcompensating the contribution of $4$ from standard fermions in our case.
    }
    Although the $\overline{3}+\overline{6}$ alternative features less new fermionic degrees of freedom, our choice of only $\overline{3}$s is simpler to analyze, since less types of (non-renormalizable) operators will have to be considered in the Yukawa sector.   
    \par
    Finally, let us note that there is a remnant $\U(1)_{\rm PQ}\times \SU(3)_{f_R}^2$ anomaly. In order for the axion to relax to zero the $\theta$ term of QCD (and not the one of $\SU(3)_{f_R}$), $\SU(3)_{f_R}$ should be spontaneously broken (effectively suppressing the contribution of $\SU(3)_{f_R}$ instantons to the axion potential due to the Higgsing of $\SU(3)_{f_R}$). This is consistent with the requirement that the flavor symmetry must be completely broken in order to give mass to SM fermions. 
    \item \underline{The scalar sector:}\\
    Once our choice of the gauge group and fermions has been made, there is surprisingly little freedom for having a realistic scalar sector, as we now discuss.
    \par
    For a realistic Yukawa sector, at least two different types of scalar representations $\Phi$ and $\Sigma$ are necessary: each will contain an admixture of the SM Higgs and will come with a different Clebsch coefficient. The necessary presence of renormalizable Yukawa operators $\overline{Q}_L Q_{R}\Phi$ and $\overline{Q}_L Q_{R}\Sigma$ leads to unique assignments of $\Phi$ and $\Sigma$ under both the vertical and horizontal gauge groups, cf.~Table~\ref{tab:PSirrep}.
    Furthermore, a suitably high Majorana mass is given to right-handed neutrinos via the operator $Q_{R}Q_{R}\overline{\Delta}$, which again essentially enforces a unique irrep assignment for $\Delta$. 
    \par
    The second requirement from the scalar sector is that it facilitates symmetry breaking from the Pati-Salam to the EW scale. We know the already introduced $\Delta$ participates, but another complex field $\chi$ must be introduced due to also breaking the rank of the group: both the $\mathrm{U}(1)_{B-L}$ of $\SU(4)_{\rm PS}$ and the accidental $\U(1)_{\rm PQ}$ symmetry at high energies~\cite{Mohapatra:1982tc} must be broken well above the EW scale, hence at least 2 complex irreps are required. Our choice of $\chi$ has been inspired by the presence of a scalar $16$ in the $\SO(10)$ model; it essentially the smallest choice containing a SM singlet, but it is not unique. 
    \par
    Finally, we introduce also an ``auxiliary'' real scalar $\xi$.
    Its function is to connect $\Delta$ and $\chi$ in the renormalizable scalar potential. This is crucial for two reasons: it prevents an enlarged global symmetry of the potential, and its presence generates only one accidental $\U(1)_{\rm PQ}$. The enlarged symmetry we are referring to would be 
    a global double copy of $G_{\text{PS}}\times \SU(3)_{f_{R}}$, with $\Delta$ and $\chi$ each transforming independently under one copy. This would result in massless but physical Goldstone modes from high-scale breaking of this enlarged symmetry. The gauge assignment of $\xi$ is tied to the assignment of $\chi$.
    \item \underline{General model building remarks:}\\
    The above points discuss the model building constraints of various aspects of the model we study. Apart from several ingredients when alternative choices are possible, the constraints predominantly determine the model contents. The proposed model thus seems one of a small number of minimalistic possibilities, and it is in that sense well motivated.  
    \par
    The largest degree of arbitrariness is seemingly in the scalar sector, in which $\chi$ and $\xi$ could be replaced by a different set of fields. Note, however, that arbitrary choices that yield the necessary symmetry breaking to the SM group generically do not provide an accidental $\mathrm{U}(1)_{\text{PQ}}$, let alone guarantee the protection of its quality. Heuristically, 
    the number of operators of a fixed dimension rises quickly with the number of representation types introduced into the scalar sector; PQ quality is thus increasingly harder to achieve with larger representation sets. This greatly restricts the number of viable models, even when forgoing any consideration of minimality.
\end{itemize}

\begin{table}[htb]
$$\begin{array}{c|c|cc|cc|c|c}
\rowcolor[HTML]{C0C0C0} 
\hbox{Field} & \hbox{Lorentz} &  
\text{Pati-Salam}
& 
\mathbb{Z}_4 
& \SU(3)_{f_R}  & 
\mathbb{Z}_3 
& \text{Generations} 
& \U(1)_{\rm PQ} \\ \hline
Q_L & (1/2,0) & (4,2,1) & +i & \1 & +1 & 3 & +3 \\ 
Q_R & (0,1/2) & (4,1,2) & +i & \3 & e^{i 2\pi/3} & 1 & +1 \\ 
\rowcolor{CGray} 
\Psi_{R} & (0,1/2) & (1,1,1) & +1 & \overline{\3} & e^{i 4\pi/3} & 8 & +2 \\
\hline
\Phi & (0,0) & (1,2,2) & +1 & \overline{\3} & e^{i 4\pi/3} & N_{\Phi} \geq 1 & +2 \\ 
\Sigma & (0,0) & (15,2,2) & +1 & \overline{\3} & e^{i 4\pi/3} & N_{\Sigma} \geq 2 & +2 \\ 
\Delta & (0,0) & (10,1,3) & -1 & \6 & e^{i 4\pi/3} & 1 & +2 \\ 
\chi & (0,0) & (4,1,2) & +i & \overline{\3} & e^{i 4\pi/3} & 1 & -1 \\ 
\xi & (0,0) & (15,1,3) & +1 & \1 & +1 & 1 & \phantom{+}0 \\ 
\end{array}$$
\caption{Field content of the Pati-Salam model and relative transformation properties under the Lorentz group, $\SU(4)_{\rm PS} \times \SU(2)_L \times \SU(2)_R \times \SU(3)_{f_R}$, its $\mathbb{Z}_4 \times \mathbb{Z}_3$ center, and the accidental $\U(1)_{\rm PQ}$. Exotic fermions, which ensure $\SU(3)_{f_R}$ anomaly cancellation, are highlighted in light gray. Note that the $\U(1)_{\rm PQ}$ charge of $\Psi_R$ is fixed by non-renormalizable operators, while $\xi$ is not charged under $\U(1)_{\rm PQ}$ being a real scalar field. 
}
\label{tab:PSirrep}
\end{table}%
 
Having established the field content of the model, we can investigate its renormalizable Lagrangian. We first write down explicitly the renormalizable Yukawa terms allowed by gauge invariance:
\begin{align}
-\mathcal{L}_Y&=\sum_{\alpha=1}^{N_{\Phi}}\sum_{I=1}^{3} \parY{\Phi^{\alpha}}{I}\,(\bar{Q}_{L})^I Q_R\;\Phi^\alpha
    +\sum_{\alpha=1}^{N_{\Sigma}}\sum_{I=1}^{3} 2\sqrt{3}\,\parY{\Sigma^{\alpha}}{I}\,(\bar{Q}_{L})^I Q_R\;\Sigma^{\alpha} \nonumber \\
    &\quad +\frac{1}{2}\; Y_R\, Q_R Q_R\,\Delta^*+ \text{h.c.} \, , 
    \label{eq:Yukawa-sector}
\end{align}
where gauge contractions are understood, $\parY{\Phi,\, \Sigma}{}$ are 3-dimensional complex vectors (with $I$ spanning over $Q_L$ families) 
for each copy of $\Phi$ or $\Sigma$ (labeled by index $\alpha$), 
and $Y_R$ is a complex number. Multiple copies of the $\Phi$- and $\Sigma$-type irrep are generically introduced to facilitate a realistic description of SM fermion masses and mixings. We argue in \sect{sec:SMflavour} that the minimal realistic case requires $(N_{\Phi},N_{\Sigma})=(1,2)$. From now on we focus on this minimal scenario and simplify the notation: we denote the irreps by $\Phi$, $\Sigma$ and $\Sigma'$, and the associated Yukawa couplings as $\parY{\Phi}{I}$, $\parY{\Sigma}{I}$ and $\parY{\Sigma'}{I}$, respectively.

Note that because of the action of the $\SU(3)_{f_R}$ flavor symmetry, it is not possible in \eq{eq:Yukawa-sector} to 
write invariants featuring the conjugate of $\Phi$ and $\Sigma$ fields, which is at the origin of the accidental 
$\U(1)_{\rm PQ}$ symmetry, inherited by the anomalous 
rephasing of the matter fields $Q_L$ and $Q_R$. 

Observe also that the Yukawa sector of \eq{eq:Yukawa-sector} does not involve $\Phi_{R}$, implying that anomalons are massless at the renormalizable level. This remarkable feature is shared with the $\SO(10) \times \SU(3)_{f}$ model \cite{DiLuzio:2020qio}, and is closely connected to the existence of the accidental $\U(1)_{\rm PQ}$. It is non-renormalizable Yukawa operators that lift the mass of anomalons $\Psi_R$, hence this observable represents a low-energy signature of our setup. We work out how their mass arises in \sect{sec:SMflavour-NandA}, while phenomenological consequences are discussed in \sect{sec:anomalon-cosmology}.   

We now turn to the scalar potential of the model. It is crucial to check whether the $\U(1)_{\rm PQ}$ survives in the renormalizable scalar potential, $\V = \V_2 + \V_3 + \V_4 + \V_{\mathbb{C}}$, whose terms read schematically 
\begin{align}
    \V_{2}&= |\Phi|^2+|\Sigma|^2+|\Delta|^2+|\chi|^2+\xi^2 \, , \label{eq:V2}\\
    \V_{3}&= \xi \(|\Sigma|^2+|\Delta|^2+|\chi|^2+\xi^2\) \, , \\
    \V_{4}&= \(|\Phi|^2+|\Sigma|^2+|\Delta|^2+|\chi|^2+\xi^2\)^2 \, , \\
    \V_{\mathbb{C}}&= \Phi\Sigma^\ast \xi + \Phi\Sigma^* \(|\Sigma|^2+|\Delta|^2+|\chi|^2+\xi^2\)+\Sigma^{*2} \(\Phi^2+\Delta^2\)+ \Delta\chi^2\xi + \text{h.c.} \, , 
    \label{eq:VC}
\end{align}
where each term specifies only field powers, while the contraction of gauge indices is suppressed. Note also that for a given term there may be multiple independent index contractions, i.e.~multiple independent invariants. We have suppressed the multiplicity of $\Phi$ and $\Sigma$ in the notation; in particular, each appearance of $\Sigma$ may be replaced by $\Sigma'$.

The accidental global symmetries 
can be read off the last set of terms, $\V_{\mathbb{C}}$, featuring complex operators. 
Including also the Yukawa terms from \eq{eq:Yukawa-sector}, 
a single accidental $\U(1)_{\rm PQ}$ emerges\footnote{
 The emergence of only a single global $\U(1)$ is crucial. Note that two critical invariants for counting accidental $
 \U(1)$'s, namely $\Sigma^{*2} \Delta^2$ and $\Delta \chi^2 \xi$, vanish on the vacuum. We clarify the subtle point of why this does not increase the number of accidental $\U(1)$ charges beyond one in \app{app:number-of-U1s}.
} 
in the renormalizable Lagrangian 
with the charge assignments (up to overall normalization) 
reported in Table~\ref{tab:PSirrep}. 
Since $\xi$ is a real scalar field, it has zero PQ charge.

Finally, we assign the PQ charge of the anomalon fields $\Psi_R$
based on lowest-dimensional (non-renormalizable) Yukawa operators 
in which they appear, so that the disturbance of PQ symmetry from that sector is minimized. In particular, all dimension $5$ Yukawa operators in which $\Psi_R$ appear preserve PQ, cf.~Table~\ref{tab:Yukawa-nonrenormalizable}.  Furthermore, anomalons are fermions and Pati-Salam singlets, so the PQ charge assignment of $\Psi_{R}$ has no impact on our subsequent study, such as the anomaly coefficients $E$ and $N$ in \sect{sec:axionembedding} or the PQ quality analysis in \sect{sec:PQquality}.

%%%%%%%%%%%%
\subsection{Symmetry breaking pattern}

In the following, we assume that the scalar potential allows for the symmetry breaking pattern 
\begin{align}
\label{eq:PSbreakpatt}
& \ \SU(4)_{\rm PS} \times \SU(2)_L \times \SU(2)_R \times \SU(3)_{f_R} \times \U(1)_{\rm PQ} \nonumber \\
& \xrightarrow[]{\langle\Delta,\chi\rangle_{3},\langle \xi\rangle}  
\SU(3)_c \times \SU(2)_L \times \U(1)_{Y} \times \SU(2)_{f_R} 
\nonumber \\
& \xrightarrow[]{\langle\Delta,\chi\rangle_{1,2}}  
\SU(3)_c \times \SU(2)_L \times \U(1)_{Y} \nonumber \\
& \xrightarrow[]{\langle \Phi,\Sigma,\Sigma'\rangle}
\SU(3)_c \times \U(1)_{\rm EM}
\, , 
\end{align}
such that the representations $\Delta$, $\chi$ and $\xi$ from \Table{tab:PSirrep} are involved in the breaking of Pati-Salam and flavor $\SU(3)_{f_{R}}$, while the weak doublets in $\Phi$, $\Sigma$ and $\Sigma'$ are responsible for the breaking of EW symmetry in the SM. 
We recall that the fields $\Delta$ and $\chi$ need to be 
simultaneously present in order to fully break the PQ symmetry. In particular, in the 
hierarchical limit $\vev{\Delta} \gg \vev{\chi}$, there is an approximate  PQ symmetry, which is a linear combination of the original PQ and the broken $B-L$ generator, and such remnant PQ symmetry is eventually broken by $\vev{\chi}$ (for more details, see Ref.~\cite{DiLuzio:2020qio}). 

In the following, we will hence consider the 
following hierarchy among VEVs:
\begin{align}
    \langle \Delta\rangle,
    \langle\chi\rangle,
    \langle \xi\rangle
    \gg
    \langle \Phi\rangle,
    \langle \Sigma\rangle,
    \langle \Sigma'\rangle.
\end{align}
As a shorthand notation, we shall denote the ``large'' and ``small'' VEVs as $V$ and $v$, respectively.  

Since scalar representations also transform under flavor, which is eventually completely broken, they obtain VEVs in multiple entries. Namely, in flavor space we denote the high scale VEVs as
\begin{align}
    \langle \Delta \rangle &= V_{\Delta^{AB}} \equiv \colorPQ{Z}^{AB} \,,\quad 
    \langle \chi \rangle = V_{\chi_{A}} \equiv \colorPQ{V}_{A}\,,\quad
    \langle \xi \rangle = V_{\xi},
\end{align}
where $A$ and $B$ are flavor indices and $\Delta$ is symmetric in $AB$. We use the \colorPQ{red} color to denote the flavor-dependent high-scale VEVs.

Note that in Eq.~\eqref{eq:PSbreakpatt} we assumed an (optional) little hierarchy in the flavor dependent VEVs, such that the $\SU(2)_{f_{R}}$ singlets $\{\colorPQ{Z}^{33},\colorPQ{V}_{3},V_{\xi}\}$ are slightly larger than the other VEV entries, which may be beneficial for explaining flavor hierarchies (hence the subscript $3$ under $\langle\Delta,\chi\rangle$ denotes the first stage of flavor breaking). The complete breaking of the $\SU(3)_{f_R}$ flavor group gives rise to 8 massive flavor gauge bosons, which we denote generically with $W_{f_R}$, leaving flavor and Lorentz indices implicit. The precise spectrum depends on the details of symmetry breaking. Within the assumption of sequential breaking, $\SU(3)_{f_R}\to \SU(2)_{f_R}\to 1$, we expect some hierarchy between the masses. Naively, the 
   masses scale as $m_{W_{f_R}}\sim g_{f_R} V$ in terms of the flavor breaking scale and the flavor gauge coupling $g_{f_R}$. 

We postpone the discussion on EW VEVs in $\Phi$, $\Sigma$ and $\Sigma'$ until \sect{sec:SMflavour-QandL}, while technical comments on the tensor notation for all irreps are gathered in \app{app:procedure-explicit}.

%%%%%%%%%%%
\subsection{Axion embedding and key properties}
\label{sec:axionembedding}

To identify the axion field, we follow a derivation similar to the one in Ref.~\cite{DiLuzio:2020qio}. 
Consider the classically conserved currents 
(keeping only the scalar components)
\begin{align}
\label{eq:JPQPhi}
J_\mu^{\rm PQ} &= -\sum_\phi
q_{\phi} \phi^\dag i \overset{\leftrightarrow}{\partial_\mu} \phi \, , \\
\label{eq:JBmLPhi}
J_\mu^{B-L} &= 
-\sum_\phi
(B-L)_{\phi} 
\phi^\dag i \overset{\leftrightarrow}{\partial_\mu} \phi \, , 
\end{align}
where $\phi \in \{ \chi, \Delta , \ldots \}$ denotes a generic scalar multiplet charged under 
both $B-L$ and the PQ symmetry. The PQ charges, $q_\phi$, are defined in 
the PQ-broken phase, 
and they generally differ from the PQ charges in the 
unbroken phase, 
which are reported in the last column of \Table{tab:PSirrep}. 
The physical axion field spans 
over EM-singlet 
polar components 
of $\phi$, 
defined via 
\beq 
\label{eq:defphiax}
\phi \supset V_{\phi} 
\exp\left(\frac{i\,a_{\phi}}{\sqrt{2}\abs{V_{\phi}}}\right) \, ,  
\eeq
so that given the kinetic term $\partial_\mu \phi^\dag \partial^\mu \phi$, the orbital mode $a_\phi$ is 
canonically normalized. 
Expanding the conserved currents
in terms of $a_\phi$ components, we obtain\footnote{Here, we 
keep only the contribution of polar components associated to 
large-scale VEVs. The generalization including the 
contribution of EW VEVs is discussed 
in \app{sec:loweaxion}.} 
\begin{align}
\label{eq:JPQ}
J_\mu^{\rm PQ} &= 
q_{\chi} \sum_{A} \sqrt{2} |\colorPQ{V}_{A}| \partial_\mu (a_\chi)_A + 
q_{\Delta} \sum_{AB} \sqrt{2} |\colorPQ{Z}^{AB}| \partial_\mu (a_\Delta)^{AB} \, , \\
\label{eq:JBmL}
J_\mu^{B-L} &= 
(B-L)_{\chi}
\sum_{A} \sqrt{2} |\colorPQ{V}_{A}| \partial_\mu (a_\chi)_A + 
(B-L)_{\Delta} 
\sum_{AB} \sqrt{2} |\colorPQ{Z}^{AB}| \partial_\mu (a_\Delta)^{AB} \, , 
\end{align}
where $(B-L)_{\chi} = -1$ and $(B-L)_{\Delta} = -2$. 
The $q$ charges instead can be expressed 
as a linear combination of the original PQ charges in \Table{tab:PSirrep} 
and the broken (Cartan) gauge generators\footnote{Since 
the VEVs have zero hypercharge and $Y = T^3_R + (B - L)/2$, 
it is not necessary to include the $T^3_R$ generator.} 
\beq 
\label{eq:generalUVIRPQ}
q = c_1 \PQ + c_2 (B-L) \, ,  
\eeq
and they 
can be fixed by requiring that the PQ 
and $B-L$ currents are orthogonal:
\beq 
\label{eq:orthogPQBmL}
q_{\chi} 
V_\chi^2
+ 2 q_{\Delta} 
V_\Delta^2
= 0 \, ,  
\eeq
where we introduced the norms of the VEVs in flavor space 
\beq
V_{\chi}^2 = \sum_{A} |\colorPQ{V}_{A}|^2 \, , \qquad
V_{\Delta}^2 = \sum_{AB} |\colorPQ{Z}^{AB}|^2 \, .
\eeq
In particular, combining \eq{eq:generalUVIRPQ} and 
\eq{eq:orthogPQBmL} we obtain 
\beq
\label{eq:defqchic1}
q_{\chi} = c_1 \frac{-8 V_\Delta^2}{V_\chi^2 + 4 V_\Delta^2} \, , 
\qquad 
q_{\Delta} = c_1 \frac{4 V_\chi^2}{V_\chi^2 + 4 V_\Delta^2} \, .
\eeq
The condition in \eq{eq:orthogPQBmL} 
ensures no kinetic mixing between 
the axion field and the $B-L$ massive gauge boson,
thus providing a canonical axion field, 
defined as 
\cite{Srednicki:1985xd}
\beq 
\label{eq:axiondef}
a = \frac{1}{V_a} \( q_{\chi} \sum_{A} |\colorPQ{V}_{A}| (a_\chi)_A + 
q_{\Delta} \sum_{AB} |\colorPQ{Z}^{AB}| (a_\Delta)^{AB} \) \, ,
\eeq
with 
\beq 
\label{eq:Vnorm}
V_a^2 = q^2_{\chi} 
V_\chi^2
+ q^2_{\Delta} 
V_\Delta^2 
= c_1 \frac{16 V_\chi^2 V_\Delta^2}{V_\chi^2 + 4 V_\Delta^2}
\, , 
\eeq
where in the last step we employed \eq{eq:defqchic1}.
In this way, 
$J_\mu^{\rm PQ} = \sqrt{2} V_a \partial_\mu a$, 
so that $\langle 0 | J_\mu^{\rm PQ} | a \rangle = i \sqrt{2} V_a p_\mu$, 
compatibly with the Goldstone theorem.   
Moreover, under a PQ transformation acting on the polar 
field components, 
$(a_\chi)_A \to (a_\chi)_A + \kappa \, q_\chi \sqrt{2} |\colorPQ{V}_{A}|$ 
and 
$(a_\Delta)^{AB} \to (a_\Delta)^{AB} 
+ \kappa \, q_\Delta \sqrt{2} |\colorPQ{Z}^{AB}|$, 
the canonical axion field, as defined in \eq{eq:axiondef},
transforms as $a \to a + \kappa \, \sqrt{2} V_a$. 
Finally, inverting the orthogonal transformation in \eq{eq:axiondef}, 
one readily 
obtains the 
projection of the angular modes on the axion field 
\beq 
\label{eq:aproj}
(a_{\chi})_A \to q_{\chi} |\colorPQ{V}_{A}| \frac{a}{V_a} \, , \qquad 
(a_\Delta)^{AB} \to q_{\Delta} |\colorPQ{Z}^{AB}| \frac{a}{V_a} \, . 
\eeq 
Note that the coefficient $c_1$ in \eq{eq:Vnorm} 
has yet to be determined.
Defining the anomalies of the PQ current as 
\beq 
\label{eq:defanomJPQ}
\partial^\mu J^{\rm PQ}_{\mu} =
\frac{\alpha_s N}{4\pi} 
G \tilde G
+ \frac{\alpha_{\rm EM} E}{4\pi}  
F \tilde F
\, , 
\eeq
the value of $c_1$ can be fixed by matching the PQ-QCD$^2$ anomaly 
between the UV and the IR theory, that is 
\begin{align} 
\label{eq:NanomalyUV}
N &= 3 \times T(3) \( 2 \, \text{PQ} (Q_L) - 2 \,  \text{PQ} (Q_R) \) = 6  \quad (\text{UV}) \, , \\
\label{eq:NanomalyIR}
N &= 3 \times  T(3) \( 2 \, q(q_L) - 2 \,  q(q_R) \) = 6 c_1 \quad (\text{IR}) \, ,
\end{align}
where we used $T(3) =1/2$, $\text{PQ}(Q_L)=+3$, $\text{PQ}(Q_R)=+1$,  
$q(q_L) = c_1 ( 3 - 1/3 \cdot (V_\chi^2 - 4 V_\Delta^2) / (V_\chi^2 + 4 V_\Delta^2) )$ 
and 
$q(q_R) = c_1 ( 1 - 1/3 \cdot (V_\chi^2 - 4 V_\Delta^2) / (V_\chi^2 + 4 V_\Delta^2) )$.  
Therefore, $c_1 = 1$ by anomaly matching, 
and the axion decay constant is
\beq 
\label{eq:axiondc}
f_a \equiv \frac{\sqrt{2} V_a}{2N} =
\frac{V_{\chi} V_{\Delta}}{3\sqrt{V^2_{\chi} + 4 V^2_{\Delta}}}
\, . 
\eeq
The PQ-QED$^2$ anomaly factor is found to be $E=16$, 
so that $E/N = 8/3$, which yields the same axion-photon 
coupling as in the DFSZ model \cite{Zhitnitsky:1980tq,Dine:1981rt}. 

The axion couplings to matter fields can be obtained by including in the previous analysis also the EW VEV components, and orthogonalizing the PQ current with respect to both the $B-L$ and hypercharge gauge currents, 
as shown in \app{sec:loweaxion}. 
In particular, we obtain 
\beq 
\label{eq:cucdce}
c_u = \frac{1}{3} \cos^2\beta \, , \qquad 
c_d = c_e = \frac{1}{3} \sin^2\beta \, , 
\eeq
where the axion couplings to fermions are defined in 
\eq{eq:defcf} and the vacuum angle, 
$\beta$, 
given in \eq{eq:deftanbeta}, 
is a function of the EW VEVs. 
Formally, the axion couplings to SM fields 
are like those in the DFSZ model, 
although the perturbativity range of $\beta$ 
depends on the details of the fit to the 
SM flavor structure. 
On the other hand, given the structures of the SM fermion 
mass matrices in \eqs{eq:MUvevs}{eq:MEvevs}, we expect the 
perturbativity range of $\beta$ to be similar to the one 
in the standard DFSZ model, that is $\tan\beta \in [0.25, 170]$ \cite{DiLuzio:2020wdo}.

\subsection{Quality of the PQ symmetry}
\label{sec:PQquality}
The $\U(1)_{\rm PQ}$ symmetry emerges accidentally within the renormalizable Lagrangian, as discussed earlier. Beyond this, one must consider potential sources of PQ symmetry breaking in the ultraviolet (UV) regime, typically characterized by effective operators suppressed by a cut-off scale $\LUV$. In this context, we assume $\LUV\sim\mpl=1.22 \times 10^{19}\,\mathrm{GeV}$, motivated by the possible connection between PQ breaking and quantum gravity effects at the Planck scale.

A straightforward estimate indicates that $\U(1)_{\rm PQ}$ should remain intact for operators up to dimension $d=9$ under the assumption $f_a\sim 10^9\,\mathrm{GeV}$. This requirement ensures that the energy density arising from UV-induced PQ breaking remains approximately $10^{-10}$ times smaller than the energy density associated with the QCD axion potential:
\begin{equation}\label{eq:PQquality}
    \left(\frac{f_a}{\LUV}\right)^{d-4}f_a^4\lesssim 10^{-10}\chi_{\rm QCD}^4 \, , 
\end{equation}
with $\chi^4_{\rm QCD} = m_u m_d / (m_u+m_d)^2 m_\pi^2 f_\pi^2 \simeq (76 \, \mathrm{MeV})^4$, 
so that the induced axion VEV displacement from zero is $\theta_{\rm eff}\equiv\langle{a\rangle}/f_a\lesssim10^{-10}$, within the bound set by the 
non-observation of the 
neutron electric dipole moment (EDM) \cite{Abel:2020pzs}. 

\begin{figure*}[t!]
    \includegraphics[width=0.485\textwidth]{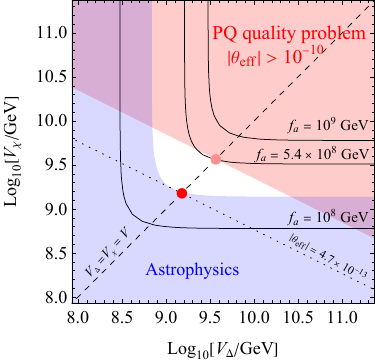} 
    \hspace{0.05\linewidth}
    \includegraphics[width=0.45\textwidth]{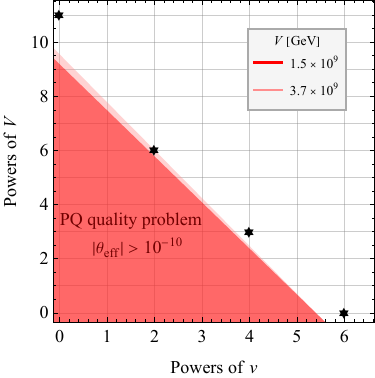}
     \caption{\textit{Left panel:} 
     Iso-contours of $f_a$ in the $(V_\Delta, V_\chi)$ 
     plane.   
     The red area signals the region of parameter space 
     where the PQ quality problem arises, with 
     $|\theta_{\rm eff}|>10^{-10}$. The blue area is disfavored by astrophysical bounds, $f_a\gtrsim 2.3\times 10^8\,\mathrm{GeV}$ (cf.~\sect{eq:astrolimits}). Along the dashed line, $V_\Delta=V_\chi\equiv V$ holds. Along the dotted line, $|\theta_{\rm eff}|$ acquires its minimal value  compatible with astrophysical constraints. \textit{Right panel:} Contribution to $|\theta_{\rm eff}|$ from operators whose projection on the VEVs is of the type $v^m V^n/\LUV^{m+n-4}$, assuming $V_\Delta=V_\chi=V=3\sqrt{5}f_a$.  
      In the red regions, $|\theta_{\rm eff}|>10^{-10}$ for the values of $V$ in the legend, which correspond color-wise to the red dots in the left panel. The black stars correspond to operators generated in our model which impose the strongest conditions for PQ quality,
      cf.~\eq{eq:contributions-axion}. 
      }
    \label{fig:plotfa}
\end{figure*}

In general, the PQ-breaking axion potential receives contributions both from high-scale ($V_\Delta$ and $V_\chi$) and EW 
scale ($v$) VEVs, and \eq{eq:PQquality} generalizes to 
\begin{align}
       \frac{v^m \;V_{\Delta}^{l} \;V_\chi^{n-l}}{\LUV^{n+m-4}}
    &\lesssim 10^{-10}\chi_{\rm QCD}^4 
    &
    \xrightarrow{V_\chi \sim V_\Delta \sim V}
    &&
    \frac{v^m \; V^n}{\LUV^{n+m-4}}
    &\lesssim 10^{-10}\chi_{\rm QCD}^4 \, , 
    \label{eq:PQqualityGen}
\end{align}
where the high-scale VEVs 
are related to $f_a$ via \eq{eq:axiondc}. Although only an admixture of the full EW VEV will be present in any 
$v$-type VEV, we can conservatively take the maximal value $v=246/\sqrt{2}\,\mathrm{GeV}$. 

As discussed later in this section, our main result is that the model has dominant contributions to the axion potential of the form
\begin{align}
    &&
    \frac{V_{\Delta}^{5}\, V_{\chi}^{6}}{\LUV^{7}}\, ,&&
    \frac{v^{2}\,V_{\Delta}^{2}\, V_{\chi}^{4}}{\LUV^{4}}\, ,&&
    \frac{v^{4}\,V_{\Delta}\, V_{\chi}^{2}}{\LUV^{3}}\, ,&&
    \frac{v^{6}}{\LUV^{2}} 
    \label{eq:contributions-axion-refined}\\[6pt]
    \xrightarrow{V_\chi \sim V_\Delta \sim V}&&
    \frac{V^{11}}{\LUV^{7}}\, ,&&
    \frac{v^{2}\,V^{6}}{\LUV^{4}}\, ,&&
    \frac{v^{4}\,V^{3}}{\LUV^{3}}\, ,&&
    \frac{v^{6}}{\LUV^{2}}\, ,
    \label{eq:contributions-axion}
\end{align}
which originate from the operators (and their Hermitian conjugates)
\begin{align}
    \Delta^{4}\Delta^{\ast} \chi^{*6}\, ,&&
    \Phi^{2-k}\Sigma^{k}\Delta^2\chi^{*4}\, ,&&
    \Phi^{4-k}\Sigma^{k}\Delta\chi^{*2}\, ,&&
    \Phi^{4-k}\Sigma^{k}\Sigma^2\, , \label{eq:operators-dominant-quality-contribution}
\end{align}
respectively, for at least one integer value $k$. 
Given the values of $v$ and $\LUV=\mpl$ stated above, the $d=8$ contribution in \eq{eq:contributions-axion-refined} (second one from the left) 
turns out to be the dominant one.

In the left panel of \fig{fig:plotfa}, we plot in red 
the region in the $V_{\Delta}$--$V_{\chi}$ plane where PQ quality is spoiled according to \eq{eq:PQqualityGen},
with the leading contribution coming from $(m,l,n-l)=(2,2,4)$, i.e., the dimension $8$ operator in \eq{eq:operators-dominant-quality-contribution}. The constraint from astrophysics (discussed in \sect{eq:astrolimits}) on the other hand demands $f_a \gtrsim 2.3\times 10^8\,\mathrm{GeV}$, computed from the VEVs via \eq{eq:axiondc}, with the excluded region 
shown in blue. The allowed region which also solves the PQ quality problem is thus the white remainder, and 
the configuration optimizing the solution to the PQ quality problem is close to the diagonal line $V_\chi \sim V_\Delta \sim V \sim f_a$ and lies on the astrophysics boundary; it is marked with a darker red dot. The upper limit of the diagonal line is $f_{a} \lesssim 5.4\times 10^{8}\,\mathrm{GeV}$ and is marked by a lighter red dot. Off the dashed diagonal the admissible $f_{a}$ can be raised slightly further to $f_a=5.6\times 10^8\,\mathrm{GeV}$, corresponding to an unmarked point $(V_\chi,V_\Delta)\simeq (4.1,2.9)\times 10^9\,\mathrm{GeV}$. 

The minimal value of $|\theta_{\rm eff}|$ sourced by the $d=8$ operator and allowed by astrophysical bounds is $|\theta^{\rm min}_{\rm eff}|\simeq4.7\times 10^{-13}$, 
shown as a dotted line in the left panel of~\fig{fig:plotfa}. This could be tested by future experiments measuring the 
neutron \cite{nEDM:2019qgk,n2EDM:2021yah} 
and proton \cite{Alexander:2022rmq}
EDM, which are expected to improve the 
current sensitivity on 
$\theta_{\rm eff}$ 
by up to 
three orders of magnitude.  

From now on we focus on the limit $V_{\Delta}\sim V_{\chi}\sim V$. The PQ quality constraint of \eq{eq:PQqualityGen} can then be plotted for a fixed value of $V$ in the plane of powers of $v$ and $V$, as shown in the right panel of \fig{fig:plotfa}. 
We show the quality constraint for two different values of $V$, corresponding to the maximal and minimal quality solutions from the left panel, where they are shown as dots of corresponding color. The light red region represents an enlargement of the forbidden darker region, while sufficient PQ-quality is achieved in the upper-right corner. The dominant contributions to the axion potential of \eq{eq:contributions-axion} are shown as stars.

We now turn to the discussion of how the operators of \eq{eq:operators-dominant-quality-contribution} have been identified as the source of the  dominant contribution to the axion potential. The salient points of discussion are given below:
\begin{itemize}
    \item Each of the representations $\Delta$, $\chi$ or $\xi$ in an operator provides one power of $V$, while each of the irreps $\Phi$ or $\Sigma$ (generically labeling any copy of $\Phi^{\alpha}$ or $\Sigma^{\alpha}$) provides one power of $v$. All EW VEVs in the model originate from weak doublets, and hence an even power of them is needed to form an $\SU(2)_L$ invariant. A contribution $v^{m} V^{n}$ must thus necessarily come with $m$ being an even non-negative integer. Since $m\geq 6$ by itself already solves the quality issue, it is sufficient to identify the dominant contributions for $m=0,2,4,6$, which have been listed in \eq{eq:operators-dominant-quality-contribution} in matching order. 
    \item The dominant contributions to the axion potential can be searched for in a systematic way:
        \begin{enumerate}
            \item \label{list:dominant-contributions-begin}
                We first obtain a list of all invariants involving scalar fields of dimension $d$. We leverage external software, in particular \texttt{GroupMath}~\cite{Fonseca:2020vke} and \texttt{LiE}~\cite{Leeuwen1992LiE}, see \app{app:procedure-invariant-enumeration} for more details. 
            \item 
                We retain only invariants with a non-vanishing PQ charge, since only these can affect the axion potential.
            \item
                The list is refined further by demanding that an operator actually gives a non-vanishing vacuum contribution. This occurs only if the invariant contains an all-VEV term, i.e., a term in which every factor is a field with a non-vanishing VEV. 
                This is hard to test for in an automated way, so we instead use a necessary (but not sufficient) condition: taking the VEV-directions in every factor of the invariant, the total charge of the product under any gauged $\mathrm{U}(1)$ must vanish. In the context of our setup, the pertinent test to perform is for the $B-L$ charge, which is part of $\SU(4)_{\rm PS}$.
            \item \label{list:dominant-contributions-end}
                Having a fully refined list of operator candidates, we find those with the largest contributions $v^{m}V^{n}$ to the vacuum, and confirming their non-vanishing contribution by computing it explicitly. Both the refined list and the explicit confirmation of contributions are relegated to the Appendix, cf.~\app{app:PSmodel-PQ-at-tree-level},
            while the final result of dominant 
            operators has been given in \eq{eq:operators-dominant-quality-contribution}. 
        \end{enumerate}
    \item To further illuminate the resulting operators in \eq{eq:operators-dominant-quality-contribution}, it is beneficial to build some intuition for which combinations of scalar fields of \Table{tab:PSirrep} can form an invariant.
    We list five rules an invariant must adhere to: 
        \begin{itemize}
        \item[\namedlabel{item:intuition1}{(i)}]
            The total power of $\Phi$ and $\Sigma$ must be even (due to $\SU(2)_L$ invariance, as already described earlier).
        \item[\namedlabel{item:intuition2}{(ii)}]
            The total power of $\chi$ and $\chi^\ast$ must be even (a restriction from the center of $\SU(2)_R$ combined with \ref{item:intuition1}).  
        \item[\namedlabel{item:intuition3}{(iii)}]
            The only non-trivial way a combination of $\Delta$ and $\chi$ enters is in powers of $\Delta \chi^{\ast 2}$ (a restriction from the center of $\SU(4)_{\rm PS}$). 
        \item[\namedlabel{item:intuition4}{(iv)}]
            The difference between the number of conjugated and un-conjugated irreps (counting $\Phi$, $\Sigma$, $\Delta$ and $\chi$ fields) must be divisible by $3$ (a restriction from the center of $\SU(3)_{f_{R}}$).
        \item[\namedlabel{item:intuition5}{(v)}]
            $\xi$ can usually be added with an arbitrary power, since it behaves as an adjoint or singlet under every group factor. In particular, it can usually also be omitted without jeopardizing the existence of the invariant.
        \end{itemize}
        Given these considerations, the list of \eq{eq:operators-dominant-quality-contribution} could in fact be guessed, the only surprise is that a pure $V^n$ contribution arises only at $n=11$ and not at $n=9$. The invariant $\Delta^{3} \chi^{\ast 6}$ does in fact exists, but it does not contribute to the vacuum, and so the first $V^{n}$ contribution comes only after adding an additional $\Delta\Delta^{\ast}$ piece to it. This is a rather subtle result, which is discussed in more detail in  \app{app:PSmodel-PQ-at-tree-level}.
\end{itemize}
We conclude with a few remarks regarding the limitations of the analysis above. First, the analysis of PQ quality in this section is based on a tree-level analysis of the effective theory, in which every operator is assumed to contribute via an (at most) $\mathcal{O}(1)$ dimensionless coefficient and a suppression with an appropriate power of $\LUV$. It is an important, albeit 
often neglected point, that quantum corrections may in fact enhance operator contributions above those of the tree-level structural set-up; this is further elaborated on conceptually in \app{app:PSmodel-PQ-at-loop-level}.

Second, we briefly comment on the effect of $\SU(4)_{\rm PS}$ and $\SU(3)_{f_R}$ instantons. As extensively discussed e.g.~in~\cite{Csaki:2023ziz}, small-scale instantons provide new contributions to the axion potential which become important if the relevant gauge coupling decreases slowly enough or  grows in the UV. Furthermore, these new potential terms are generically not aligned with the usual QCD-vacuum, thus inducing a shift on the value of $\theta_{\rm eff}$, in analogy to the $\Lambda_{\rm UV}$-suppressed operators discussed above. Small-scale instantons may therefore also introduce an axion quality problem. In our model, this issue is particularly relevant, since gauge couplings grow quickly and develop Landau poles soon above the PS scale due to a proliferation of many scalar degrees of freedom. This is discussed in Appendix~\ref{app:PSpert}, where we quantify the issue by showing that the Landau poles arise around one order of magnitude above the unification scale. As a result, given the strength of the gauge couplings, we expect that the contribution of small-scale instantons may become important, even though an explicit computation is not possible in the non-perturbative regime (indeed, the computation rules outlined in~\cite{Csaki:2023ziz} are only applicable in the perturbative regime). Even a perturbative calculation would not be straightforward,  since the contribution of a $\SU(4)_{\rm PS}\,(\SU(3)_{f_R})$ instanton requires the insertion of 12 (16) fermionic zero modes, which must be combined by means of  effective fermionic operators with the appropriate quantum numbers. 

The proximity of Landau poles to the Pati-Salam scale thus presents a severe challenge to the question of PQ quality in our model, or indeed to any model in which nearby Landau poles arise. In light of this fact, our analysis is necessarily incomplete, and one could imagine the next step being one of the following:
\begin{itemize}[leftmargin=0.5cm,itemsep=0cm]
    \item One might attempt to construct a UV completion of our model where the Landau pole problem is resolved. Adding new fermion or scalar mediators exacerbates the Landau pole issue, so the model-building solution here would have to be non-trivial.
    \item It is in principle still possible that non-perturbative dynamics in the model as-is saves the theory and PQ quality. Such a conclusion, however, is inaccessible via a perturbative calculation.
    \item 
    One accepts the presented model does not fully resolve the question of PQ quality. Our analysis identifies, however, many salient features of vertical-horizontal models and investigates how to achieve quality protection from higher-dimensional operators. These lessons, along with model building limitations learned from the main part of \sect{sec:PS-model}, should then be applied to constructing a model without a Landau pole, in which the question of PQ quality can then be unambiguously addressed.
\end{itemize}
The pursuit of any of these strategies is ultimately beyond the scope of this paper.

%%%%%%%%%%%%%%%%%%%%%%%%%%%%%%%%%%%
%%%%%%%%%%%%%%%%%%%%%%%%%%%%%%%%%%%
\section{Reproducing the SM flavor structure}
\label{sec:SMflavour}

The fermions in our model consist of those in the SM, as well as the right-handed neutrinos and anomalons, introducing $3$ and $24$ new Weyl fermions, respectively. The right-handed neutrinos
are in a flavor-triplet, while the anomalons form $8$ generations of flavor-antitriplets, cf.~\Table{tab:PSirrep}. 

Since both right-handed neutrinos and anomalons are singlets under the SM group, they are also singlets in the EW broken phase $\SU(3)_C\times\mathrm{U}(1)_{\rm EM}$, thus mixing with left-handed neutrinos in the mass matrix. We analyze this complicated neutrino-anomalon sector in detail in Section~\ref{sec:SMflavour-NandA}, while we start the analysis with the simpler quark and charged-lepton sectors in Section~\ref{sec:SMflavour-QandL}.

%%%%%%%%%%%%%%%%%%%%%%%%%%%%%%%%%%%
\subsection{Quark and charged-lepton sector}
\label{sec:SMflavour-QandL}

The operators from which the quark and charged-lepton mass matrices dominantly arise are the $\parY{\Phi^{\alpha}}{I}$ and $\parY{\Sigma^{\alpha}}{I}$ terms in Eq.~\eqref{eq:Yukawa-sector}. Note that these terms are renormalizable. Higher-order non-renormalizable operators will provide corrections, but they structurally connect the same fields as the renormalizable ones, hence their $V/\LUV$ suppression yields only a numerically negligible correction. The analysis for quarks and charged leptons, unlike the neutrinos and anomalons, can thus be performed by considering the renormalizable operators only.

The requirements for a good fermion fit will dictate the needed number of copies of the scalar representations $\Phi$ and $\Sigma$ containing the SM Higgs, denoted by $N_{\Phi}$ and $N_{\Sigma}$,  respectively. Since each type of operator with either $\Phi^\alpha$ or $\Sigma^\alpha$ comes with its own Clebsch-Gordan ratio between the down- and charged-lepton sectors, at least one of each type is required to avoid the bad mass relation $M_D\propto M_E^T$, hence $N_{\Phi},N_{\Sigma}\geq 1$. A further consideration on the number of copies comes from redundancies in the Yukawa matrices, namely rotations in family space and $\SU(3)_{f_R}$ gauge rotations. 
We shall discuss this in more detail in \sect{sec:PSmodel-Yukawa-realistic}; for now it suffices to say that the minimal realistic setup requires at least $N_{\Phi}=1$ and $N_{\Sigma}=2$, since $N_{\Sigma}=1$ would enable a difference between the down- and charged-lepton sectors only in one family.  

We focus below on the minimal scenario and use the simplifying notation introduced already below Eq.~\eqref{eq:Yukawa-sector}: the representations are denoted by $\Phi$, $\Sigma$ and $\Sigma'$, and the corresponding Yukawa couplings by $\parY{\Phi}{I}$, $\parY{\Sigma}{I}$ and $\parY{\Sigma'}{I}$, where the family index $I$ runs from $1$ to $3$. Each of the scalar irreps contains six SM weak doublets; three transform as $(1,2,+\tfrac{1}{2})$ and three transform as $(1,2,-\tfrac{1}{2})$ under the SM group, 
with the corresponding VEVs respectively equipped with the label $u$ or $d$, 
and the multiplicity of $3$ originating from these irreps being antitriplets under $\SU(3)_{f_R}$. Given the weak-doublet content of $\Phi$, $\Sigma$ and $\Sigma'$ described above, we label their VEVs by the self-evident notation
\begin{align}
    \vevv{u\Phi}{A},
        \quad
    \vevv{u\Sigma}{A},
        \quad
    \vevv{u\Sigma'}{A}, 
        \quad
    \vevv{d\Phi}{A},
        \quad
    \vevv{d\Sigma}{A},
        \quad
    \vevv{d\Sigma'}{A}, \label{eq:labels-EW-VEVs}
\end{align}
where $A$ is the flavor index. The EW VEVs (excluding the flavor indices that they carry) will be colored in \colorEW{blue} for better visual clarity.

We relegate the bulk of the computational details to \app{app:PSmodel-Yukawa-computational-details}, and here merely state the results. The up-quark ($U$), down-quark ($D$) and charged-lepton ($E$) mass matrices take the form
\begin{align}
\label{eq:MUvevs}
    (M_U)_{IA}&= \parY{\Phi}{I} \vevv{u\Phi}{A}
        +\parY{\Sigma}{I} \vevv{u\Sigma}{A}
        +\parY{\Sigma'}{I} \vevv{u\Sigma'}{A},\\
\label{eq:MDvevs}
    (M_D)_{IA}&= 
        \parY{\Phi}{I} \vevv{d\Phi}{A}
        +\parY{\Sigma}{I} \vevv{d\Sigma}{A}
        +\parY{\Sigma'}{I} \vevv{d\Sigma'}{A},\\
\label{eq:MEvevs}
    (M_E)_{IA}&=
        \parY{\Phi}{I} \vevv{d\Phi}{A}
        -3\,\parY{\Sigma}{I} \vevv{d\Sigma}{A}
        -3\,\parY{\Sigma'}{I} \vevv{d\Sigma'}{A},
\end{align}
where the LR convention was used, i.e.~the Lagrangian contains terms 
\begin{align}
    \mathcal{L}\supset -\sum_{I,A}(M_U)_{IA}(\bar{u}_L)^{I}(u_R)^{A},
\end{align}
and analogously for other sectors. A choice of redundancy removal (see \app{app:PSmodel-Yukawa-computational-details}), which yields a particularly transparent form for the mass matrices, is
\begin{align}
    \vevv{d\Sigma}{1}=\vevv{d\Sigma}{2}=\vevv{d\Sigma'}{1}&=0 \, ,
    &
    \parY{\Sigma}{1}=\parY{\Sigma}{2}=\parY{\Sigma'}{1}&=0 \, ,
    \label{eq:redundancy-removal-choice}
\end{align}
resulting in the explicitly result
\begin{align}
    M_U&=
        \Matrix{
        \parY{\Phi}{1} \vevv{u\Phi}{1} & \parY{\Phi}{1} \vevv{u\Phi}{2} & \parY{\Phi}{1} \vevv{u\Phi}{3} \\
    %%%
        \parY{\Phi}{2}\vevv{u\Phi}{1}+
        \parY{\Sigma'}{2} \vevv{u\Sigma'}{1} 
        & 
        \parY{\Phi}{2} \vevv{u\Phi}{2} +
        \parY{\Sigma'}{2} \vevv{u\Sigma'}{2}
        & 
        \parY{\Phi}{2} \vevv{u\Phi}{3} +
        \parY{\Sigma'}{2} \vevv{u\Sigma'}{3} \\
    %%%
        \parY{\Phi}{3} \vevv{u\Phi}{1} +
        \parY{\Sigma}{3} \vevv{u\Sigma}{1} +
        \parY{\Sigma'}{3} \vevv{u\Sigma'}{1}
        & 
        \parY{\Phi}{3} \vevv{u\Phi}{2} +
        \parY{\Sigma}{3} \vevv{u\Sigma}{2} +
        \parY{\Sigma'}{3} \vevv{u\Sigma'}{2} 
        & 
        \parY{\Phi}{3} \vevv{u\Phi}{3} +
        \parY{\Sigma}{3} \vevv{u\Sigma}{3} + 
        \parY{\Sigma'}{3} \vevv{u\Sigma'}{3} \\
        },
    \label{eq:mass-mu}\\[6pt]
    M_D&=
        \Matrix{
        \parY{\Phi}{1} \vevv{d\Phi}{1} 
        & 
        \parY{\Phi}{1} \vevv{d\Phi}{2}
        & 
        \parY{\Phi}{1} \vevv{d\Phi}{3} \\
    %%%
        \parY{\Phi}{2} \vevv{d\Phi}{1}
        & 
        \parY{\Phi}{2} \vevv{d\Phi}{2} +
        \parY{\Sigma'}{2} \vevv{d\Sigma'}{2}
        &
        \parY{\Phi}{2} \vevv{d\Phi}{3} +
        \parY{\Sigma'}{2} \vevv{d\Sigma'}{3} \\
    %%%
        \parY{\Phi}{3} \vevv{d\Phi}{1}
        &
        \parY{\Phi}{3} \vevv{d\Phi}{2} +
        \parY{\Sigma'}{3} \vevv{d\Sigma'}{2} 
        & 
        \parY{\Phi}{3} \vevv{d\Phi}{3} +
        \parY{\Sigma}{3} \vevv{d\Sigma}{3} +
        \parY{\Sigma'}{3} \vevv{d\Sigma'}{3}
         \\
        },
    \label{eq:mass-md}\\[6pt]
    M_E&=
        \Matrix{
        \parY{\Phi}{1} \vevv{d\Phi}{1}
        & 
        \parY{\Phi}{1} \vevv{d\Phi}{2}
        &
        \parY{\Phi}{1} \vevv{d\Phi}{3} \\
    %%%
        \parY{\Phi}{2} \vevv{d\Phi}{1} 
        &
        \parY{\Phi}{2} \vevv{d\Phi}{2}
        -3 \parY{\Sigma'}{2} \vevv{d\Sigma'}{2} 
        &
        \parY{\Phi}{2} \vevv{d\Phi}{3}
        -3 \parY{\Sigma'}{2} \vevv{d\Sigma'}{3} \\
    %%% 
        \parY{\Phi}{3} \vevv{d\Phi}{1} 
        &
        \parY{\Phi}{3} \vevv{d\Phi}{2}
        -3 \parY{\Sigma'}{3} \vevv{d\Sigma'}{2} 
        &
        \parY{\Phi}{3} \vevv{d\Phi}{3}
        -3 \parY{\Sigma}{3} \vevv{d\Sigma}{3} 
        -3 \parY{\Sigma'}{3} \vevv{d\Sigma'}{3}
        \\
        }.
        \label{eq:mass-me}
\end{align}
The parameter values $\parY{\Sigma}{3}$ and $\parY{\Sigma'}{2}$, as well as the EW VEVs $\vevv{d\Sigma}{3}$ and $\vevv{d\Sigma'}{2}$, can be taken to be real without loss of generality.

The advantage of the choice in Eq.~\eqref{eq:redundancy-removal-choice} is that the mass matrices for the $D$ and $E$ sectors become as simple as possible, allowing to scrutinize whether the $D$ and $E$ masses can be made sufficiently different. The main driver are the $\Sigma$ and $\Sigma'$ terms, due to the Clebsch coefficient $-3$ for charged leptons. \eqs{eq:mass-mu}{eq:mass-me} should yield a realistic Yukawa sector, with further justification given in \sect{sec:PSmodel-Yukawa-realistic}.

%%%%%%%%%%%%%%%%%%%%%%%%%%%%%%%%%%%
%%%%%%%%%%%%%%%%%%%%%%%%%%%%%%%%%%%
\subsection{The neutrino-anomalon sector}
\label{sec:SMflavour-NandA}

Under $\SU(3)_C\times\mathrm{U}(1)_{\rm EM}$, the left-handed neutrinos $(\bar{\nu}_L)^{I}$, right-handed neutrinos $(\nu_R)^{A}$, and anomalons $(\Psi_R)^{K}{}_{A}$ are all singlets, hence they mix in the mass matrix. In short-hand notation,  
we shall refer to them by $L$, $R$ and $\Psi$, respectively. 

The index $I$ is a family index originating from $Q_{L}$, $A$ is a flavor index (for $\SU(3)_{f_{R}}$), while $K$ denotes the anomalon multiplicity index; $I,A\in\{1,2,3\}$ while $K\in\{1,\ldots,8\}$. Note that anomalons carry two types of indices and the full space of considered fields $L\oplus R\oplus \Psi$ has dimension $3+3+8\times 3=30$.

The incorporation of anomalons into the neutrino sector greatly complicates the analysis, since their masses and mixings arise from non-renormalizable Yukawa operators. This is in contrast with considerations of anomalons in the neutrino sector in prior works~\cite{Berezhiani:1983obe,Berezhiani:1989fp,Berezhiani:1990wn}, where their masses are heavy and arise directly from flavon scalars. Furthermore, the anomalon analysis in our model is made even more intricate due to their split spectrum: $8$ of the $24$ (dominantly) anomalon states turn out to have a parametrically smaller mass, and thus special care is needed to consider them properly. To improve the clarity of presentation, we shall break up our considerations into two parts: the neutrino-anomalon mass matrix is presented and discussed in \sect{sec:neutrinos-matrix}, while the resulting spectrum and mixing are analyzed in \sect{sec:neutrinos-mass-mixing}; the derivation and computational details are relegated to \app{app:PSmodel-Yukawa-computational-details}.

In order to avoid confusion in terminology between the flavor and mass eigenstates of this sector, we explicitly state our naming convention already now at the onset:
\begin{itemize}[leftmargin=0.5cm,itemsep=0.0cm]
    \item The flavor states $L\oplus R\oplus\Psi$  are referred to as \textit{L-neutrinos}, \textit{R-neutrinos} and \textit{anomalons}.
    \item The mass eigenstates which contain a dominant admixture of the above flavor eigenstates are in order named \textit{active neutrinos}, \textit{sterile neutrinos} and \textit{massive anomalons}. Anticipating the $24=16+8$ parametric split in the latter's masses, we dub the two respective groups \textit{heavy anomalons} and \textit{light anomalons}.
\end{itemize}

%%%%%%%%%%%%%%%%%%%%%%%%%%%%%%%%%%%
\subsubsection{The neutrino-anomalon mass matrix
\label{sec:neutrinos-matrix}
}

The neutrino-anomalon mass matrix can be written in block form using the basis split $L\oplus R\oplus\Psi$:
\begin{align}
    \mathcal{L}&\supset \frac{1}{2}\, 
    \begin{pmatrix}
        (\bar{\nu}_L)^{I}\\[2pt] (\nu_R)^{A} \\[2pt](\Psi_{R})^{K}{}_{A}\\ 
    \end{pmatrix}^T
    \begin{pmatrix}
        (M_{LL})_{IJ} & 
        (M_{LR})_{IB} & 
        (M_{L\Psi})_{IN}{}^{B} \\[2pt]
        %%%
        (M_{RL})_{AJ}&
        (M_{RR})_{AB}&
        (M_{R\Psi})_{AN}{}^{B}\\[2pt]
        %%%
        (M_{\Psi L})_{K}{}^{A}{}_{J}&
        (M_{\Psi R})_{K}{}^{A}{}_{B}&
        (M_{\Psi\Psi})_{K}{}^{A}{}_{N}{}^{B}\\
    \end{pmatrix}
    \begin{pmatrix}
        (\bar{\nu}_L)^{J}\\[2pt] (\nu_R)^{B} \\[2pt] (\Psi_{R})^{N}{}_{B}\\
    \end{pmatrix}\, . \label{eq:mass-matrix-LRA}
\end{align}
Each block carries the appropriate type of indices for the fields it connects. The double index notation of the basis of anomalons 
carries over to all anomalon-related blocks, e.g.~$M_{\Psi\Psi}$ has $4$ indices.

The first step in the determination of each block is to identify the operators which contribute in a relevant way. The list of sources, along with the label for its corresponding coefficient, is conveniently compiled in \Table{tab:Yukawa-nonrenormalizable}.   
This is a non-trivial result, which requires careful consideration of a number of subtleties, which we are about to discuss. With the computational details of each contribution relegated to \app{app:PSmodel-Yukawa-computational-details}, we directly state here the final result:
\begingroup
\allowdisplaybreaks
\begin{align}
        (M_{LL})_{IJ}&\approx 0 \, , 
                \label{eq:Nblock-LL}\\[4pt]
        %%%%%%%%%%%
        (M_{LR})_{IB}&=
                \parY{\Phi}{I}\,\vevv{u\Phi}{B}
                -3\, \parY{\Sigma}{I}\,\vevv{u\Sigma}{B}
                -3\, \parY{\Sigma'}{I}\,\vevv{u\Sigma'}{B}\, ,
                \label{eq:Nblock-LR}\\[4pt]
        %%%%%%%%%%%
        (M_{RR})_{AB}&= Y_{R}\,\vevZc{AB} \, ,
            \label{eq:Nblock-RR}\\[4pt]
        %%%%%%%%%%%
        (M_{L\Psi})_{IN}{}^{B}&= \frac{1}{\LUV}\,\epsilon^{BCD}\,\vevV{C}\,\left(
        \parA_{IN}\,\vevv{u\Phi}{D}
        -3\,\parB_{IN}\,\vevv{u\Sigma}{D}
        -3\,\parBp_{IN}\,\vevv{u\Sigma'}{D}
        \right) \nonumber\\
%%%
        	&\quad + 
				\frac{1}{\LUV^{2}}\;\vevZ{BC}\,\left(
				\parAt_{IN}\,\vevvc{u\Phi}{D}
				+\parBt_{IN}\,\vevvc{u\Sigma}{D}
				+\parBt'_{IN}\,\vevvc{u\Sigma'}{D}
			\right)			
		\,\vevVc{E}\,\epsilon_{CDE}
        	\, ,
            \label{eq:Nblock-LA}\\[4pt]
        %%%%%%%%%%%
        (M_{R\Psi})_{AN}{}^{B}&=\frac{\parC_{N}}{\LUV}\,\epsilon^{BCD} \vevZc{AC}\,\vevV{D}
		+
		\frac{\parCt_{N}}{\LUV^{3}}\;\vevZ{EF}\,\vevZc{ED}\,\vevZc{AC}\,\vevV{F}\;\epsilon^{BCD}        
        \, ,
            \label{eq:Nblock-RA}\\[4pt]
        %%%%%%%%%%%
        (M_{\Psi\Psi})_{K}{}^{A}{}_{N}{}^{B}&=
            \frac{\parD_{KN}}{\LUV^2}\,\epsilon^{ACD}\,\epsilon^{BEF} \vevZc{CE} \,\vevV{D}\, \vevV{F} \nonumber\\[4pt]
            %%%
            &\quad +
            \frac{\parDt_{KN}}{\LUV} 
            \left(
                \vevv{u\Phi}{}^{\ast A}
                \vevv{d\Phi}{}^{\ast B} +
                \vevv{d\Phi}{}^{\ast A}
                \vevv{u\Phi}{}^{\ast B}
            \right) \nonumber\\[4pt]
            %%% 
            &\quad +
            \frac{\parEt_{KN}}{\LUV} 
            \left(
                \vevv{u\Sigma}{}^{\ast A}
                \vevv{d\Sigma}{}^{\ast B} +
                \vevv{d\Sigma}{}^{\ast A}
                \vevv{u\Sigma}{}^{\ast B}
            \right)\nonumber \\[4pt]
            %%%
            &\quad +
            \left(
            \frac{1}{2}\,\frac{\parEt'_{KN}}{\LUV} 
            \left(
                \vevv{u\Sigma}{}^{\ast A}
                \vevv{d\Sigma'}{}^{\ast B} +
                \vevv{d\Sigma}{}^{\ast A}
                \vevv{u\Sigma'}{}^{\ast B}
            \right)
            +
            \left[{}^{K\leftrightarrow N}_{A\leftrightarrow B}\right]
            \right) \nonumber\\[4pt]
%%%
            &\quad +
            \frac{\parEt''_{KN}}{\LUV} 
            \left(
                \vevv{u\Sigma'}{}^{\ast A}
                \vevv{d\Sigma'}{}^{\ast B} +
                \vevv{d\Sigma'}{}^{\ast A}
                \vevv{u\Sigma'}{}^{\ast B}
            \right)\nonumber \\[4pt]
%%%
		&\quad +            
			\frac{\parFt_{KN}}{\Lambda^{4}_{\rm UV}}\;\vevZ{GH}\,\vevZc{CE}\,\vevZc{DF}\,\vevV{G}\,\vevV{H}\;\epsilon^{ACD}\epsilon^{BEF}            
			\nonumber \\[4pt]
%%%
		&\quad +            
			\frac{\parGt_{KN}}{\Lambda^{4}_{\rm UV}}\;\vevZ{GH}\,\vevZc{CE}\,\vevZc{DG}\,\vevV{F}\,\vevV{H}\;\epsilon^{ABC}\epsilon^{DEF}
        \, .
        \label{eq:Nblock-AA}
    \end{align}
\endgroup
We remind the reader of our color coding: \colorPQ{red} for the Pati-Salam braking scale, and \colorEW{blue} for the EW scale VEVs.
The neutrino-anomalon mass matrix is symmetric, implying
    \begin{align}
        (M_{RL})_{AJ}&=(M_{LR})_{JA} \, , &
        (M_{\Psi L})_{K}{}^{A}{}_{J}&=(M_{L\Psi})_{JK}{}^{A} \, , 
            \nonumber\\
        (M_{\Psi R})_{K}{}^{A}{}_{B}&=(M_{R\Psi})_{BK}{}^{A} \, , &
        (M_{\Psi\Psi})_{K}{}^{A}{}_{N}{}^{B}&=(M_{\Psi\Psi})_{N}{}^{B}{}_{K}{}^{A} \, ,
    \end{align}
from where the remaining blocks (below the diagonal) can be determined.
\begin{table}[t!]
\begin{center}
\begin{tabular}{llllll}
	\toprule
	type&operator $\mathcal{O}$&$d$&$\langle M_{\mathcal{O}}\rangle$&$\#$ of $\mathcal{O}$&coefficients\\
	\midrule
		$LL$&
			$\overline{Q}_{L}\overline{Q}_{L}\,
			\Delta (\Phi^{2}+\Phi\Sigma+\Sigma^2)$&
            $6$&
			$v^2V/\LUV^2$&
			$1+2+5$&---
			\\[6pt]
%%%%%%
		$LR$&
			$\overline{Q}_L Q_{R}
			\,(\Phi+\Sigma+\Sigma')$&
			$4$&
            $v$&
			$1+1+1$&
			$\parY{\Phi}{I}$,
			$\parY{\Sigma}{I}$,$\parY{\Sigma'}{I}$
			\\[6pt]
%%%%%%
		$RR$&
			$Q_R Q_R\,\Delta^\ast$&
            $4$&
			$V$&
			$1$&
			$Y_{R}$\\[6pt]
%%%%%%
		$L\Psi$&
			$\overline{Q}_{L}\Psi_{R}\,
				\chi(\Phi+\Sigma+\Sigma')$&
            $5$&
			$vV/\LUV$&
			$1+1+1$&
			$\parA_{IN}$, $\parB_{IN}$, $\parB'_{IN}$
			\\[2pt]
		$L\Psi$&
			$\overline{Q}_{L}\Psi_{R}\,
				\Delta\chi^* (\Phi^*+\Sigma^*+\Sigma'^*)$&
            $6$&
			$vV^2/\LUV^2$&
			$1+2+2$&
			$\parAt_{IN}$, $\parBt_{IN}$, $\parBt'_{IN}$
			\\[6pt]
%%%%%
		$R\Psi$&
			$Q_{R}\Psi_{R}\,\Delta^\ast \chi$&
            $5$&
			$V^{2}/\LUV$&
			$1$&
			$\parC_{N}$\\[2pt]
		$R\Psi$&
			$Q_R\Psi_R\Delta \Delta^{*2}\chi$&
            $7$&
			$V^{4}/\LUV^{3}$&
			$56$&
			$\parCt_{N}$\\[6pt]
%%%%%%%
		$\Psi\Psi$&
			$\Psi_{R}\Psi_{R}\,\Delta^\ast\chi^{2}$&
            $6$&
			$V^{3}/\LUV^{2}$&
			$1$&
			$\parD_{(KN)}$\\[2pt]
		$\Psi\Psi$&
			$\Psi_{R}\Psi_{R}\,\Phi^{\ast 2}$&
            $5$&
			$v^{2}/\LUV^{}$&
			$1$&
			$\parDt_{(KN)}$\\[2pt]
		$\Psi\Psi$&
			$\Psi_{R}\Psi_{R}\,
			(\Sigma^{\ast 2}
				+\Sigma^{\ast}\Sigma'^{\ast}
                +\Sigma'^{\ast 2})$&
            $5$&
			$v^{2}/\LUV^{}$&
			$1+1+1$&
			$\parEt_{(KN)},\parEt'_{KN},\parEt''_{(KN)}$\\[2pt]
		$\Psi\Psi$&
			$\Psi_{R}\Psi_{R}\Delta \Delta^{*2}\chi^2$&
            $8$&
			$V^{5}/\LUV^{4}$&
			$42$&
			$\parFt_{(KN)},\parGt_{[KN]}$\\[3pt]
	\bottomrule
\end{tabular}
\caption{List of the operators contributing to the neutrino-anomalon mass matrix in a relevant way. The operators $\mathcal{O}$ are arranged according to type, i.e.~which block in the basis $L\oplus R\oplus\Psi$ they contribute to. For each operator we also give its mass dimension $d$, its contribution $\langle M_{\mathcal{O}}\rangle$ to the mass in terms of VEVs ($v$ and $V$ are the EW and PS-breaking VEVs, respectively), the number $\#$ of independent invariants (index contractions) of the involved fields, and finally the labeling of coefficients. The subdominant contributions have a tilde in the coefficient label. 
\label{tab:Yukawa-nonrenormalizable}}
\end{center}
\end{table}
The main points to note about this result are the following:
\begin{enumerate}[label=(\arabic*),ref=\arabic*,leftmargin=0.5cm]
    \item \textit{Regarding the coefficients:}
        \begin{itemize}[leftmargin=0.0cm]
            \item 
            For every block entry all dominant contributions are provided, while for anomalon-related blocks we also add certain subdominant contributions, which as we shall later see are critical for describing the light anomalons. The coefficients of subdominant contributions are labeled with a tilde, while primes are used when $\Sigma$-irreps are replaced by $\Sigma'$ in an operator. 
            \item All coefficients in \Table{tab:Yukawa-nonrenormalizable} are complex, with $I\in \{1,\ldots,3\}$ denoting $Q_{L}$-family indices and $K,N\in\{1,\ldots,8\}$ $\Psi$-family indices. The (anti-)symmetry of matrix coefficients is denoted in \Table{tab:Yukawa-nonrenormalizable} by parenthesis (square brackets). The parameter $Y_{R}$ has no indices and is simply a complex number.
       \end{itemize}
    \item \textit{Determining the dominant contributions:}
        \begin{itemize}[leftmargin=0.0cm]
            \item For each block in Eq.~\eqref{eq:mass-matrix-LRA}, we identify the dominant contribution of the form $v^{m}V^{n}/\LUV^{m+n-1}$ by listing all operators up to dimension $d=9$ involving the relevant fermion fields (using the methods described in \app{app:procedure-invariant-enumeration}), and ordering them according to the size of their contribution. We assume the VEV scales as 
            \begin{align}
                v&\sim 10^2\,\mathrm{GeV}\, , &
                V&\in [10^9,10^{14}]\,\mathrm{GeV}\, , &
                \LUV &\sim \mpl \sim 10^{19}\,\mathrm{GeV} \, ,
                \label{eq:vev-scaling}
            \end{align}
            where importantly we do not commit to a predetermined scale $V$. We instead consider a broader range of interesting scenarios, from the $V=10^{9}\,\mathrm{GeV}$ high-quality PQ symmetry scenario (according to \sect{sec:PQquality}) on the lower end, 
            to 
            $V=10^{14}\,\mathrm{GeV}$ 
            on the higher end (this value being set by the upper bound on the seesaw scale --- cf.~also \sect{eq:axionsearches}). This range of $V$ is of interest both for axion phenomenology and anomalon cosmology, cf.~\sect{sec:axionpheno} and \sect{sec:anomalon-cosmology}, respectively.
            \item   Regardless of choice of $V$ in \eq{eq:vev-scaling}, it turns out that dominant operators can always be identified unambiguously. This is not the case for all subdominant operators: a comparison of $\parDt$- and $\parEt$-contributions to those of $\parFt$-type show that the former are larger at small $V$, and the latter at large $V$. Furthermore, the operators (as specified by field powers) yielding dominant contributions always have a single independent contraction in their gauge indices, thereby requiring only one coefficient label for each. This is unlike  some subdominant contributions with $\# >1$ in \Table{tab:Yukawa-nonrenormalizable}, which in any case require an altogether different treatment. Finally, one has to take care in determining the dominant contribution: due to the involvement of two scales, namely $v$ and $V$, the dominant contributions in a block entry are not always those of lowest dimension $d$. In the $\Psi\Psi$-block, the $d(\parD)=6$ is larger than the subdominant case with $d(\parDt)=5$.
            \item The $M_{LL}$ block is parametrically small, so we set it to zero in \eq{eq:Nblock-LL} --- hence no associated coefficient in \Table{tab:Yukawa-nonrenormalizable}. The blocks $M_{RR}$ and $M_{LR}$ have the renormalizable Majorana and Dirac contributions, respectively, as in 
            type-I seesaw; they are renormalizable contributions from \eq{eq:Yukawa-sector}. The block $M_{LR}$ takes an analogous form to $M_{U,D,E}$ in \eqs{eq:mass-mu}{eq:mass-me} and with the redundancy-removal ansatz of \eq{eq:redundancy-removal-choice} takes the explicit form
            \begin{align}\label{eq:MLR}
            M_{LR}&=
                \Matrix{
                \parY{\Phi}{1} \vevv{u\Phi}{1} 
                & 
                \parY{\Phi}{1} \vevv{u\Phi}{2}
                & 
                \parY{\Phi}{1} \vevv{u\Phi}{3} \\
            %%%
                \parY{\Phi}{2} \vevv{u\Phi}{1}
                -3 \parY{\Sigma'}{2} \vevv{u\Sigma'}{1} 
                &
                \parY{\Phi}{2} \vevv{u\Phi}{2}
                -3 \parY{\Sigma'}{2}  \vevv{u\Sigma'}{2} 
                &
                \parY{\Phi}{2} \vevv{u\Phi}{3}
                -3 \parY{\Sigma'}{2} \vevv{u\Sigma'}{3} \\
            %%% 
                \parY{\Phi}{3} \vevv{u\Phi}{1}
                -3 \parY{\Sigma}{3} \vevv{u\Sigma}{1}
                -3 \parY{\Sigma'}{3} \vevv{u\Sigma'}{1}
                &
                \parY{\Phi}{3} \vevv{u\Phi}{2} 
                -3 \parY{\Sigma}{3} \vevv{u\Sigma}{2}
                -3 \parY{\Sigma'}{3} \vevv{u\Sigma'}{2}
                &
                \parY{\Phi}{3} \vevv{u\Phi}{3}
                -3 \parY{\Sigma'}{3} \vevv{u\Sigma'}{3}
                -3 \parY{\Sigma}{3} \vevv{u\Sigma}{3} \\
                } \,.
            \end{align}
        All contributions in anomalon-related blocks of \eqs{eq:Nblock-LA}{eq:Nblock-AA} come, however, from non-renormalizable operators; for further explicit details, see \app{app:PSmodel-Yukawa-computational-details}. 
        \end{itemize}
    \item \textit{The need for subdominant contributions --- a split in the massive-anomalon spectrum:}
        \begin{itemize}[leftmargin=0.0cm]
            \item Suppose we switch on only the dominant contributions in the anomalon-related blocks of \eqs{eq:Nblock-LA}{eq:Nblock-AA}. Then the $30\times 30$ mass matrix of \eq{eq:mass-matrix-LRA} has $8$ massless modes. These massless anomalon modes are related to the direction of the VEV of $\chi$ in flavor space: $\vevV{A}$. Explicitly, defining
                \begin{align}
                    (\Psi_{0})^{K}{}_{A}&=\mathcal{N}^{K}\,\vevV{A} \, , \label{eq:Psi0-definition}
                \end{align}
            for arbitrary coefficients $\mathcal{N}^{K}$ (hence $8$ such independent states), it is easily seen that by retaining only dominant terms (which we abbreviate by $\tilde{\mathcal{X}}\to 0$) we have
            \begin{align}
                (M_{L\Psi})_{IN}{}^{B}
                    \Big|_{\tilde{\mathcal{X}}\to 0}
                    \,(\Psi_{0})^{N}{}_{B} &=0 \, ,\\[5pt]
                (M_{R\Psi})_{AN}{}^{B}
                    \Big|_{\tilde{\mathcal{X}}\to 0}
                    \,(\Psi_{0})^{N}{}_{B} &=0 \, ,\\[5pt]
                (M_{\Psi\Psi})_{K}{}^{A}{}_{N}{}^{B}
                    \Big|_{\tilde{\mathcal{X}}\to 0}
                    \,(\Psi_{0})^{N}{}_{B} &=0 \, ,
            \end{align}
            hence $(0,0,\Psi_{0})$ are null eigenmodes of the full neutrino-anomalon mass matrix of \eq{eq:mass-matrix-LRA}. Indeed, the reason for the massless anomalon modes in the $\langle\chi\rangle$-direction is that all dominant operators  form invariants, in which the flavor indices of anomalons and the actual VEV $\langle\chi\rangle=\vevV{A}$ are contracted through the 3D Levi-Civita tensor, see expressions for non-tilde contributions in \eqs{eq:Nblock-LA}{eq:Nblock-AA}.
            This is apparently an accidental feature in our model. 
            \item The $\Psi_{0}$ direction is properly described only once contributions beyond the dominant ones are considered. Crucially, the tilde contributions are those, which contribute dominantly to the $\Psi_{0}$ direction. Their derivation is sufficiently subtle that we have relegated it to \app{app:PSmodel-Yukawa-computational-details}. Since the tilde contributions are subdominant in directions orthogonal to $\Psi_{0}$, there is a split in the massive-anomalon spectrum. Note, however, that in the presence of subdominant contributions, $\Psi_{0}$ is no longer an exact eigenmode; the light anomalons align close to (but not exactly in) the $\Psi_{0}$ direction, while heavy anomalons align closely along directions orthogonal to $\Psi_{0}$.
            In conclusion, the contributions in \eqs{eq:Nblock-LL}{eq:Nblock-AA} thus properly describe, to parametrically leading order, the entire space of states in the neutrino-anomalon sector.
        \end{itemize}
\end{enumerate}

%%%%%%%%%%%%%%%%%%%%%%%%%%%%%%%%%%%
\subsubsection{Understanding the neutrino-anomalon spectrum and mixing
\label{sec:neutrinos-mass-mixing}
}

We now turn to analyzing the masses and mixings in the neutrino-anomalon sector.
Although \eqs{eq:Nblock-LL}{eq:Nblock-AA} provide a complete description of the $30\times 30$ mass matrix of \eq{eq:mass-matrix-LRA} up to parametrically leading order, 
we aim for a simplified analytic understanding of its features.

A convenient approach is to start with the block split $L\oplus R\oplus\Psi$, and then take into account the additional heavy-light split in the anomalon sector. Splitting the space $\Psi=\Psi_{0}\oplus\Psi_{\perp}$ (with a rotation relative to the original basis $\Psi^{K}{}_{A}$), where $\Psi_{\perp}$ represents the orthogonal complement in $\Psi$-space of the directions $\Psi_0$, we obtain the basis $L\oplus R\oplus \Psi_{\perp}\oplus \Psi_{0}$ ($3+3+16+8$ in dimensions), in which the mass matrix can be parametrically written
as
\begin{align}
	\mymatrix{
		M_{LL}&M_{LR}&M_{L\Psi_\perp}&M_{L\Psi_0}\\
		M_{RL}&M_{RR}&M_{R\Psi_\perp}&M_{R\Psi_0}\\
		M_{\Psi_\perp L}&M_{\Psi_\perp R}&M_{\Psi_\perp \Psi_\perp}&M_{\Psi_\perp \Psi_0}\\
		M_{\Psi_0 L}&M_{\Psi_0 R}&M_{\Psi_0\Psi_\perp}&M_{\Psi_0\Psi_0}\\
	}&\sim
	\mymatrix{
		\tfrac{v^2 V}{\LUV^{2}}&yv&
		l\,\tfrac{vV}{\LUV}&
		\tilde{l}\,\tfrac{vV^2}{\LUV^2}
		\\[3pt]
		%%%%%%%%%%
		..&V&r\,\tfrac{V^2}{\LUV}&
			\tilde{r}\,\tfrac{V^4}{\LUV^3}
		\\[3pt]
		%%%%%%%%%%	
		..&..&
			\tfrac{V^{3}}{\LUV^{2}}&
			\tfrac{v^2}{\LUV}+
			\tfrac{V^{5}}{\LUV^{4}}\\[3pt]
		%%%%%%%%%%	
		..&..&..&
			\tfrac{v^2}{\LUV}+
			\tfrac{V^{5}}{\LUV^{4}}
			\\
	}
	\, ,\label{eq:mass-LRA-block4}
\end{align}
where the entries denoted by dots are filled-in so that the matrix is symmetric.
The VEVs $V$ and $v$, along with the UV scale $\LUV$, take the values from \eq{eq:vev-scaling}, and the parametric sizes of each entry have been looked up in \Table{tab:Yukawa-nonrenormalizable}. We omit the coefficients; we assume they are order one in anomalon-only blocks, but their effect in off-diagonal blocks mixing anomalons and neutrinos is mimicked 
by the parameters $\{l,r,\tilde{l},\tilde{r}\}$. Finally, $y$ represents the Yukawa coupling; assuming the interference from mixing terms with anomalons is subdominant, the correct mass range for active neutrinos $m_{\nu}\sim 0.1\,\mathrm{eV}$ is automatically obtained by expressing $y$ from the seesaw formula:
\begin{align}
    y&\sim \sqrt{m_{\nu} V/v^2} \, . \label{eq:fix-y}
\end{align}
A seesaw scale $V=10^{9}\,\mathrm{GeV}$ is rather low, so we get a suppression from $y\sim 10^{-2.5}$, while the ideal seesaw scale $V=10^{14}\,\mathrm{GeV}$ yields $y\sim 1$. 

To reiterate the connection of the $4\times 4$ block form  with the explicit terms in \eqs{eq:Nblock-LL}{eq:Nblock-AA}, note that switching off the tilde (subdominant) terms results in the last row and column in \eq{eq:mass-LRA-block4} vanishing, since the dominant terms do not connect to $\Psi_{0}$. Switching on the tilde terms induces contributions to both the $\Psi_{0}$ and $\Psi_\perp$ rows/columns, but we omit writing them for the latter 
case since the dominant contributions there are already parametrically larger. In particular, the $\parDt,\parEt,\parFt$-type contributions are present in all $\Psi\Psi$-blocks, i.e.~the entire bottom-right $2\times 2$ part of the block matrix.

\begin{figure}[t!]
    \centering
    \includegraphics[width=0.48\linewidth]{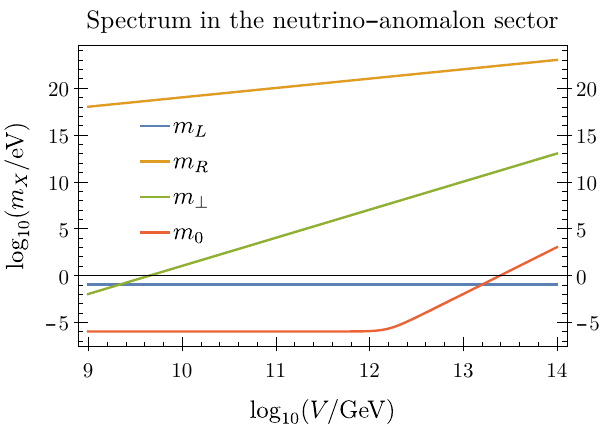}
    \includegraphics[width=0.49\linewidth]{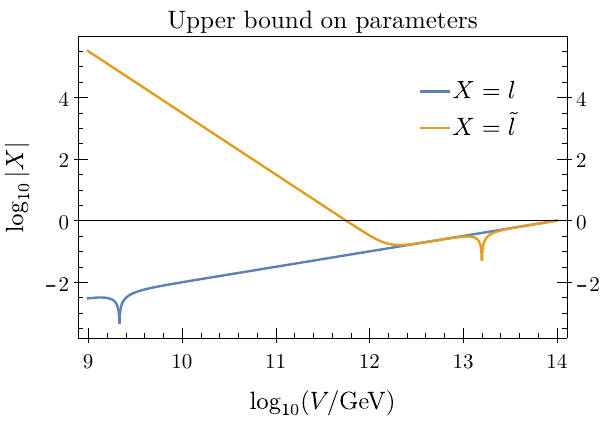}
    \caption{The spectrum of neutrinos and anomalons as a function of the Pati-Salam breaking scale $V$ (left panel), and the upper bounds on parameters $l$ and $\tilde{l}$ from \eq{eq:parlimit-l} and \eqref{eq:parlimit-lt} for mixing not to interfere with the spectrum approximation (right panel). }
    \label{fig:NA-eigenvalues}
\end{figure}

Given the hierarchical nature of the $4\times 4$ mass matrix in \eq{eq:mass-LRA-block4}, we can apply perturbative methods in its analysis, see \app{app:PSmodel-Yukawa-computational-details} for a quick overview. Assuming a small enough interference from off-diagonal couplings other than $M_{LR}$ --- to be quantified later --- the mass eigenvalues are simply the diagonal entries (except $m_L$ coming from the seesaw formula):
\begin{align}
    m_{L}&\sim \frac{y^2 v^2}{V}\sim m_{\nu} \, , &
	m_{R}&\sim V \, , &
	m_{\perp} &\sim \frac{V^{3}}{\LUV^2} \, , &
	m_{0}&\sim \frac{v^2}{\LUV}+\frac{V^{5}}{\LUV^{4}} \, . \label{eq:LRA-masses}
\end{align}
Given the VEV and parameter values in \eq{eq:vev-scaling} and \eqref{eq:fix-y}, the following hierarchies become apparent:
\begin{align}
 	m_{L},m_{\perp},m_{0} &\ll m_{R} \, , &
 	m_{0} &\ll m_{\perp} \, .
\end{align}
The only ambiguity is how the active neutrino mass $m_L$ is placed relative to anomalon masses $m_0$ and $m_\perp$. We plot the eigenvalues as functions of the PS-breaking scale $V$ in the left panel of \fig{fig:NA-eigenvalues}. We see that in the given range of scale $V$, heavy anomalons have masses between $0.01\,\mathrm{eV}$ and $10\,\mathrm{TeV}$, while light anomalons are between $1\,\mathrm{\mu eV}$ and $1\,\mathrm{keV}$.

The approximate expressions in \eq{eq:LRA-masses} are valid as long as that quantity dominates in the approximation for $m_{i}$, cf.~\eq{eq:mass-angle-approximation2} in the Appendix. Since $m_{R}\approx M_{RR}$ is very large, terms involving this quantity are negligible, so we get the main restrictions in the form 
\begin{align}
	M_{L\perp}^2 &\ll m_{L}\,|m_{L} - m_{\perp}| \, , &
    M_{L0}^2 &\ll m_{L}\,|m_{L} - m_{0}| \, , \\
	M_{L\perp}^2 &\ll m_{\perp}\,|m_{L} - m_{\perp}| \, , &
    M_{L0}^2 &\ll m_{0}\,|m_{L} - m_{0}| \, .
\end{align}
These can be rearranged to derive upper bounds on parameters $l$ and $\tilde{l}$:
\begin{align}
    l^2&\ll \min\left(
            \left|y^2-\frac{v^2 y^4 \LUV^2}{V^4}\right|,
            \left|y^2-\frac{V^4}{v^2 \LUV^2} \right|
        \right)\, ,
            \label{eq:parlimit-l}\\[6pt]
    \tilde{l}^2 &\ll 
        \left| V-\frac{y^2 v^2 \LUV^4}{V^5}\right|\cdot
        \min\left(
            \frac{y^2}{V}, \frac{V^5}{v^2 \LUV^4} + \frac{1}{\LUV}
        \right) \, .
             \label{eq:parlimit-lt}
\end{align}
The two upper bounds for magnitudes of $l$ and $\tilde{l}$ are shown graphically in the right panel of \fig{fig:NA-eigenvalues}, and their actual values should preferably be an order of magnitude below that. The results show mild restrictions on the parameters, e.g.~$|\tilde{l}|\lesssim 10^{-2}$ around $V=10^{12}\,\mathrm{GeV}$,  and $|l|\lesssim 10^{-3}$ in the small-$V$ regime. There are two apparent cusps at $V$ around $10^{9.4}\,\mathrm{GeV}$ and $10^{13.2}\,\mathrm{GeV}$, corresponding respectively to the cross-over of heavy and light anomalons with the mass of active neutrinos, cf.~left panel of \fig{fig:NA-eigenvalues}. \eq{eq:mass-angle-approximation2} namely has a pole around those values of $V$, so the parameters have to be suppressed accordingly in order for the approximation not to break down. Finally, one might wonder about restrictions on $r$ and $\tilde{r}$; as alluded to earlier, a large $m_{R}$ implies the approximation is always good for $r,\tilde{r}\sim \mathcal{O}(1)$ or below.

We now turn to the question of the mixing itself. The mixing between a flavor state $i$ and a mass eigenstate $j$ is given by $U_{ij}$, where the mixing matrix $\mathbf{U}$ diagonalizes the mass matrix $\mathbf{M}$ via $\mathbf{M}=\mathbf{U}\mathbf{m}\mathbf{U}^{T}$ (with $\mathbf{m}$ being diagonal). The general perturbative expression up to second order is again given in the Appendix, see Eqs.~\eqref{eq:mass-angle-approximation2} and \eqref{eq:perurbative-U}. Phenomenologically the most interesting are $U_{L\perp}$ and $U_{L0}$, i.e.~the admixtures of heavy and light anomalons in the L-neutrino flavor state that is involved in SM weak interactions. Their approximation from \eq{eq:mass-angle-approximation2} gives
\begin{align}
    U_{L\perp}&\sim \frac{1}{m_{\perp}-m_L} \left(
        M_{L\perp} 
        -\frac{M_{LR}M_{R\perp}}{m_{R}-m_\perp}
        -\frac{M_{L0}M_{\perp 0}}{m_{0}-m_\perp}
        \right)+\delta U_{L\perp}\, ,\\[3pt]
        U_{L0}&\sim \frac{1}{m_{0}-m_L} \left(
        M_{L0} 
        -\frac{M_{LR}M_{R0}}{m_{R}-m_0}
        -\frac{M_{L\perp}M_{\perp 0}}{m_{\perp}-m_0}
        \right) +\delta U_{L0}\, ,
\end{align}
where $\delta U$ denote higher order effects. Given the VEV hierarchies, it turns out there is one mildly-relevant third-order effect in $U_{L0}$ given by
\begin{align}
    \delta U_{L0}&\sim
        \frac{1}{m_{0}-m_{L}}\;
        \frac{M_{LR}\,M_{R\perp} M_{\perp 0}}{(m_\perp-m_{R})(m_{0}-m_\perp)} \, ,
\end{align}
and the mixing angles then simplify to
\begin{align}
    U_{L\perp}&\sim \frac{v}{V^{4}\LUV^{2}-y^2 v^2 \LUV^{4}} \left(
        (l-r\,y)\,V^{2}\LUV^{3}+\tilde{l}\,(V^{5}+v^2 \LUV^3)
        \right) \, , \label{eq:terms-LP}\\[3pt]
    U_{L0}&\sim  \frac{v\LUV}{V^{7}-y^2 v^2 V \LUV^4}\,\left(
            (\tilde{l}\,\LUV - \tilde{r}\,yV) V^{4}
            - (l+r\,y)\,(V^{5}+v^2 \LUV^{3})
        \right) \, . \label{eq:terms-L0}
\end{align}
In $U_{L\perp}$, the $l$-term is first-order and $r$- and $\tilde{l}$-terms are second-order. In $U_{0L}$ the $\tilde{l}$-term is first order, the $\tilde{r}$- and $l$-terms are second order, and the $r$-term is third order. Notice that all terms are linear in the parameters $\{l,r,\tilde{l},\tilde{r}\}$, further motivating their inclusion in the analytic approximation.

\begin{figure}[t!]
    \centering
    \includegraphics[width=0.49\linewidth]{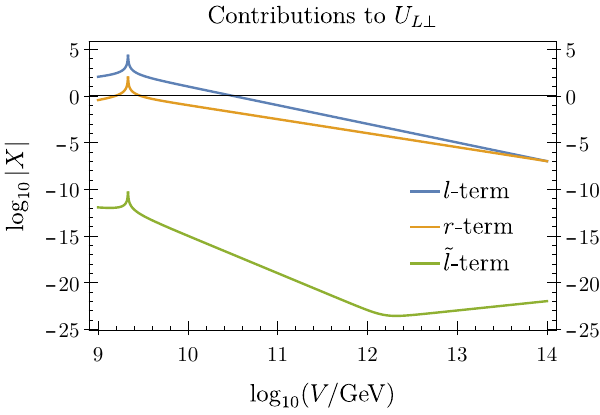}
    \includegraphics[width=0.49\linewidth]{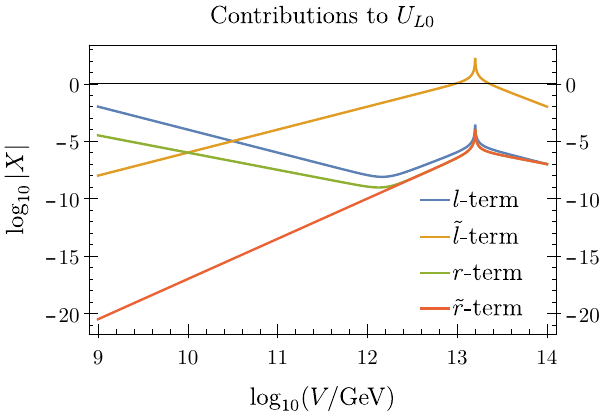}
    \caption{Comparison of the size of contributions
    to $U_{L\perp}$ from \eq{eq:terms-LP} (left panel)
    and to $U_{L0}$ from \eq{eq:terms-L0} (right panel).
    Each contribution is proportional to one of the coefficients $c\in\{l,r,\tilde{l},\tilde{r}\}$. If $c\neq 1$, a contribution shifts by $\log_{10}|c|$; for $|c|<1$ the shift is downward.
    }
    \label{fig:NA-mixing-terms}
\end{figure}

We plot the various contributions in the mixings $U_{L\perp}$ and $U_{L0}$ from \eqs{eq:terms-LP}{eq:terms-L0} as a function of the PS-scale $V$ in \fig{fig:NA-mixing-terms}. Note that the plot shows contributions assuming the parameter values $\{l,r,\tilde{l},\tilde{r}\}$ to be equal $1$, otherwise the associated curves are shifted by an appropriate amount, cf.~caption of \fig{fig:NA-mixing-terms}. We see that $U_{L\perp}$ depends on $l$ and $r$, while the $\tilde{l}$ effect can be neglected. 
For $U_{L0}$ the situation is more complex: at small $V$ the $l$-term is the most important, followed by the $r$-term; at large $V$, the $\tilde{l}$-term dominates, while the other contributions are comparable.
We observe cusps representing poles in the mixing angle, again due to mass cross-over as in \fig{fig:NA-eigenvalues}. The coefficients $l,r,\tilde{l},\tilde{r}$ need to be taken sufficiently small so that $U_{L\perp},U_{L0}\ll 1$, otherwise we are outside the regime of \eq{eq:LRA-masses}.

\def\SINGLEWIDTH{0.32\linewidth}
\begin{figure}[t!]
    \centering
        \hspace{-0.9cm}
        \begin{tabular}{l@{ }l@{ }l@{ }l}
            \includegraphics[width=\SINGLEWIDTH]{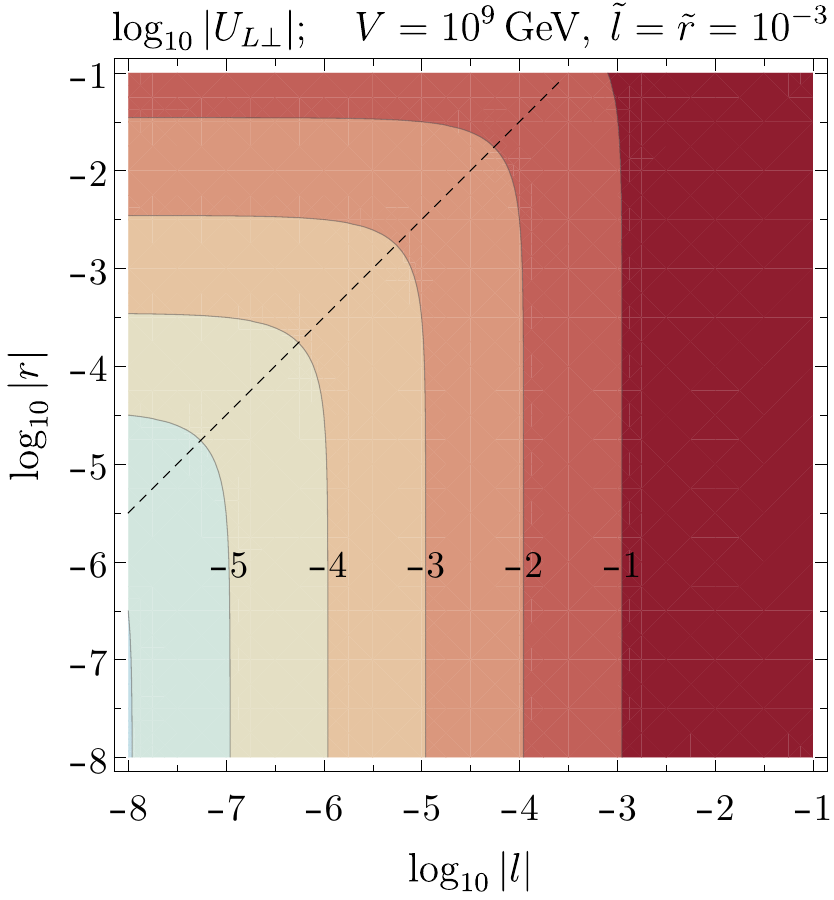}
            &
            \includegraphics[width=\SINGLEWIDTH]{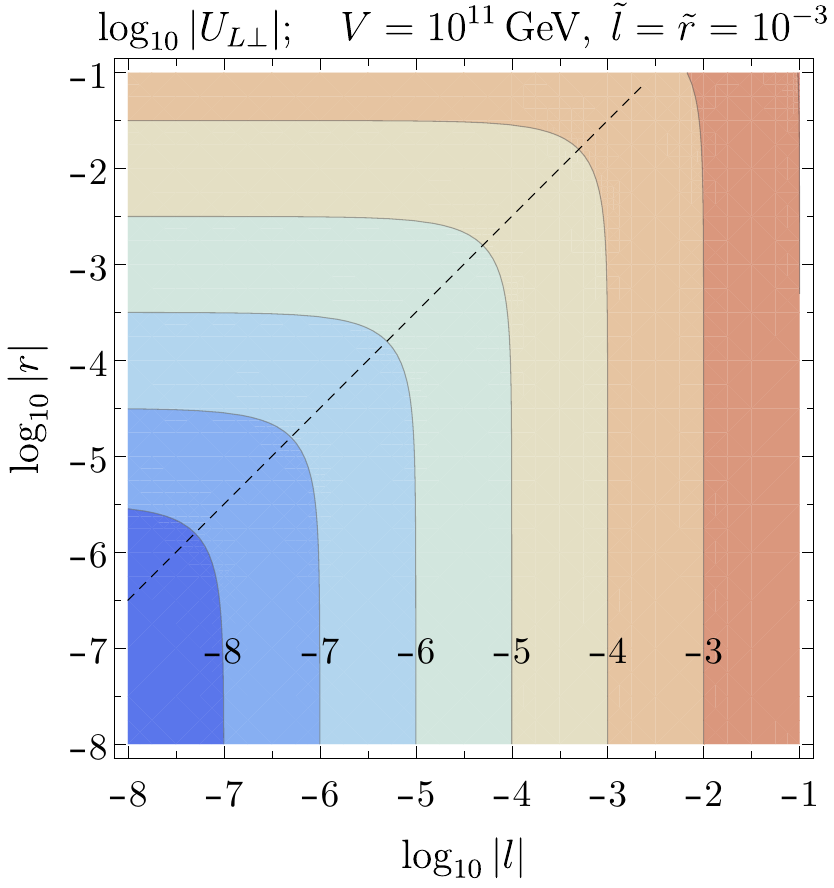}
            &
            \includegraphics[width=\SINGLEWIDTH]{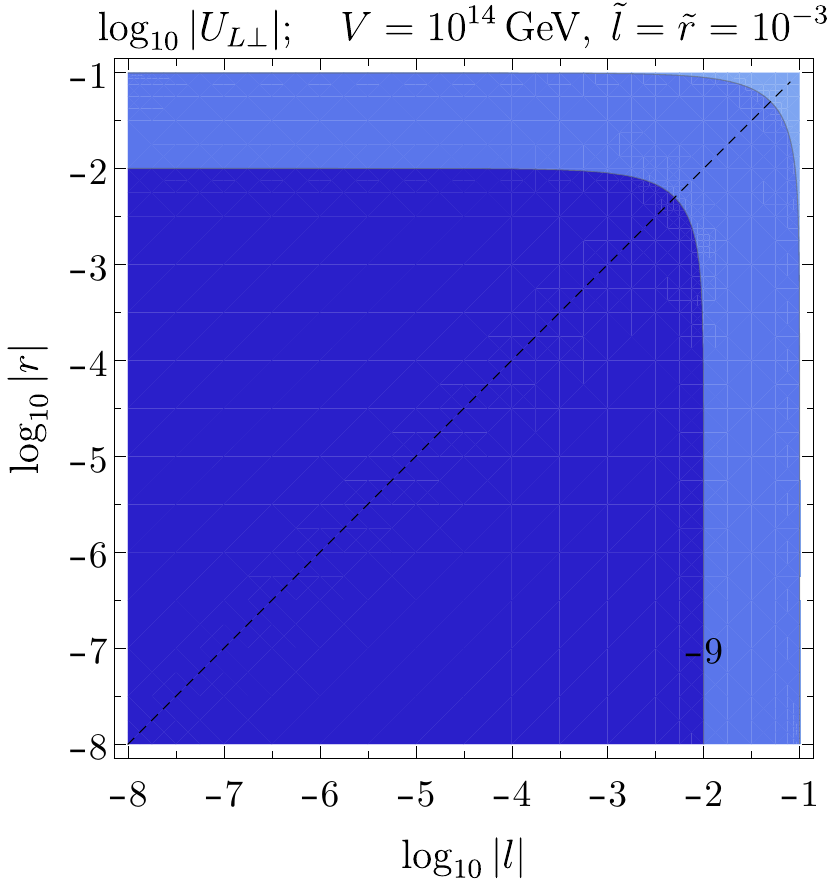}&
            \includegraphics[height=0.32\linewidth]{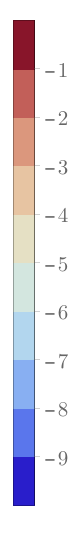}\\
            \includegraphics[width=\SINGLEWIDTH]{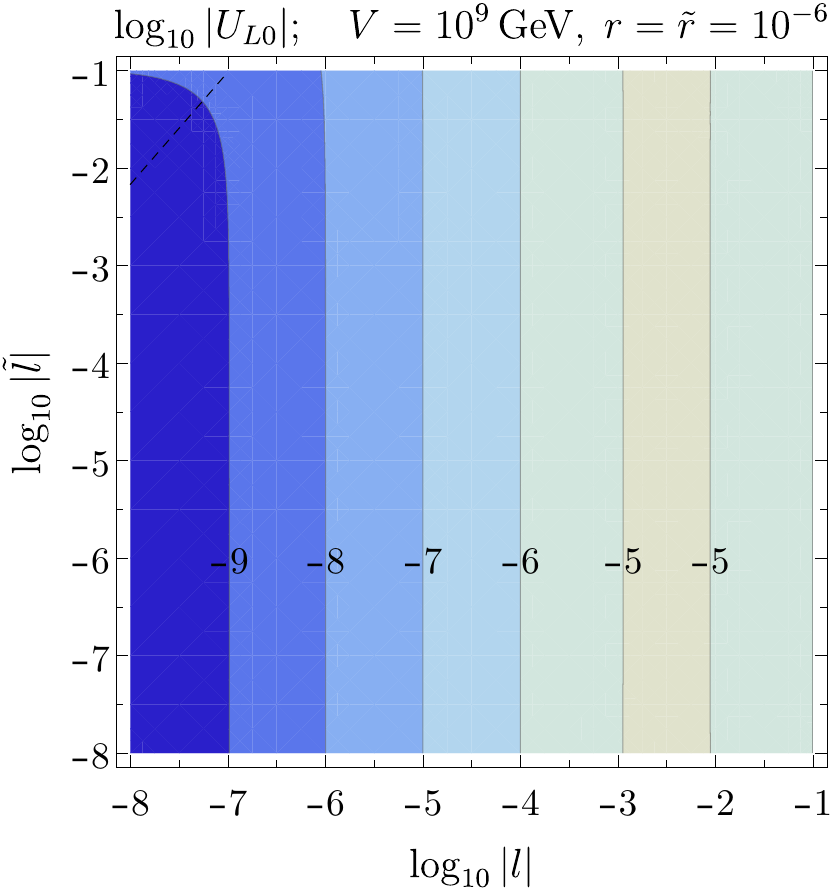}
            &
            \includegraphics[width=\SINGLEWIDTH]{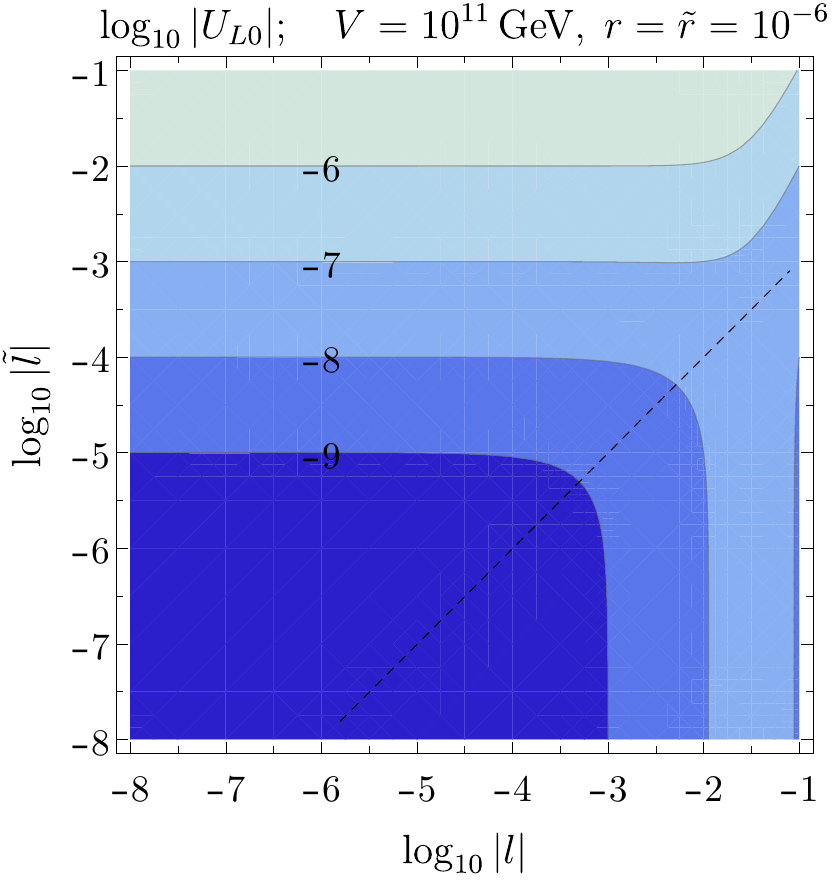}
            &
            \includegraphics[width=\SINGLEWIDTH]{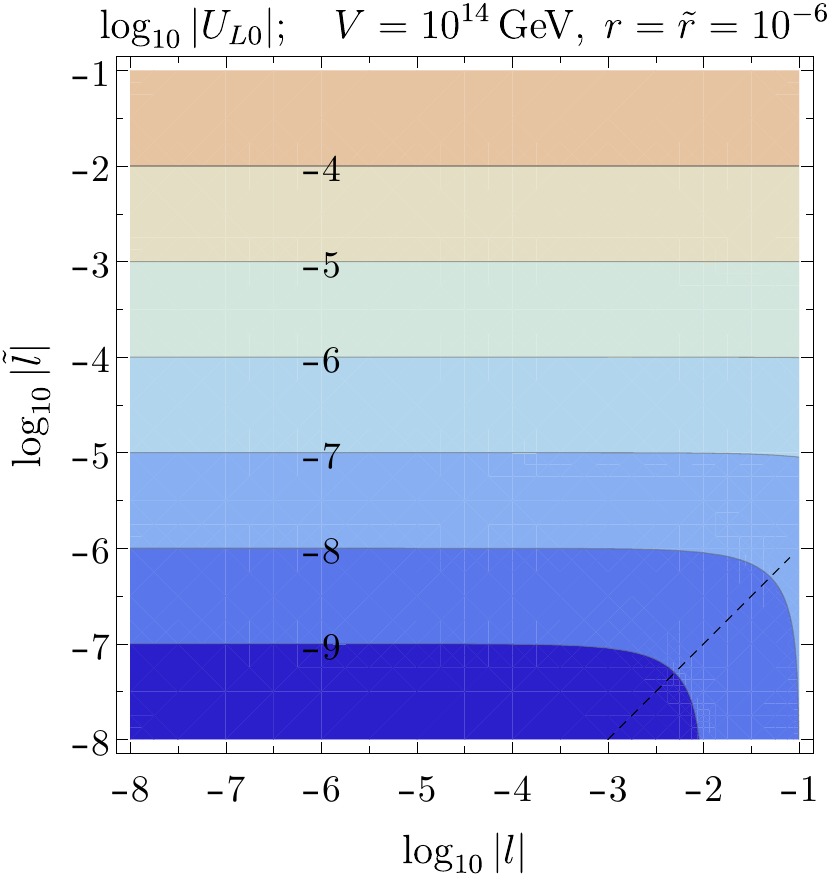}&
            \includegraphics[height=0.32\linewidth]{plots/plot-mix-legend.pdf}\\
        \end{tabular}
    \caption{Contour plots for $\log_{10}$ of the size of mixing of L-neutrinos with anomalons. The upper row shows mixing with heavy anomalons $U_{L\perp}$ in the $l$-$r$ plane and benchmark values $\tilde{l}=\tilde{r}=10^{-3}$. The lower row  shows mixing with light anomalons $U_{L0}$ in the $l$-$\tilde{l}$ plane and benchmark values $r=\tilde{r}=10^{-6}$. 
    The columns represent three different values of the PS-breaking scale: $V=10^{9,11,14}\,\mathrm{GeV}$. 
    The color coding is consistent for all plots, and the dashed lines represent where the contributions of the two terms from the $x$- and $y$-axis are equal, see main text.
    \label{fig:NA-mixing}
    }
\end{figure}

We now have a good analytic understanding of the spectrum and mixing angles. The latter are rather involved, so it is useful to consider some benchmark scenarios. Contour plots of $\log_{10}|U_{L\perp}|$ and $\log_{10}|U_{L0}|$ for three different scales $V$ are shown in \fig{fig:NA-mixing}, where the former are drawn in the $l$-$r$ plane and the latter in the $l$-$\tilde{l}$ plane. More features are elaborated on in the points below: 
\begin{itemize}[itemsep=0cm,leftmargin=0.5cm]
    \item Since the situation for light mixing $U_{L0}$ is quite involved, we take small benchmark values for the remaining parameters $r$ and $\tilde{r}$ in order for their contributions not to interfere at any scale $V$.
    \item Numerical diagonalization of \eqref{eq:mass-LRA-block4} rather than analytic formulae from \eqs{eq:terms-LP}{eq:terms-L0} was used for the plots. Consequently there is no upper bound on parameters limiting applicability, i.e.~the right panel of \fig{fig:NA-eigenvalues} does not apply. Note, however, that in such regions the spectrum may be modified relative to \eq{eq:LRA-masses} and non-linear features in the angles appear, specifically the inflection around $|l|\sim 10^{-2.5}$ in the bottom-left panel. 
    \item The parameters associated to the two axes in each panel were taken with opposite signs. The region where their contributions to the mixing angle become equal are visually shown with a dashed line.
    If same-sign parameters were taken, the two contributions can be canceled along the dashed line, as can be seen also analytically from \eqs{eq:terms-LP}{eq:terms-L0}, and thus the mixing angle is suppressed relative to the values in the plot.
\end{itemize}
We conclude the analysis of mixing angles by saying that we cross-checked the consistency of both the analytic expressions in \eqs{eq:terms-LP}{eq:terms-L0} as well as numeric behavior of \eq{eq:mass-LRA-block4} with the numeric behavior of the full $30\times 30$ neutrino-anomalon mass matrix of  \eqs{eq:mass-matrix-LRA}{eq:Nblock-AA}. The spectrum and mixing angles from the full matrix 
indeed show analogous behavior, with some further subtleties:
\begin{enumerate}[itemsep=0cm,label=(\alph*)]
    \item There are $3$ active neutrinos, $16$ heavy- and $8$ light anomalons in the spectrum. Therefore there are more regions in parameter space of mass cross-over and thus enhancement of the mixing angle.
    \item Whenever there is a sum of intermediate states in analytic expressions, the large number of states can easily cause an enhancement of one order of magnitude relative to the $4\times 4$ block case if most terms have the same sign.
    \item Beside the cancellation of contributions to mixing angles from different blocks (the dashed lines in \fig{fig:NA-mixing}), cancellations can also happen between the many parameters in the same block. For example, contributions from the $L\Psi_{\perp}$ block to mixing with some specific heavy anomalon state are suppressed if tuning of the right combination of parameters $\parA_{IN}$, $\parB_{IN}$, and $\parB'_{IN}$ is performed.
\end{enumerate}

%%%%%%%%%%%%%%%%%%%%%%%%%%%%%%%%%%%
\subsection{How realistic is the Yukawa sector?}
\label{sec:PSmodel-Yukawa-realistic}

We now address the question whether the Yukawa sector is realistic, i.e., it should reproduce the masses, mixing angles and CP-violating phases of the SM (including the neutrino oscillation data). A full numeric fit is beyond the scope of this paper, but we shall argue by counting independent parameters.

The fermion mass matrices $M_{U}$, $M_{D}$ and $M_{E}$ of the up-, down-, and charged lepton sectors are given in \eqs{eq:mass-mu}{eq:mass-me}. For neutrinos, we consider a simplified case where no mixing with anomalons occurs\footnote{Once the anomalon mixing is switched on, the number of available parameters for fitting neutrino observables increases, and hence the case we consider is a conservative one.}, and we thus consider only the Dirac and Majorana type mass blocks $M_{LR}$ and $M_{RR}$ in \eq{eq:Nblock-LR} (and further expanded in Eqs.~\eqref{eq:MLR}) and \eqref{eq:Nblock-RR}, respectively. Crucially, all these expressions already have the gauge and family redundancies removed, cf.~\app{app:PSmodel-Yukawa-computational-details}, by making the choice in \eq{eq:redundancy-removal-choice}.

Before proceeding into the somewhat tedious procedure of parameter counting, we note that fermion fits have been investigated before in similar contexts, in particular in models with $\SU(5)\times \SU(3)_{f}$ symmetry, see e.g.~\cite{Berezhiani:1983rk,Berezhiani:1982rr}. In our present Pati-Salam model, however, the manifestly left-right asymmetric implementation of flavor symmetry results in a somewhat unusual structure for mass matrices. Instead of being a linear combination of Yukawa matrices (as when flavor is not gauged), or an outer product of EW VEV vectors with an overall Yukawa coefficient (when both L- and R-fermions transform non-trivially under gauged flavor), we have a sum of terms, each consisting of an outer product of a vector of Yukawa parameters in L-flavor space and a vector of EW VEVs in R-flavor space, with no correlation between the two vectors. We postpone the discussion on how flavor hierarchy can arise within this structure to the end of this section.

Given our choice of redundancy removal, the mass matrices depend on the following set of parameters and VEVs:
\begin{align}
    \parY{\Phi}{1,2,3},&&
    \parY{\Sigma}{3},&&
    \parY{\Sigma'}{2,3},&&
    Y_{R},&\nonumber\\
    \vevv{u\Phi}{1,2,3},&&
    \vevv{u\Sigma}{1,2,3},&&
    \vevv{u\Sigma'}{1,2,3},&&
    \vevv{d\Phi}{1,2,3},&&
    \vevv{d\Sigma}{3},&&
    \vevv{d\Sigma'}{2,3},&&
    \colorPQ{Z}^{\ast}_{11,22,33,12,13,23},
    \label{eq:fit-parameters}
\end{align}
where we explicitly listed the available values for the flavor or family index in the subscript. This amounts in total to $7$ Yukawa parameters, $15$ EW VEVs and $6$ PS-breaking VEVs. Note that EW VEVs are sourced by the SM Higgs VEVs, hence the normalization condition
\begin{align}
        \sum_{A} \left(
        |\vevv{u\Phi}{A}|^{2}+
        |\vevv{u\Sigma}{A}|^{2}+
        |\vevv{u\Sigma'}{A}|^{2}+
        |\vevv{d\Phi}{A}|^{2}+
        |\vevv{d\Sigma}{A}|^{2}+
        |\vevv{d\Sigma'}{A}|^{2}\right)
        &=
        \tfrac{1}{2}\,v_{\rm EW}^{2}, \label{eq:normalization-EWvevs}
\end{align}
where $v_{\rm EW}=246\,\mathrm{GeV}$. The EW VEVs depend on the admixture of the SM Higgs in every doublet component, which is determined from their mass matrix, which in turn depends on the parameters of the scalar potential of \eqs{eq:V2}{eq:VC}. The structure of the scalar potential should be rich enough that the EW VEVs can be considered as independent, see \app{app:PSmodel-Yukawa-computational-details} for technical details and additional discussion.

The parameters in \eq{eq:fit-parameters} are complex, with the exception of $\parY{\Sigma}{3}$, $\parY{\Sigma'}{2}$, $\vevv{d\Sigma}{3}$ and $\vevv{d\Sigma'}{2}$, which are made real as part of redundancy removal, see~\app{app:PSmodel-Yukawa-computational-details} for details. With a large number of complex phases among the parameters, the fit of complex phases of the CKM and PMNS matrix are of least concern, and we thus focus on masses and mixing angles by use of real-valued parameters.

The real observables to be fit are $3$ masses each in the up-, down-, charge-lepton and neutrino\footnote{Although only $2$ differences of squared neutrino masses are currently measured, we consider all $3$ neutrino masses as requiring a hypothetical fit, demonstrating the viability of our scenario in generality.} sectors, $3$ CKM mixing angles and $3$ PMNS angles. This amounts to a total of $18$ values, which is less than the $7+15+6-1=27$ parameters of \eq{eq:fit-parameters}. Simple parameter counting thus suggests a fit should be possible. 

Parameter counting arguments can fail if there are any flavor structure limitations arising from the mass matrices. We therefore make further refinements to our argument, by considering which sets of mass matrix entries are independent functions\footnote{Whether entries $f_{i}$ are independent functions of parameters $x_j$ is determined by appealing to the implicit function theorem: the number of independent entries $f_i$ is equal to the matrix rank of the Jacobian matrix of derivatives $\partial f_{i}/\partial x_{j}$.} of the parameters listed in \eq{eq:fit-parameters}.
The following observations are strongly indicative of no obstructions to a successful flavor fit:
\begin{enumerate}[itemsep=0cm,leftmargin=0.7cm,label=(\textit{\roman*})]
    \item The mass matrix of active neutrinos reduces in the limit of no mixing with anomalons to the standard seesaw expression:
    \begin{align}
        M_{\nu}&= - M_{LR} M_{RR}^{-1} M_{LR}^{T}. \label{eq:nu-typeI}
    \end{align}
    Since $M_{RR}$ of Eq.~\eqref{eq:Nblock-RR} 
    is an arbitrary (complex) symmetric matrix, since it depends on PS-breaking VEVs $\colorPQ{Z}^{\ast}_{AB}$ appearing nowhere
    else in the Yukawa sector, $M_{\nu}$ can also be set to be a symmetric complex matrix of choice independently from the $U$, $D$ and $E$ sectors. The neutrino masses and PMNS mixing angles are thus taken care of. 
    \item The remaining mass matrices $M_{U}$, $M_{D}$ and $M_{E}$ should thus fit the $9$ masses (of the $U$,$D$,$E$ sectors) and $3$ CKM mixing angles. The use of implicit function theorem reveals which entries in $M_{U,D,E}$ are independent and can be used for which observables; we shall argue based on counting results given in \Table{tab:fermion-independent entries}:
    \begin{itemize}[leftmargin=0.5cm,itemsep=0cm]
        \item The number of independent diagonal entries in $M_{U,D,E}$ is $8$, namely $(M_{D})_{11}=(M_{E})_{11}$. This indicates $8$ of the $9$ masses can be fit independently by use of these entries, while the breaking of one relation requires involving off-diagonal entries.
        \item In the LR convention and limit of small mixings, the left mixing angles dominantly arise from the strictly-upper-triangular entries of a matrix. Enlarging the set of diagonal entries by the upper-triangular part in $M_{U}$, we see that the $3$ new entries are independent from prior ones, enabling a description of the $3$ CKM angles. 
        \item Taking also the upper triangular parts of $M_{D}$ and $M_{E}$, we gain $3$ more mixings, providing the up-to-now missing parameter to separate all the masses, thus completing the list of SM observables in the Yukawa sector. The overhead of $2$ leftover independent parameters could even conceivably be used to help with the PMNS fit that we already established to be possible by using only the VEVs $\colorPQ{Z}^{\ast}_{AB}$ in the neutrino sector.
        \item Including the lower triangular parts, the full matrices $M_{U,D,E}$ have in total $18$ independent parameters. This enlargement includes some
        right mixing angles that are unobservable in the SM.
    \end{itemize}
\end{enumerate}

\begin{table}[t!]
\begin{center}
    \begin{tabular}{lr}
    \toprule
    set of entries& $\#$ independent\\
    \midrule
    $\{\diag M_U,\diag M_D,\diag M_E\}$&$8$\\[3pt]
    $\{\upp M_U,\diag M_D,\diag M_E\}$&$11$\\[3pt]
    $\{\upp M_U,\upp M_D,\upp M_E\}$&$14$\\[3pt]
    $\{M_U,M_D,M_E\}$&$18$\\
    \bottomrule
    \end{tabular}
    \caption{ The number $\#$ of independent entries with respect to the parameters of \eq{eq:fit-parameters} in various subsets of mass-matrix entries from $M_{U,D,E}$ in \eqs{eq:mass-mu}{eq:mass-me}. The diagonal and upper-triangular part of a matrix are denoted by $\diag$ and $\upp$, respectively.\label{tab:fermion-independent entries}}
\end{center}
\end{table}

In summary, by all indications a fit of the SM masses, mixing angles and phases should be possible already in the case $N_{\Phi}=1$ and $N_{\Sigma}=2$, as has been alluded to in \sect{sec:SMflavour-QandL}. 

We conclude with a discussion on how the hierarchical masses of the SM can arise in our setup and their possible relation to a horizontal hierarchy of PS-breaking VEVs. As was already alluded to, the up-, down-, and charged lepton-sector mass matrices have a particular structure: a sum of outer products of vectors, with one vector being Yukawa parameters $\parY{}{}$ and the second vector being the EW VEVs $\vevv{}{}$, cf.~\eqs{eq:mass-mu}{eq:mass-me}. The flavor hierarchy can arise from either type of object, or a mixture of both, i.e.~from hierarchy conditions of the type $\parY{\Phi}{3}\gg \parY{\Phi}{2}\gg \parY{\Phi}{1}$ or $\vevv{u\Phi}{3}\gg \vevv{u\Phi}{2}\gg \vevv{u\Phi}{1}$ (and similar for other Yukawa and VEV EW vectors). The hierarchy in EW VEVs, however, is dynamically linked to the PS-breaking VEVs through the Higgs doublet mass matrix, cf.~\app{app:PSmodel-Yukawa-computational-details}. More specifically, some of the contributions in the doublet mass matrix are flavor non-universal, involving flavor-dependent VEVs $\vevV{A}$ and $\vevZ{AB}$; if these dominate over the flavor-universal ones, the doublet mass matrix may become hierarchical with respect to R-flavor. The EW VEVs represent the effectively null mass eigenmode of that matrix (after fine-tuning) subject to the normalization condition in~\eq{eq:normalization-EWvevs}. Smaller entries in this mass matrix typically signify a larger admixture of that component in the null eigenmode, and hence the flavor hierarchy in EW VEVs should be inverted relative to the horizontal hierarchy in $
\vevV{A}$ and $\vevZ{AB}$ (in a fixed basis of $R$-flavor). The inversion is also present in the type I seesaw expression of \eq{eq:nu-typeI} due to the inversion $M_{RR}^{-1}$. To which degree the hierarchy can come from the EW VEVs compared to the Yukawa parameters would ultimately be answered by a numeric fit, which is, however, beyond the scope of this paper.

%%%%%%%%%%%%%%%%%%%%%%%%%%%%%%%%%%%%%%%%%%%%%
%%%%%%%%%%%%%%%%%%%%%%%%%%%%%%%%%%%%%%%%%%%%%
\section{Phenomenological profile of the accidental/high-quality axion}  
\label{sec:axionpheno} 

In this section, we present a phenomenological overview 
of the accidental Pati-Salam axion, touching upon 
astrophysical axion limits, axion cosmology and direct searches. 
In summary, the key features closely resemble those of the benchmark DFSZ model \cite{Zhitnitsky:1980tq,Dine:1981rt}, 
albeit with significant constraints on the 
axion decay constant, 
which narrow down the allowed ranges 
for the axion mass: 
$m_a \in [2 \times 10^{-8}, 10^{-3}]\,\mathrm{eV}$ 
(accidental $\U(1)_{\rm PQ}$, pre-inflationary PQ breaking scenario)
and $m_a \gtrsim 0.01\,\mathrm{eV}$ 
(high-quality $\U(1)_{\rm PQ}$, post-inflationary PQ breaking scenario).

\subsection{Astrophysical constraints}  
\label{eq:astrolimits}

The goal of this section is to provide a
lower bound on the axion decay constant, 
stemming from astrophysical limits on the axion couplings 
(for reviews on axion astrophysics, see e.g.~\cite{DiLuzio:2021ysg,Caputo:2024oqc,Carenza:2024ehj}). 
The main constraints within our model arise from the axion couplings 
to electrons and nucleons. The latter can be expressed in terms 
of the axion-quark couplings provided in \eq{eq:cucdce} 
(see e.g.~\cite{DiLuzio:2024vzg,diCortona:2015ldu}), 
obtaining
\begin{align} 
\label{eq:axionnuclcoupl}
c_p &= -0.45\phantom{0} + 0.29 \cos^2\beta-0.15 \sin^2\beta \, , \\ 
c_n &= +0.013 - 0.14 \cos^2\beta + 0.27 \sin^2\beta \, . 
\end{align}
Defining $g_{aX} = c_X m_X / f_a$, with $X=p,n,e$, we consider the stringent limits imposed by 
Supernova (SN) 1987A \cite{Carenza:2019pxu,Carenza:2020cis},
as well as by the observed evolution of red giants \cite{Capozzi:2020cbu,Straniero:2020iyi}
and white dwarfs \cite{MillerBertolami:2014rka,Corsico:2019nmr}, 
yielding respectively
$g_{an}^2+0.61 g_{ap}^2+0.53 g_{an} g_{ap} \lesssim 8.3 \times 10^{-19}$ and 
$g_{ae} \lesssim 2 \times 10^{-13}$. 
Combining these two limits, the
lowest value for $f_a$ is obtained 
by saturating the lower bound on $\beta$ set by perturbativity (i.e.~$\tan\beta = 0.25$), for which we obtain 
$f_a \gtrsim 2 \times 10^8 \, \mathrm{GeV}$.   
We observe that, for $\tan\beta \lesssim 0.5$, 
this limit is basically 
constant and is dominated by the SN 1987A bound. 

Given the structural uncertainties related to the SN bound (see e.g.~\cite{Fiorillo:2023frv}), 
it is worth to mention that for $\tan\beta = 0.25$ 
the bound on the axion decay constant stemming from the 
(tree-level)
axion-electron coupling 
is $f_a \gtrsim 5 \times 10^7\,\mathrm{GeV}$. 
Note, however, 
that 
in the small $\tan\beta$ limit
the axion-electron coupling (cf.~\eq{eq:cucdce}) 
receives large 
radiative corrections from the axion coupling to the 
top-quark \cite{Choi:2017gpf,Chala:2020wvs,Bauer:2020jbp,Bonilla:2021ufe,Choi:2021kuy,DiLuzio:2022tyc,DiLuzio:2023tqe}. Including the latter correction, 
one obtains $c_e \simeq \frac{1}{3} \sin^2\beta 
+ r^t_e (m_{\rm BSM}) \frac{1}{3} \cos^2\beta$, 
where $m_{\rm BSM}$ denotes the energy scale of the 
non-SM Higgs doublets. 
Assuming e.g.~that the effective theory below $f_a$ is 
the SM, i.e.~$m_{\rm BSM} = f_a \simeq 10^8\,\mathrm{GeV}$, 
one finds $r^t_e (m_{\rm BSM}) \simeq 0.23$ 
(cf.~Table B.4 in Ref.~\cite{DiLuzio:2023tqe}), 
from which we obtain 
\beq 
\label{eq:astrobound}
f_a \gtrsim 2.3 \times 10^8 \, \mathrm{GeV} \, ,  
\eeq
that is slightly stronger than the SN bound derived above. 
The 
lower limit in \eq{eq:astrobound} 
plays 
an important role for the issue of 
PQ quality, as discussed in \sect{sec:PQquality}.

\subsection{Axion cosmology}  
\label{eq:axioncosmo}

The cosmological evolution of the axion field 
and the associated cosmological observables 
crucially 
depend on
the interplay between the inflationary scale and the PQ symmetry-breaking scale, with standard considerations applicable 
also to our model (for reviews on axion cosmology see e.g.~\cite{Marsh:2015xka,OHare:2024nmr}). 

If the PQ is broken after inflation or restored afterwards (post-inflationary PQ breaking scenario), one also has topological defects, 
i.e.~cosmic strings and domain walls, that contribute to the axion 
dark matter relic density, on top of the usual misalignment mechanism \cite{Dine:1982ah,Abbott:1982af,
Preskill:1982cy}. In particular, the contribution from topological defects can be significant and difficult to assess (see e.g.~\cite{Gorghetto:2020qws,Buschmann:2021sdq,Benabou:2024msj}), 
so that a robust prediction for the axion dark matter mass is not 
possible at the moment. In any case, one can derive a 
lower bound on the axion dark matter mass, $m_a \gtrsim 28 \, \mu\text{eV}$ (equivalently $f_a \lesssim 2.0 \times 10^{11}\,\mathrm{GeV}$) \cite{Borsanyi:2016ksw},  
from the requirement of not overshooting the dark matter relic density. 
Note, however, that the entire dark matter 
abundance can be reproduced also for 
$m_a \gg 28 \, \mu\text{eV}$, 
either through the contribution of topological defects 
(see e.g.~\cite{Gorghetto:2020qws,Kawasaki:2014sqa})
or non-standard axion production mechanisms
(see e.g.~\cite{DiLuzio:2024fyt}). 

Specifically for our model, from \eq{eq:NanomalyUV} it follows that $N_{\rm DW} \equiv 2N = 12$, so the model has a domain wall problem in the post-inflationary PQ-breaking scenario. 
The Lazarides-Shafi solution \cite{Lazarides:1982tw,Barr:1982bb}, 
in which the $\mathbb{Z}_{N_{\rm DW}}$ 
discrete symmetry connecting the $N_{\rm DW}$ 
degenerate minima of the 
axion potential is
fully embedded in the center of a non-abelian symmetry group, applies neither to the present 
Pati-Salam model nor to the original 
$\SO(10)$ model of 
Ref.~\cite{DiLuzio:2020qio}. 
More precisely, denoting the generator of the $\mathbb{Z}_{N_{\rm DW}}$ 
symmetry as $r_{12}$ and that of the centers of the 
$G_{\rm PS} \times \SU(3)_{f_R}$
gauge symmetries as $r_4r_2r_2r_3$,  
it can be checked that there is no solution 
for the system of equations 
$r_{12}=r_4^mr_3^nr_2^kr_2^l$, 
for integer values of $m,n,k,l$,
when applied to the field content 
in \Table{tab:PSirrep}. 
Hence, to address the domain wall problem, one needs 
to invoke a small breaking of the $\U(1)_{\rm PQ}$ via Planck suppressed operators, which could remove the degeneracy among the 
vacua of the axion potential, effectively 
causing the domain walls to 
decay before they dominate the energy density of the universe \cite{Sikivie:1982qv}. While the latter scenario is consistent with the general approach discussed here, 
in terms of an approximate $\U(1)_{\rm PQ}$ symmetry, 
the parameter space for such a solution is somewhat limited \cite{Beyer:2022ywc}. 
The aforementioned aspects 
reflect the tension between cosmology and the PQ 
quality problem, associated with the post-inflationary PQ breaking scenario \cite{Lu:2023ayc}. 

Conversely, if the PQ symmetry is broken before inflation and not restored afterwards (pre-inflationary PQ-breaking scenario), the axion 
dark matter relic density depends on the (random) initial 
misalignment angle and hence it cannot be predicted. 
Nonetheless, dark matter axions are allowed to have masses of at most  
$m_a \lesssim 1\,\mathrm{meV}$ (equivalently $f_a \gtrsim 5.7 \times 10^{9}\,\mathrm{GeV}$), 
because quantum fluctuations during inflation 
would imply too large iso-curvature fluctuations \cite{Wantz:2009it}.

\subsection{Axion searches}  
\label{eq:axionsearches}

\begin{figure}[t!]
\centering
\includegraphics[width=16.cm]{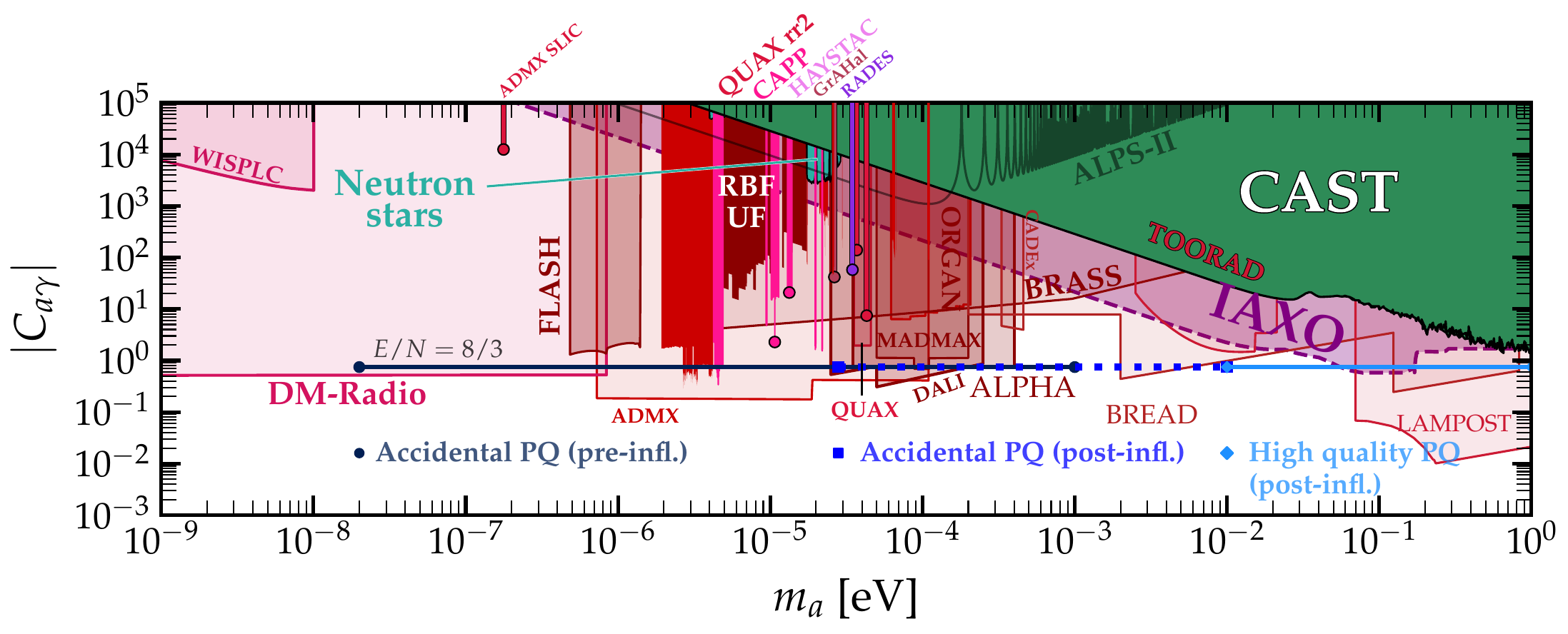}
\caption{Axion-photon coupling (\eq{eq:axionph} with $E/N = 8/3$) as a 
function of the axion mass. 
Current limits and future sensitivities 
are represented by filled and shaded regions, 
respectively. 
A high-quality PQ symmetry is obtained for $m_a \gtrsim 0.01\,\mathrm{eV}$ (light blue line)
in the post-inflationary PQ-breaking scenario. 
An accidental PQ symmetry 
(not addressing the PQ-quality problem) 
corresponds to the intervals 
$m_a \in [2 \times 10^{-8}, 10^{-3}]\,\mathrm{eV}$
in the pre-inflationary PQ-breaking scenario 
(black line)
and $m_a \in [2.8 \times 10^{-5}, 0.01]\,\mathrm{eV}$
in the post-inflationary PQ-breaking scenario 
(blue/dashed line). 
Axion limits adapted from~\cite{AxionLimits}. 
}
\label{fig:axionphoton}       
\end{figure}

Axion couplings to photons and SM matter 
fields have been computed in 
\app{sec:loweaxion} (in \eq{eq:axionph} and \eqs{eq:axioncu}{eq:axioncde}, respectively). 
For the present scenario, 
the axion coupling to photons 
(displayed in \fig{fig:axionphoton} as a function of the axion mass)
represents the main experimental probe in the context of axion physics. 
The mass range of the accidental Pati-Salam axion 
is constrained by various considerations. 

An absolute lower bound on $m_a$ is obtained 
within the model, since $f_a$ is bounded from above by the seesaw scale. 
From \eq{eq:axiondc} we find the upper limit 
$f_a \leq V_{B-L}  / 12$,
with $V_{B-L} \equiv \sqrt{V^2_{\chi} + 4 V^2_{\Delta}}$. 
Here, $V_{B-L}$ 
assumes the
meaning of $B-L$ breaking scale, 
proportional to the mass of the $B-L$ gauge boson.  
This is  
subject to the constraints 
from the neutrino mass spectrum.   
In particular, the light neutrino masses are mainly controlled 
by $V_\Delta$ through the 
right-handed neutrino 
mass contributions,   
$M_{RR} \sim Y_{R} V_{\Delta}$. 
In contrast, $V_{\chi}$ has only a minor effect on the neutrino spectrum, as its contribution arises solely through the mixing with anomalons via non-renormalizable operators 
(cf.~\sect{sec:SMflavour-NandA}).
Setting a robust upper limit on $V_{\Delta}$ 
would necessitate a fit to fermion masses and mixings, 
which goes beyond the scopes of the present work. 
However, barring fine-tuned regions in the global fit, 
one requires $V_\Delta \lesssim 10^{15}\,\mathrm{GeV}$ 
not to undershoot the active neutrino mass scale. 
Given this bound on $V_{\Delta}$, with $V_{\chi}$ unconstrained, 
the upper limit on $f_a$ is obtained in the $V_{\chi} \gg V_{\Delta}$ limit, corresponding to $f_a \approx V_{\Delta}/3$. 
Hence, we obtain $f_a \lesssim 3 \times 10^{14}\,\mathrm{GeV}$ 
(equivalently $m_a \gtrsim 2 \times 10^{-8}\,\mathrm{eV}$).
Other constraints on the axion mass arise from cosmology, astrophysics and the requirement of PQ quality, as previously discussed and recalled in the caption of 
\fig{fig:axionphoton}, which shows 
the prediction of the model (blue line)
in the axion-photon vs.~axion-mass plane, 
including also present constraints 
and future sensitivities \cite{DePanfilis:1987dk,Wuensch:1989sa,CAST:2007jps,CAST:2017uph,McAllister:2017lkb,Baryakhtar:2018doz,ADMX:2018gho,Alesini:2019ajt,IAXO:2019mpb,Lawson:2019brd,Crisosto:2019fcj,HAYSTAC:2020kwv,Ortiz:2020tgs,Alesini:2020vny,Schutte-Engel:2021bqm,CAST:2020rlf,ADMX:2021nhd,Grenet:2021vbb,Zhang:2021bpa,BREAD:2021tpx,Foster:2022fxn,DMRadio:2022pkf,Alesini:2022lnp,Yi:2022fmn,HAYSTAC:2023cam,DeMiguel:2023nmz,Alesini:2023qed,CAPP:2024dtx,Quiskamp:2024oet,ADMX:2024xbv,Garcia:2024xzc}. 
In particular, we introduce two main scenarios for the allowed axion mass range in \fig{fig:axionphoton}: 
\begin{itemize}
\item[\text{(I)}] {\bf High-quality PQ.} 
This scenario,  
corresponding to a post-inflationary PQ breaking,  
is obtained for $m_a \gtrsim 0.01\,\mathrm{eV}$ 
(light blue line in the right side of \fig{fig:axionphoton}). 
This setup is not directly testable at the moment, 
since axion experiments 
(both helioscopes and haloscopes) lose sensitivity 
in the 
$m_a \sim 0.01\,\mathrm{eV}$ ballpark, while 
for $m_a \gtrsim 0.025\,\mathrm{eV}$ 
astrophysical constraints
prevent the full visibility of this parameter space. On the other hand, given the remnant $|\theta^{\rm min}_{\rm eff}|\simeq4.7\times 10^{-13}$ in correspondence with
astrophysical limits (cf.~\sect{sec:PQquality}),
an indirect way to test 
the high-quality PQ scenario  
is provided by the search for nucleon EDMs, 
which will improve the 
current sensitivity on 
$\theta_{\rm eff}$ 
by up to 
three orders of magnitude \cite{nEDM:2019qgk,n2EDM:2021yah,Alexander:2022rmq}. 
\item[\text{(II)}] {\bf Accidental PQ.} 
This scenario, characterized by an accidental PQ symmetry that does not resolve the PQ-quality problem, can occur in either the pre- or post-inflationary PQ-breaking regime (in the latter case only for $m_a \gtrsim 28 \, \mu\text{eV}$  \cite{Borsanyi:2016ksw}). 
The two cases correspond respectively to the black line 
on the left side of \fig{fig:axionphoton} 
($m_a \in [2 \times 10^{-8}, 10^{-3}]\,\mathrm{eV}$)
and to the blue/dashed line ($m_a \in [2.8 \times 10^{-5}, 0.01]\,\mathrm{eV}$). 
Barring a hole in sensitivity around $m_a \sim 10^{-3}\,\mathrm{eV}$, 
the parameter space of the accidental Pati-Salam axion 
will be mostly explored 
in the coming decades, under the crucial assumption that the axion comprises the entirety of dark matter.

\end{itemize}

%%%%%%%%%%%%%%%%%%%%%%%%%%%%%%%%%%%
%%%%%%%%%%%%%%%%%%%%%%%%%%%%%%%%%%%
\section{Anomalon cosmology}
\label{sec:anomalon-cosmology}

The presence of parametrically light anomalons 
in the spectrum is a direct consequence of the gauging of the $\SU(3)_{f_R}$ flavor group. In this regard, the 
dynamics of anomalons 
represents a possible low-energy signature
of the UV mechanism addressing the accidental origin of the PQ symmetry and the solution to the PQ quality problem.
The mass spectrum of anomalons strongly depends on the PQ breaking scale $f_a$, or equivalently on the scale $V = 3 \sqrt{5} f_a$ (cf.~Eq.~\eqref{eq:axiondc}), 
as shown in the left panel of Fig.~\ref{fig:NA-eigenvalues}. The main result of Sect.~\ref{sec:SMflavour-NandA} is that anomalons split into two sets: 16 \textit{heavy} anomalons with mass $m_{\perp}$ and 8 \textit{light} anomalons with mass $m_0$. The mass spectrum directly determines the cosmological history of the anomalon sector. We discuss this in two particular scenarios, 
following the classification introduced in Sect.~\ref{eq:axionsearches}:
\begin{itemize}
     \item[\text{(I)}] 
     \textbf{High-quality PQ.} The PQ scale lies in the interval $f_a\in [2.3,5.6]\times 10^8\,\mathrm{GeV}$, corresponding roughly to $V\sim 10^9\,\mathrm{GeV}$. There are $N_\Psi=24$ ultra-light anomalons with masses below $0.1\,\mathrm{eV}$, as we observe from Fig.~\ref{fig:NA-eigenvalues} (the distinction between heavy and light anomalons here is mostly irrelevant). The PQ symmetry is broken after the end of inflation, $T_{\rm RH}>f_a$. 

     \item[\text{(II)}] 
    \textbf{Accidental PQ.} The PQ scale lies in the interval $f_a\in [5.7\times 10^9, 3\times 10^{14}]\,\mathrm{GeV}$, corresponding to $V\in [4\times 10^{10} , 2 \times 10^{15}]\,\mathrm{GeV}$. The reheating temperature could satisfy either $T_{\rm RH}<f_a$ (pre-inflationary) or $T_{\rm RH}>f_a$ (post-inflationary).
    We discuss the following two benchmark 
    sub-cases, referring to the results shown in Fig.~\ref{fig:NA-eigenvalues}:
     \begin{itemize}
        \item[\text{(IIa)}] for $V=10^{14}\,\mathrm{GeV}$ ($f_a\sim 10^{13}\,\mathrm{GeV}$), we have $m_{\perp}\sim 10$ TeV and $m_{0}\sim\mathrm{keV}$;
        \item[\text{(IIb)}] for $V=10^{11}\,\mathrm{GeV}$ ($f_a\sim 10^{10}\,\mathrm{GeV}$), we have $m_{\perp}\sim 10\,\mathrm{keV}$ and $m_0\sim \mu\mathrm{eV}$. 
     \end{itemize} 
\end{itemize}
Following standard notation in cosmology, 
we define $T$ as the temperature of the photon thermal bath (or more generically the SM thermal bath for $T\gtrsim\mathrm{MeV}$, prior to neutrino decoupling). We introduce the Hubble expansion rate, $H(T)$, and the entropy density of the universe $s(T)$ as
\begin{equation}
H(T)=\sqrt{\frac{4\pi^3 g_*(T)}{45}}\frac{T^2}{M_{\rm Pl}} \, , \qquad s(T)=\frac{2\pi^2}{45}g_s(T)T^3 \, ,
\end{equation} 
with the expression for the Hubble rate holding
in a radiation-dominated universe.
The functions $g_s(T)$ and $g_*(T)$ count the number of relativistic degrees of freedom (DOF) in the thermal bath
and only differ at temperatures below $\sim 0.5\,\mathrm{MeV}$.
For instance, at temperatures above EW symmetry breaking (EWSB) the number of SM DOF is $g_*^{\rm SM}=g_s^{\rm SM}\equiv g_{\rm SM}=106.75$, while for $T\ll 0.5\,\mathrm{MeV}$, $g_s^{\rm SM}=43/11\simeq3.91$ and $g_*^{\rm SM}\simeq 3.36$. Extra particles (the axion, the anomalons, etc.) also contribute as long they are relativistic.
Finally, the equilibrium interaction rate density for a 2-to-2 fermion annihilation $\bar{F}F\to \bar{\psi}\psi$ is defined as 
\begin{equation}
\gamma_{{\rm eq}}(\bar{F}F\to \bar{\psi}\psi) =g_{F}^2\frac{T}{32\pi^4}\int_{4m_F^2}^\infty ds \,s^{3/2}\,\sigma_{\bar{F}F\to\bar{\psi}\psi}(s)\left(1-\frac{4m_F^2}{s}\right)K_1(\sqrt{s}/T) \, ,
\end{equation}
where $\sqrt{s}$ is the center-of-mass (c.o.m.) energy, $K_1$ is a Bessel function, $g_{F}$ is the number of polarization DOF of the initial states ($g_F=2$ for a Weyl fermion), and $\sigma(s)$ is the cross section of the process. 
The corresponding (equilibrium) interaction rate $\Gamma_{\rm eq}$ is defined in terms of the equilibrium number density of the Weyl fermions $n_{F,\rm eq}$ as
\begin{equation}
    \Gamma_{\rm eq}=\frac{\gamma_{\rm eq}}{n_{F,\rm eq}} \, , \qquad n_{F,\rm eq}\simeq \frac{3\,g_F\,\zeta(3)}{4\pi^2}T^3 \text{\ \ for \ \ }T\gg m_F \, .
\end{equation}
The rest of this section is organized as follows: we present a general discussion on the production of anomalons in the early universe and their cosmological evolution in~\sect{sec:production}, and then discuss the phenomenology of anomalons as a component of dark radiation or dark matter in Sects.~\ref{sec:anomalonsDR} and \ref{sec:anomalonsDM}, respectively.

\subsection{Anomalon production in the early universe}\label{sec:production}

Anomalons communicate to the SM sector in different ways: $i)$ via the exchange of massive $\SU(3)_{f_R}$ flavor gauge bosons $W_{f_R}$, which interact at the renormalizable level  with both anomalons and SM right-handed fermions, $ii)$ through neutrino-anomalon mixings induced by the the non-renormalizable operators of Table~\ref{tab:Yukawa-nonrenormalizable} (cf.~discussion in section~\ref{sec:SMflavour-NandA}), which lead to neutrino-anomalon conversions, $iii)$ through effective Yukawa interactions, induced again by the non-renormalizable operators of Table~\ref{tab:Yukawa-nonrenormalizable}. 
We define the total anomalon-SM interaction rate as $\Gamma_\Psi=\Gamma_{\Psi}^{\rm flavor}+\Gamma_{\Psi}^{\nu-\rm mix}+\Gamma_{\Psi}^{\rm Yuk}$. Anomalons are kept in thermal equilibrium with the SM bath whenever
$\Gamma_{\Psi}(T) \gtrsim H(T)$. If this occurs, equilibrium is maintained until decoupling (or freeze-out) when $\Gamma_{\Psi}(T_{\rm dec})\simeq H(T_{\rm dec})$. 

On the other hand, if the interactions are sufficiently weak such that $\Gamma_{\Psi} \ll H$ holds at any time of cosmological evolution, anomalons never reach thermal equilibrium. Assuming a vanishing anomalon abundance just after the end of inflation, a non-thermal population is then produced from scatterings (i.e.~multiple scattering processes) of SM particles in the thermal bath, which occasionally produce anomalons as final states. This mechanism is generically referred to as freeze-in in the dark matter literature (see e.g.~\cite{Hall:2009bx}). 

As we will motivate extensively in the following, we are particularly interested in this second possibility (freeze-in). In such a case, the following independent populations of anomalons are produced:
\begin{enumerate}[label=(\textit{\roman*}),leftmargin=0.7cm]
    \item First, via flavor gauge interactions, namely fermion annihilations $\bar{F}^AF^B\to (W_{f_R})\to \overline{\Psi}_K^C\Psi_K^D$, where $F$ are right-handed SM fermions and the intermediate  step represents the $s$-channel exchange of a virtual flavor gauge boson. The anomalons are produced in the flavor basis: $A,B,C,D$ are $\SU(3)_{f_R}$ gauge indices and $K$ denotes the anomalon family, cf.~\sect{sec:SMflavour-NandA}, while all $W_{f_R}$ indices are left implicit.
    \item A second population is produced via neutrino mixing, for instance processes such as $e^+e^-\to \bar{\nu}_f\nu_f\to\bar{\nu}_f\Psi_m$, where $\nu_f$ is a left-handed neutrino in the flavor basis ($f=e,\mu,\tau$), $\Psi_m$ an anomalon in the mass basis ($m=1,\dots,24)$  and the last step involves a neutrino-anomalon mixing angle $\theta_{\nu_f\Psi_m}$. 
    This production mechanism is active only after EWSB (the mixing angles vanish before) and above $T\sim\,\mathrm{MeV}$ (neutrino decoupling). 
    Notice that we use a slightly different and more compact notation compared to Sect.~\ref{sec:SMflavour-NandA}: $(\nu_L)_I$ is replaced by $\nu_f$, while we denote the mixing between a flavor eigenstate $\nu_f$ and a mass eigenstate $\Psi_m$ as $\theta_{\nu_f\Psi_m}$.
    \item A third population is produced through effective Yukawa interactions generated via the non-renormalizable operators of Table~\ref{tab:Yukawa-nonrenormalizable} (also responsible for the anomalon-neutrino mixings discussed above). These include  effective $\bar{\nu}_L\Psi_R h$ and $\Psi_R\Psi_R h$ interactions ($h$ being the SM Higgs boson), as well as interactions of anomalons with heavy-scalar radial modes. 
    Thus, anomalons are produced through decays of heavy-scalar radial modes, Planck-suppressed scalar annihilations and 2-to-2 annihilation of SM particles mediated by the Higgs boson.
\end{enumerate}  

\subsubsection{Production via flavor interactions}

We provide simple estimates for the interaction rate of flavor-mediated annihilations and the corresponding anomalon population. The $\SU(3)_{f_R}$ gauge bosons are massless at temperatures above the flavor breaking scale $T>V$. At lower temperatures they acquire a mass, which scales as $m_{W_{f_R}}\sim g_{f_R}V< V$, in terms of the flavor gauge coupling $g_{f_R}$ (we assume $g_{f_R}<1$). The production cross section scales as (up to $\mathcal{O}(1)$  factors and phase space)
\begin{equation}
\label{eq:sigmaif}
\sigma_{\bar{F}F\to\bar{\Psi}\Psi}(s) \propto g_{f_R}^4\times \begin{cases}
    1/s \hspace{0.1cm} &\quad \text{ if } T> V \\
   s/\left((s-m_{W_{f_R}}^2)^2+m_{W_{f_R}}^2\Gamma_{W_{f_R}}^2\right) \hspace{0.1cm} &\quad \text{ if } T\leq V \\
\end{cases} \, , 
\end{equation}
with $\Gamma_{W_{f_R}}\propto g_{f_R}^2m_{W_{f_{R}}}\ll m_{W_{f_R}}$ 
being the decay width of $W_{f_R}$, while $F\in\{u_R,d_R,\nu_R,e_R\}$ generically denotes the right-handed fermions (with flavor indices left implicit). 
In the lower case of \eq{eq:sigmaif}, 
we recognize the Breit-Wigner formula, so that the cross-section gets a resonant enhancement when the c.o.m.~energy of the process is $\sqrt{s}=m_{W_{f_{R}}}$. 

In all the parameter space of our interest, $m_\Psi\ll m_{W_{f_R}}$ holds, and thus anomalons can be treated as effectively massless. 
We first focus on those annihilation channels which satisfy $m_{W_{f_R}}\gg m_F$.
Far from the resonant region, the interaction rate can be estimated as
\begin{equation}\label{eq:GvsHflavor1}
    \frac{\Gamma_{\Psi}^{\rm flavor}}{H}\sim \frac{\mpl}{g_{*}^{1/2}T}\times 
    \begin{cases}
    (g_{f_R}^2/4\pi)^2 &\text{if } T\gg m_{W_{f_R}} \\
    (T/V)^4  & \text{if }T\ll m_{W_{f_R}}
    \end{cases} \, .
\end{equation}
In the resonant peak we can adopt the small-width approximation, so that the cross section is
$
\sigma(s\sim m_{W_{f_R}}^2)\sim g_{f_R}^4\pi s\,\delta(s-m_{W_{f_R}}^2)/({m_{W_{f_R}}\Gamma_{W_{f_R}}})
$
(see also~\cite{Heeck:2014zfa}).
The resonant peak corresponds to $T\simeq m_{W_{f_R}}$, and the interaction rate is enhanced as
\begin{equation}
   \frac{\Gamma_{\Psi}^{\rm flavor}}{H}\bigg|_{\rm peak}\equiv \frac{\Gamma_{\Psi}^{\rm flavor}}{H}({T\simeq m_{W_{f_R}}})\sim \sqrt{\frac{g^2_{f_R}}{g_*(m_{W_{f_R}})}}\frac{M_{\rm Pl}}{V} \, .
\end{equation}
The same result can equivalently be derived by considering the direct decay of $W_{f_R}$ to anomalons, along with its inverse processes.
Flavor interactions are weak enough to avoid anomalon thermalization if  $(\Gamma_{\Psi}^{\rm flavor}/H)|_{\rm peak}\ll1$, which requires very small values for the flavor gauge coupling 
\begin{equation}\label{eq:gcoupling}
    g_{f_R}\ll 10^{-9}\left(\frac{V}{10^9\,\mathrm{GeV}}\right)\left(\frac{g_*(m_{W_{f_R}})}{106.75}\right)^{1/2} \, .
\end{equation}
In such a scenario, the flavor gauge bosons can be quite light ($\lesssim \mathcal{O}(\mathrm{GeV})$ for $V=10^9\,\mathrm{GeV}$), but very weakly coupled to SM matter (the lighter, the less coupled). As long as $g_{f_R}\gtrsim10^{-12}(10^9\,\mathrm{GeV}/V)$, there is always at least one resonant channel (electron-positron scattering).

Such small values of the gauge coupling can be avoided if the reheating temperature $T_{\rm RH}$ is low enough. Indeed, if  $T_{\rm RH}\ll m_{W_{f_R}}$, the resonant enhancement is avoided, the cross-section always scales as $s/m_{W_{f_R}}^4$, and the production is maximal at $T=T_{\rm RH}$. Using the estimate in Eq.~\eqref{eq:GvsHflavor1}, anomalon thermalization is avoided even for an $\mathcal{O}(1)$ gauge coupling as long as
\begin{equation}\label{eq:TRH}
    T_{\rm RH}\ll V \left(\frac{g_*^{1/2}(T_{\rm RH})V}{\mpl}\right)^{1/3}\simeq 10^6\,\mathrm{GeV}\left(\frac{V}{10^9\,\mathrm{GeV}}\right)^{4/3}\left(\frac{{g_{*}(T_{\rm RH})}} {106.75}\right)^{1/6}\, .
\end{equation}
The allowed values of the reheating temperatures belong to 
the range $4\,\mathrm{MeV} \leq T_{\rm RH} \lesssim 10^{16}\,\mathrm{GeV}$ \cite{Planck:2018vyg,Kawasaki:2000en,Hannestad:2004px,DeBernardis:2008zz}. 
However, Eq.~\eqref{eq:TRH} is only compatible with the pre-inflationary scenario, and can thus only be realized in the accidental PQ scenario (II).

Finally, we comment on those annihilation channels satisfying $m_F\gg m_{W_{f_R}}$, or equivalently $g_{f_R}\ll m_F/V$, which are thus not resonantly enhanced. Notice that this condition is satisfied for sterile neutrinos, as their mass is $m_{\nu_R}\equiv m_{R}\sim V$. If $m_F\ll T_{\rm RH}$, the cross section for anomalon production grows with decreasing temperature, and the production is maximal at $T\simeq m_F$. Non-thermalization implies the conservative constraint\footnote{This is obtained as follows: the cross section is maximal at $T\sim m_F$, at which $\sigma\sim g_{f_R}^4/m_F^2$. The condition $(\Gamma/H){|_{T\simeq m_F}}\ll 1$ implies $g_{f_R}\ll (\sqrt{g_*}m_F/M_{\rm Pl})^{1/4}$. Since $m_F>g_{f_R}V$ by assumption, we get the expression in the main text as a conservative estimate.} $g_{f_R}\ll (\sqrt{g_*}V/M_{\rm Pl})^{1/3}$, which is always weaker than the constraints from resonant annihilation channels and $W_{f_R}$ (inverse) decays. If $m_F\gg T_{\rm RH}$, the production channel is exponentially suppressed and irrelevant. 

In summary, the anomalons never reach thermal equilibrium if either of the following conditions are met: 
Eq.~\eqref{eq:gcoupling}, which allows for a large 
reheating temperature up to $T_{\rm RH}\lesssim 10^{16}\,\mathrm{GeV}$, or Eq.~\eqref{eq:TRH} with $g_{f_R}\lesssim \mathcal{O}(1)$. The freeze-in population of anomalons can be estimated in terms of the yield, $Y_\Psi\equiv n_\Psi/s$, as $Y_\Psi^{\rm flavor}\sim \gamma_{\rm eq}^{\rm flavor}/(Hs)$, giving
\begin{equation}\label{eq:anomalonAbundanceFlavor}
    Y_\Psi^{\rm flavor}\sim
  \begin{cases}
      \frac{0.3\,g_{f_R}M_{\rm Pl}}{g_s^{3/2}V}\simeq 3\times 10^{-4}\,\(\frac{g_{f_R}}{10^{-10}}\)\(\frac{10^9\,\mathrm{GeV}}{V}\)\left(\frac{106.75}{g_s}\right)^{3/2} &{\text{if Eq.}~\eqref{eq:gcoupling}}
        \\
        \frac{0.9\mpl T_{\rm RH}^3}{g_s^{3/2}V^4}\simeq 9.8\times10^{-6}\left(\frac{T_{\rm RH}}{10^5\,\mathrm{GeV}}\right)^3\left(\frac{10^9\,\mathrm{GeV}}{V}\right)^4\left(\frac{106.75}{g_{s}}\right)^{3/2}&{\text{if Eq.}~\eqref{eq:TRH}} \\
    \end{cases} \, ,
    \end{equation}
where $g_s$ is computed at $T=\min\{m_{W_{f_R}},T_{\rm RH}\}$, and we assumed $g_*=g_s$ for simplicity.

As a final comment, we notice that fermion decays induced by flavor interactions, such as $F_R\to F'_R\Psi_R\Psi_R$, are irrelevant. If $m_F\ll m_{W_{f_R}}$
the decay rate is suppressed as $\Gamma\propto m_F(m_F/V)^4$; if $m_F\gg m_{W_{f_R}}$, the rate is suppressed by the small gauge coupling $\Gamma\propto g_{f_R}^4m_F$. In both cases, the abundance of the decaying fermion is negligibly small at the time at which the decay should take place.
Additionally, if $m_{F}\gg T_{\rm RH}$, the population of $F_R$ is exponentially suppressed.

\subsubsection{Production via neutrino mixing}

A second population of anomalons is produced via their mixing with the neutrino sector after EWSB. The relevant quantities are the mixing angles between the left-handed neutrino flavor eigenstates, $\nu_f$ ($f=e,\mu,\tau)$, and the anomalon mass eigenstates, $\Psi_m$ ($m=1,\ldots,24$), collectively denoted by $\theta_{\nu_f\Psi_m}$. These  are induced by the non-renormalizable operators listed in Table~\ref{tab:Yukawa-nonrenormalizable} and have been discussed extensively in Sect.~\ref{sec:neutrinos-mass-mixing}. We focus on the region of the parameter space where $\theta_{\nu_f\Psi_m}\ll1$. More precisely, we require that the mixing angles are small enough to avoid anomalon thermalization from neutrino-anomalon conversions. If the anomalons are heavier than active neutrinos, $m_{\Psi_m} \gtrsim m_\nu$, this corresponds roughly to 
$\theta_{\nu_f\Psi_m}\lesssim [0.15-0.25]\,(0.1\,\text{eV}/m_\Psi)^{1/2}$(as extrapolated from Refs.~\cite{Strumia:2006db,Hannestad:2012ky,Dasgupta:2021ies}). If instead $m_{\Psi_m} \lesssim m_\nu$, resonant effects are important and even smaller mixing angles are needed. Furthermore, bounds on dark radiation may even require stronger constraints on the mixing angles, as we discuss in more details below.

In Fig.~\ref{fig:NA-mixing} the mixings 
of L-neutrinos with massive anomalons
are plotted as a function of the coefficients of the non-renormalizable operators of Table~\ref{tab:Yukawa-nonrenormalizable} (denoted by $U_\perp$ and $U_0$ for heavy and light anomalon mixings, respectively). As we can observe, for very large values of $V$ (accidental PQ scenario), small mixing angles can be achieved for both heavy and light anomalons without  suppressing the coefficients of the operators. Conversely, in the high-quality PQ scenario, some of  the coefficients need to be suppressed in order to avoid large mixings, in particular for the heavy anomalon sector ( the coefficients $\mathcal{A,B,B',C}$ being the most relevant).
 
The physical picture is analogous to the production of $\mathrm{keV}$ sterile neutrino dark matter 
via oscillations (also referred to as Dodelson-Widrow mechanism)~\cite{Dodelson:1993je,Boyarsky:2009ix,Lesgourgues:2012uu,Drewes:2016upu,Abazajian:2017tcc,Dasgupta:2021ies,Strumia:2006db}, in which active neutrino species produced in the thermal bath convert to $\mathrm{keV}$ sterile states via mixing.\footnote{In our setup, the sterile neutrinos are heavy with masses  $\sim V$, and hence 
they do not play any role for the 
production mechanism via oscillations.} 
Applying this analogy, the $\mathrm{keV}$ sterile states correspond in our picture to massive anomalons.

There are, however, two important differences compared to the standard scenario. First, the anomalons may be lighter than active neutrinos. Moreover, most of the studies focus on the production of a single sterile state, while we must consider the simultaneous production of several anomalon species. A precise computation of the anomalon abundance  would require the numerical solution of a set of quantum kinetic equations for neutrino-anomalon oscillations~\cite{Hannestad:2012ky,Gariazzo:2019gyi,Dasgupta:2021ies}. In the regime in which the anomalons are sufficiently heavy, $m_\Psi>m_\nu$, it is possible to derive approximate analytic estimates.

Let us first review the simple physical picture. As long as the temperature of the universe is $T\gtrsim\mathrm{MeV}$, active neutrinos are kept in thermal equilibrium with the photon bath through weak interactions, which produce neutrinos in the flavor basis, for instance from electron-positron annihilations, $e^+e^-\to\bar{\nu}_f\nu_f$, with rate $\Gamma_{\nu}\sim G_F^2T^5$, where $G_F\simeq 1.166\times 10^{-5}\,\mathrm{GeV}^{-2}$ is the Fermi constant. Some of the L-neutrinos convert to massive anomalons via mixing. The production rate $\Gamma_m$ for an anomalon mass eigenstate $\Psi_m$ depends on the (zero-temperature and zero-density) mixing angle 
$\theta_{\nu\Psi_m}^2\equiv \sum_{f=e,\mu,\tau}|\theta_{\nu_f\Psi_m}|^2$, where the proper sum over neutrino flavors is performed. These mixing angles receive corrections by thermal and density effects when neutrinos propagate in a  thermal plasma~\cite{Dodelson:1993je,Boyarsky:2009ix,Cirelli:2024ssz}. We denote the finite temperature (and density) corrected angles by  $\overline{\theta}_{\nu\Psi_m}$. The production rate scales then as $\Gamma_m\sim G_F^2 T^5\overline{\theta}_{\nu\Psi_m}^2$.
In the regime $m_\Psi\gg m_\nu$, one finds~\cite{Notzold:1987ik,Boyarsky:2009ix}
\begin{equation}
\overline{\theta}_{\nu\Psi_m}\simeq \frac{\theta_{\nu\Psi_m}}{(1+\mathcal{O}(1) \frac{T^6G_F^2}{\alpha_{\rm em}m_{\Psi_m}^2})} \, .
\end{equation}
The temperature at which the production is maximized 
depends on the
anomalon mass eigenstate,  
$T_{\text{FI},m}\sim \max\,\{130\,\mathrm{MeV}(m_{\Psi_m}/\mathrm{keV})^{1/3},\mathrm{MeV}\}$.
The resulting population can be estimated as $Y_\Psi^{\nu-{\rm mix}}\sim \gamma_{\rm eq}^{\nu-{\rm mix}}/(sH)\sim \sum_m(Y_{e,\rm eq}\Gamma_m/H)|_{T=T_{\text{FI},m}}$, where $Y_{\rm eq}$ is the equilibrium distribution for electrons/positrons. We find
\begin{equation}\label{eq:anomalonsYmixing}
Y_\Psi^{\nu-{\rm mix}}\big|_{m_\Psi\gg m_\nu}\sim {2.5\times 10^4}\left(\frac{10.75}{g_{\rm SM}}\right)^{3/2}\sum_{m=1}^{N_\Psi}\left( |\theta_{\nu\Psi_m}|^2\,\frac{m_{\Psi_m}}{\text{keV}}\,\right) \, .
\end{equation}
The picture changes significantly if $m_\Psi \lesssim m_\nu$, which is the most typical situation we expect to be realized in the post-inflationary scenario. In such a case, the anomalon-neutrino conversion rate gets a resonant enhancement due to the presence of a pole in $\overline{\theta}_{\nu\Psi_m}$~\cite{Hannestad:2012ky}. This arises whenever the neutrino propagating in the thermal plasma acquires a momentum $p^2\sim \alpha_{\rm em}(m_\nu^2-m_\Psi^2)/(G_F^2 T^4)$. Thus, the resonance propagates to higher momenta as the temperature decreases. This implies that the resonance eventually covers the full momentum distribution of neutrinos. As a result, resonant neutrino-anomalon conversions lead to very efficient thermalization of the anomalons if the mixing angle is not sufficiently small. To avoid this, the mixing angles $\theta_{\nu_f\Psi_m}$ must be significantly smaller compared to the non-resonant case. The precise condition as well as an accurate computation of the anomalons abundance in this regime is rather involved and requires a numerical solution of the quantum kinetic equations~\cite{Hannestad:2012ky,Gariazzo:2019gyi,Dasgupta:2021ies}, with no simple analytic expression available. A full numerical computation is also very challenging and therefore  many studies focus on the production of only one sterile state (often making simplifying assumption on the flavor mixings among active states). A complete computation within our framework lies beyond the scope of this study and is deferred to future work. 

\subsubsection{Production from effective Yukawa interactions}
\label{sec:Higgsproduction}

The non-renormalizable operators of Table~\ref{tab:Yukawa-nonrenormalizable} provide an extra production source for anomalons, namely through effective Yukawa interactions with the scalar fields of the model. 
We distinguish two classes of operators: those involving EW-breaking scalar fields $\Phi, \Sigma, \Sigma'$ which after EWSB generate interactions with the SM Higgs boson $h$, and those involving only the heavy scalar fields $\Delta$ and $\chi$.
We begin our discussion by considering the first class. 
The operator $\mathcal{A}\,\bar{Q}_L\Psi_R\chi \Phi/M_{\rm Pl}$ generates the effective Yukawa coupling $\mathcal{A}(V/M_{\rm Pl})\,\bar{\nu}_L \Psi_R h$ at scales below the EWSB. The same applies to the operators $\mathcal{B}\,\bar{Q}_L\Psi_R\chi\Sigma/M_{\rm Pl}$ and $\mathcal{B}'\,\bar{Q}_L\Psi_R\chi\Sigma'/M_{\rm Pl}$. Concurrently, the operators $\tilde{\mathcal{D}}\,\Psi_R\Psi_R\Phi^{*2}/M_{\rm Pl}$ (and analogously, the $\mathcal{E}$-type operators) generate 
the Yukawa interaction
$(v/M_{\rm Pl})\,\Psi_R\Psi_R h$ as well as $\Psi_R\Psi_R h^{2}/M_{\rm Pl}$. 

The $\Psi_R\Psi_R h$ and $\Psi_R\Psi_R h^2$ interactions are irrelevant in view of the extreme suppression factor $v/M_{\rm Pl}\sim 10^{-17}$. On the other hand, the $\bar{\nu}_L\Psi_R h$ coupling gives rise to Higgs decays $h\to \nu\Psi$ as well as $s$-channel 
Higgs-boson-mediated annihilations $PP\to (h)\to \nu \Psi$, where $PP\in\{\bar{F}F,W^+W^-,ZZ,hh\}$ are the SM particles interacting with the Higgs boson. We checked that $(i)$ these new channels are unable to produce a population of anomalons in thermal equilibrium with the SM bath as long as $\mathcal{A,B,B'}\ll 10^{-8}(M_{\rm Pl}/V)\simeq 10^{-3}(10^{14}\,\mathrm{GeV}/V)$, which is satisfied even for $\mathcal{A,B,B'}\sim\mathcal{O}(1)$ as long as $V\lesssim 10^{11}\,\mathrm{GeV}$, and $(ii)$, the corresponding population of anomalons produced by freeze-in is typically small in most of the parameter space, so we neglect it. Notice that the branching ratio of the Higgs boson into $\nu\Psi$ is so small that there are basically no 
relevant constraints from Higgs invisible decays \cite{ATLAS:2023tkt}. 

The second class of operators only involves heavy scalar fields. The most relevant ones are $\mathcal{C}\,Q_R\Psi_R\Delta^*\chi/M_{\rm Pl}$ and $\mathcal{D}\,\Psi_R\Psi_R\Delta^*\chi^2/M_{\rm Pl}^2$. These induce the decays of radial modes of $\Delta$ and $\chi$ to $\Psi_R\nu_R$ or $\Psi_R\Psi_R$(+ heavy scalar mode). Their rates are suppressed at least by a factor $(V/M_{\rm Pl})^2$, so that the decays are completely irrelevant (they would occur at $T\ll V$, when the heavy scalars have already disappeared from the thermal bath). Annihilations mediated by heavy scalars give weaker constraints compared to the ones mediated by the Higgs boson. 

Finally, all the operators discussed in this section induce annihilations of scalar particles at large temperature. The most relevant ones are the operators of dimension 5, which induce processes such as $\chi\Phi\to Q_L\Psi_R$ or $\Phi\Phi\to \Psi_R\Psi_R$ or $\Delta\chi\to Q_R\Psi_R$. The corresponding population of anomalons is controlled by the reheating temperature, $Y\propto \mathcal{X}^2(T_{\rm RH}/M_{\rm Pl})$ with $\mathcal{X}\in\{\mathcal{A,B,B',C,\tilde{D},E}\}$.

\subsubsection{Anomalon decay}\label{sec:anomalonsDecay}

The decay modes  of the anomalons and their rates crucially depend on their mass. We can anticipate the general result of this section referring to Fig.~\ref{fig:NA-eigenvalues} and our different scenarios: the heavy anomalons of mass $m_\perp\sim 10$ TeV decay in the early universe, well before the onset of BBN. The ultra-light anomalons with mass $m\lesssim$ 0.1 eV are stable on cosmological scales. The $\mathrm{keV}$-ish anomalons are cosmologically stable for relatively small mixings. Stronger constrains on the mixings are provided by X-ray searches.

We now briefly discuss each one of these cases.
Anomalons heavier than the TeV scale can decay through (some of) the operators of Table~\ref{tab:Yukawa-nonrenormalizable} involving EW-breaking fields. For instance, the  dimension-5 operator $\mathcal{A}\,\bar{Q}_L\Psi_R\Phi\chi$ allows for $\Psi_{R}\to\Phi\mathcal{Q}_L$ (before EWSB), or analogously $\Psi_{R}\to h\nu_L$ (after EWSB). Additionally, extra decay channels are open, such as $\Psi_R\to \nu\bar{f}f$ after EWSB, which are mediated by a tree-level exchange of a Higgs boson or an EW gauge boson. These decays are fast enough to safely avoid cosmological constraints.

Anomalons with mass in the $1\div100\,\mathrm{keV}$ range dominantly decay via neutrino mixing to $3\nu_L$, with decay width 
\begin{equation}\label{eq:Psito3nu}
  \Gamma_{3\nu}=\frac{G_F^2 m_\Psi^5\theta_{\nu\Psi}^2}{96\pi^3} \, .
\end{equation}
 They are cosmologically stable if $\theta_{\nu\Psi}\lesssim 0.018(10\,\mathrm{keV}/m_\Psi)^{5/2}$. The decay channel $\Psi_{R}\to \nu_L\gamma$ is also open with rate 
 \begin{equation}
\Gamma_{\nu\gamma}=\frac{9\alpha_{\rm em}G_F^2 m_\Psi^5\theta_{\nu\Psi}^2}{256\pi^4} \, .
\end{equation}
Although subdominant, the latter process actually imposes the strongest constraints on the mixing; X-ray searches exclude $\theta_{\nu\Psi}\gtrsim2\times 10^{-6}(10\,\mathrm{keV}/m_\Psi)^{5/2}$ (see~\cite{An:2023mkf}).

Ultra-light anomalons could  decay via the same channels of $\mathrm{keV}$-ish ones if $m_\Psi>m_\nu$, but they are cosmologically stable, as their decays are suppressed by their small mass $\Gamma\propto(m_\Psi/v)^4$. The same also applies to the decays of active neutrinos into anomalons when $m_\Psi<m_\nu$.

Finally, notice that the decays of heavy anomalons to lighter ones are suppressed, as they involve the exchange of a flavor gauge boson. These decays are completely irrelevant in all scenarios: heavy anomalons $m_\perp\gtrsim $ TeV decay much faster via the other decay channels discussed above; ultra-light anomalons are still cosmologically stable; for $\mathrm{keV}$-ish anomalons we should compare these decays with \eq{eq:Psito3nu}. The exchange of a flavor gauge boson implies a decay rate, which is suppressed at least by $1/(G_F^2V^4\theta_{\nu\Psi}^2)\sim(v/V)^4(1/\theta_{\nu\Psi})^2$ compared to the decay channel into $3\nu_L$. Thus, even in the most pessimistic case ($V\sim 10^9\,\mathrm{GeV}$), the decays to active neutrinos are dominant unless $\theta_{\nu\Psi}\lesssim 10^{-13}$.

\subsection{Anomalons as dark radiation}\label{sec:anomalonsDR}

We focus here on the case of ultra-light sub-$\mathrm{eV}$ anomalons,
arising mainly in the post-inflationary scenario and possibly as a subset of states in the pre-inflationary one (see Fig.~\ref{fig:NA-eigenvalues}).
Being fermions, such low masses prevent them from being suitable dark matter candidates in view of the Tremaine-Gunn bound~\cite{Tremaine:1979we}. More precisely, a recent study of  dwarf spheroidal galaxies finds the model-independent lower bound 
$m_{\Psi}\gtrsim0.1\,\mathrm{keV}$
on the mass of a fermionic dark matter species~\cite{Alvey:2020xsk} (assuming 100\% of dark matter is made by that fermion). 

On the other hand, such light particles are relativistic at the time at which the Cosmic Microwave Background (CMB) forms, $T_{\rm CMB}\sim$ 0.1 eV, as well as at the onset of Big Bang Nucleosynthesis (BBN), $T_{\rm BBN}\sim 0.1\,\mathrm{MeV}$, thereby contributing to the energy density of the universe in the form of extra radiation (in addiction to photons and neutrinos), commonly referred to as dark radiation. This is parameterized in terms of an effective number of neutrinos as (see e.g.~\cite{ParticleDataGroup:2022pth})
\begin{equation}
    \Delta N_{\rm eff}=\frac{8}{7}\left(\frac{11}{4}\right)^{4/3}\frac{\rho_{\Psi}}{\rho_\gamma} \, ,
\end{equation}
where $\rho_{\Psi}$ and $\rho_\gamma$ are 
respectively 
the anomalon and photon energy densities ($\rho_\gamma=2\pi^2 T^4/30$), both evaluated at $T\sim T_{\rm CMB}$ (or $T\sim T_{\rm BBN}$). 
The current upper bound on $\Delta N_{\rm eff}$ has been set by the Planck collaboration: $\Delta N_{\rm eff}\leq 0.285$ (at $95\%$ C.L.)~\cite{Planck:2018vyg}. Forthcoming experiments aim to improve this bound up to roughly one order of magnitude with more precise CMB data: 
the Simon Observatory (SO) claims a sensitivity of
$\Delta N_{\rm eff}(2\sigma)\leq 0.06$~\cite{SimonsObservatory:2019qwx,SimonsObservatory:2018koc}, while EUCLID~\cite{EUCLID:2011zbd} and future generation experiments such as CMB-HD~\cite{CMB-HD:2022bsz} are expected to measure values down to $\Delta N_{\rm eff}\leq 0.014$. See also Ref.~\cite{Li:2023puz} for a review on current constraints and future sensitivities. 

The current upper bound is badly violated if the light anomalons were ever in thermal equilibrium with the SM bath and decoupled at some temperature $T_{\rm dec}$. Indeed, in such a case, their contribution 
to $\Delta N_{\rm eff}$ is 
\begin{equation}\label{eq:DeltaNeq}
    \Delta N_{\rm eff}^{\rm TH}=N_\Psi\left(\frac{11}{4}\frac{g_s(T_{\rm CMB})}{g_s(T_{\rm dec})}\right)^{4/3}
    \simeq 1.13\,\frac{N_\Psi}{24}\left(\frac{106.75}{g_s(T_{\rm dec})}\right)^{4/3}.
\end{equation}
Hence, this scenario is excluded even if only 8 anomalons are ultra-light and decoupled before EWSB. 

On the other hand, the picture changes completely if the anomalons never reach thermal equilibrium and are produced by freeze-in, as we discussed previously.
The computation of $\Delta N_{\rm eff}$ for species not in thermal equilibrium is a non-trivial task, since it requires accounting for non-thermal distributions for  dark radiation, which strongly depend on the specific  production mechanism. See e.g.~Ref.~\cite{DEramo:2023nzt} which discusses the scenario where decays or 2-to-2 scatterings of bath particles produce one relativistic dark particle in the final state. More recently, Refs.~\cite{Badziak:2024qjg,DEramo:2024jhn} performed a detailed computation of the contribution of the QCD axion to $\Delta N_{\rm eff}$ using the full Boltzmann equations for momentum distribution functions.
A detailed computation in our scenario is beyond the scope of this paper. To simplify the analysis, we briefly discuss two possible approaches sometimes adopted in the literature.

The first possibility is to assume that the anomalon number density, $n_\Psi$, and the corresponding energy density, $\rho_\Psi$, are related as $\rho_\Psi\propto n_\Psi^{4/3}$, i.e.~the same as for species in thermal equilibrium. Eq.~\eqref{eq:DeltaNeq} can be generalized and expressed in terms of the anomalon yield, $Y_\Psi=n_\Psi/s$, as~\cite{DEramo:2021lgb,Caloni:2024olo,Badziak:2024szg}
\begin{equation}\label{eq:Delta1}
    \Delta N_{\rm eff}^{\rm FI1}\simeq N_\Psi\left(\frac{\frac{11}{4}\frac{2\pi^4}{45\xi(3)}\frac{2g_s^{\rm SM}(T_{\rm CMB})Y_\Psi}{3N_\Psi}}{1-\frac{2\pi^4}{45\xi(3)}\frac{7}{6}Y_\Psi}\right)^{4/3}\simeq 76.28\,\frac{Y_\Psi^{4/3}}{N_\Psi^{1/3}} \, .
\end{equation}
Alternatively, we can drop the assumption $\rho_\Psi\propto n_\Psi^{4/3}$ and estimate $\Delta N_{\rm eff}$ as follows: the freeze-in production of the anomalons is maximally efficient at some  temperature $T_{\rm FI}$ (which depends on the specific  production mechanism).
Since anomalons are produced by scatterings of SM bath particles, each one carrying energy $E\sim T$, their average energy immediately after production is also of order $\sim T$.
Therefore, we can parametrize their energy density as $\rho_\Psi(T_{\rm FI})\simeq \xi T_{\rm FI}n_\Psi(T_{\rm FI})=\xi T_{\rm FI}s(T_{\rm FI})Y_\Psi$, where $\xi$ is an $\mathcal{O}(1)$ number. The energy density of a relativistic particle redshifts as a function of the scale factor $a_T\equiv a(T)$ as  $\rho_{\Psi}(T)=\rho_{\Psi}(T_{\rm FI})(a_{T_{\rm FI}}/a_{T})^4=\rho_{\Psi}(T_{\rm FI})(T/T_{\rm FI})^4(g_s(T)/g_s(T_{\rm FI}))^{4/3}$. At the time of the CMB, this corresponds to 
\begin{equation}\label{eq:Delta2}
    \Delta N_{\rm eff}^{\rm FI2}\simeq\frac{16}{21}\left(\frac{11}{4}\right)^{4/3}\frac{g_s(T_{\rm CMB})^{4/3}}{g_s(T_{\rm FI})^{1/3}}\xi Y_{\Psi}\simeq 56.96\left(\frac{\xi}{3.151}\right)\frac{Y_\Psi}{g_s(T_{\rm FI})^{1/3}} \, , 
\end{equation}
where in equilibrium $\rho_\Psi^{\rm eq}/n_\Psi^{\rm eq}=3.151\,T$ for a fermionic species. This expression is the equivalent of the one derived in Appendix C of~\cite{Badziak:2024szg}, for the case of one spin-0 particle (the axion), generalized to our case with $N_\Psi$ Weyl fermions contributing to $\Delta N_{\rm eff}$.  See also~\cite{Berbig:2022nre} for a numerical computation of $\Delta N_{\rm eff}$ in the freeze-in regime.

The expression in Eq.~\eqref{eq:Delta2} relies on a more precise (though still approximate) estimate of the anomalon energy density, while Eq.~\eqref{eq:Delta1} is an excellent approximation only when the anomalons follow their equilibrium distribution. Furthermore, a detailed computation of $\Delta N_{\rm eff}$ in the freeze-in regime (for the production of the QCD axion) can be found in Ref.~\cite{Badziak:2024qjg}, where the result was obtained upon numerical integration of the full Boltzmann equations for momentum distribution. The authors found that the equation for $\Delta N_{\rm eff}$ given in the Appendix C of \cite{Badziak:2024szg} (and thus its generalized version, Eq.~\eqref{eq:Delta2}) is a very good approximation to the precise numerical solution, with an error below
$\mathcal{O}(10\%)$. Therefore, we will employ  Eq.~\eqref{eq:Delta2} for our estimates. More precisely, Eq.~\eqref{eq:Delta1} underestimates the contribution to dark radiation by a factor $\Delta N_{\rm eff}^{\rm FI1}/\Delta N_{\rm eff}^{\rm FI2}\simeq 0.23\,(g_s(T_{\rm FI})/10.75)^{4/9}(\Delta N_{\rm eff}^{\rm FI2}/0.285)^{1/3}$.

We can now estimate the anomalon contribution to $\Delta N_{\rm eff}$ in our model, 
discussing separately the two scenarios introduced above. 

\subsubsection{High-quality PQ}

The abundance of anomalons produced via renormalizable flavor interactions is given by the upper case of Eq.~\eqref{eq:anomalonAbundanceFlavor}, requiring also
$g_{f_R}\ll10^{-9}$.
The corresponding values of $\Delta N_{\rm eff}$ are generically below the experimental sensitivity of future experiments unless we saturate the freeze-in condition close to $g_{f_R}\sim 10^{-9}$, where our estimates are not reliable. Indeed, at the border between the freeze-in and thermalization regimes, a more precise computation of the abundance would be needed. As we know that thermalized anomalons are already excluded by the Planck '18 result, we expect that, in order to interpolate between the two regimes, $\Delta N_{\rm eff}$ could reach values testable in the near future in that corner of the parameter space.

On the other hand, the production of anomalons via neutrino-anomalon conversion generically requires small mixing angles to satisfy the constraints on $\Delta N_{\rm eff}$.  In the non-resonant scenario, $m_\Psi \gtrsim m_\nu$, the population of anomalons can be roughly estimated as in Eq.~\eqref{eq:anomalonsYmixing}, which saturates the Planck '18 bound for $\theta_{\nu\Psi_m}\simeq 0.014\,(0.1\,\mathrm{eV}/m_{\Psi_m})^{1/2}$ 
in a simplified scenario where all the anomalons have the same masses and mixing angles, or $\theta_{\nu\Psi_1}\simeq 0.066\,(0.1\,\mathrm{eV}/m_{\Psi_1})^{1/2}$ in the limit in which only one anomalon 
($\Psi_1$) has a sizable mixing angle $\theta_{\nu\Psi_1}\gg\theta_{\nu\Psi_j}$. 
However, the typical expectation from the discussion of the spectrum in \sect{sec:SMflavour-NandA} is that anomalons are lighter than active neutrinos.
In such a case, resonant effects enhance the production, inducing more severe constraints on the mixing angles. These imply a set of non-trivial constraints on the parameters of the model. In particular, the coefficients of the non-renormalizable operators of Table~\ref{tab:Yukawa-nonrenormalizable} need to be suppressed enough to avoid thermalization via resonant conversion. A detailed numerical analysis would be required in this case to accurately compute the anomalon abundance and 
its impact on restricting the model's parameter space. 

Finally, we note that for relatively low values of $f_a$ (as derived from the analysis of the PQ quality in \sect{sec:PQquality}, where $f_a \lesssim 5 \times 10^8\,\mathrm{GeV}$), axions are efficiently produced at temperatures around or below the QCD phase transition through scattering of pions (see e.g.~\cite{Chang:1993gm,Hannestad:2005df,DiLuzio:2021vjd,Notari:2022ffe,DiLuzio:2022gsc,Bianchini:2023ubu}). Consequently, disentangling the axion contribution to $\Delta N_{\rm eff}$ from that of the anomalons is a 
non-trivial task, necessitating a comprehensive global analysis, which we defer to future work.

\subsubsection{Accidental PQ}

The results depend strongly on the scales $V$ and $T_{\rm RH}$. 
For large values of $V\sim 10^{14}\,\mathrm{GeV}$ (IIa), all the anomalons are heavier than $1\,\mathrm{keV}$ and either decay before BBN (cf.~\sect{sec:anomalonsDecay}) or contribute to dark matter (cf.~\sect{sec:anomalonsDM}). For smaller values of $V\sim 10^{11}\,\mathrm{GeV}$ (IIb), 8 anomalons are ultra-light and contribute to dark radiation. Given that $f_a\sim 10^{10}\,\mathrm{GeV}$, anomalon thermalization via renormalizable flavor interactions is inefficient  if either $T_{\rm RH}\lesssim V(V/M_{\rm Pl})^{1/3}\simeq 10^8\,\mathrm{GeV}$ (pre-inflationary) or $g_{f_R}\ll 10^{-7}$ (post-inflationary). 
Their abundance is given by the lower (upper) case of Eq.~\eqref{eq:anomalonAbundanceFlavor} in the first (second) scenario.
The contribution to $\Delta N_{\rm eff}$ is again below current and future experimental sensitivities as long as Eq.~\eqref{eq:TRH} is satisfied. Once again, close to the thermalization regime, corresponding to $T_{\rm RH}\sim V(V/M_{\rm Pl})^{1/3}$ in the pre-inflationary scenario, we expect a larger and possibly testable value of $\Delta N_{\rm eff}$. Production from neutrino conversions is typically subdominant in light of the small mixing angles, see Fig.~\ref{fig:NA-mixing} (lower center). A detailed scan of the parameter space would be needed to provide more precise results.

\subsection{Anomalons as dark matter}\label{sec:anomalonsDM}

In the accidental PQ scenario, the model predicts a set of $\mathrm{keV}$-ish anomalons for some values of $V$, see Fig.~\ref{fig:NA-eigenvalues}.
In some region of the parameter space, these can be cosmologically stable and not excluded by X-ray searches. Thus, they can contribute to the dark matter abundance of the universe, together with the axion. The physics is essentially the same as that of the well-studied $\mathrm{keV}$ sterile neutrino DM scenario, already briefly discussed in the previous section. While a detailed scan of the parameter space would be needed to provide precise results, we summarize here the main qualitative features.

For instance, if $V=10^{14}\,\mathrm{GeV}$ ($f_a\sim 10^{13}\,\mathrm{GeV}$), the 8 light anomalons have mass $m_{0}\sim 1 \div 10\,\mathrm{keV}$. As long as $T_{\rm RH}\lesssim V(V/M_{\rm Pl})^{1/3}$ $\sim10^{12}\,\mathrm{GeV}$ (pre-inflationary) or $g_{f_R}\ll 10^{-4}$ (post-inflationary), they are produced by freeze-in. The mixing with active neutrinos depends mostly on the parameters $\parAt$ and $\parBt$, as shown in Fig.~\ref{fig:NA-mixing} (lower right). Mixings larger than roughly $10^{-(6\div5)}$ are already excluded by X-ray searches, while smaller angles are allowed. 
The population produced via mixing can still be the larger one even if mixings are small, $Y_{\rm light}^{\nu-{\rm mix}}\lesssim 10^{-8}(\theta_{\nu\Psi}/10^{-7})^2(m_0/10\,\mathrm{keV})$. This corresponds to a very small fraction of the total dark matter abundance, $\Omega_{\Psi,0}/\Omega_{\rm DM}\sim 10^{-3}$ (potentially up to $1\%$ if we push $T_{\rm RH}$ up to $\sim 10^{12}\,\mathrm{GeV}$, where the computation is not fully reliable), which is not constrained by warm dark matter limits. 

On the other hand, if $V=10^{11}\,\mathrm{GeV}$, the picture reverses, as the 
16 heavy anomalons are now the $\mathrm{keV}$-ish ones, $m_\perp\sim 10\,\mathrm{keV}$. They possibly contribute to a small fraction of dark matter, analogous to the discussion above. The mixing angles are mostly determined by the coefficients $\parA,\parB,\parC$, see Fig.~\ref{fig:NA-mixing} (upper center), and tend to be slightly larger than in the previous case, 
so that X-ray searches would 
impose more severe constraints on the model parameters.

\section{Conclusions and outlook}  
\label{sec:concl}  

In this study, we have investigated a category of axion models where the accidental $\U(1)_{\rm PQ}$ symmetry naturally arises from the interplay between vertical (grand-unified) and horizontal (flavor) gauge forces. Concentrating on a Pati-Salam implementation, which shares certain parallels with a prior SO(10) model examined in Ref.~\cite{DiLuzio:2020qio}, we analyzed the model's ability 
to address the PQ quality problem and 
we showed that it can reproduce the SM flavor structure in a non-standard way.

The main predictions of the model, regarding axion phenomenology, 
are summarized in \fig{fig:axionphoton}, where three 
distinct 
axion mass 
windows are identified: 
$i)$ $m_a \in [2 \times 10^{-8}, 10^{-3}]\,\mathrm{eV}$, 
$ii)$ $m_a \in [2.8 \times 10^{-5}, 0.01]\,\mathrm{eV}$ 
(corresponding to an accidental $\U(1)_{\rm PQ}$, 
respectively in the 
pre-inflationary and post-inflationary PQ breaking scenario), and $iii)$ $m_a \gtrsim 0.01\,\mathrm{eV}$
(associated with a high-quality $\U(1)_{\rm PQ}$ in the post-inflationary PQ breaking scenario).
In this context, the predictions of the 
Pati-Salam model regarding axion physics 
are largely degenerate with those of the original 
SO(10) model discussed in Ref.~\cite{DiLuzio:2020qio}.\footnote{For completeness, relevant aspects of the present analysis with implications for the model of Ref.~\cite{DiLuzio:2020qio} are included in \app{sec:SO10model}.} While the axion-photon coupling matches that of the benchmark DFSZ model \cite{Zhitnitsky:1980tq,Dine:1981rt}, the non-trivial embedding of the axion into the scalar multiplets of the extended gauge symmetry establishes a connection between the axion decay constant and intermediate mass scales associated with $B-L$ and flavor breaking. 

A key feature of this setup is the presence of parametrically light fermions, known as anomalons, that were introduced to cancel the gauge anomalies of the flavor symmetry. For the most favorable PQ-breaking scales required to resolve the PQ quality problem, these anomalons are expected to have masses below the eV scale. As a result, they contribute to the dark radiation of the universe. While for intermediate-scale 
values of $f_a$, the anomalons can be heavier (keV scale) 
and thus contribute to dark matter. 

We studied the cosmological production of anomalons in the early universe, considering their production via flavor interactions, neutrino mixing and effective Yukawa-like operators. Current constraints on the effective number of relativistic species, $\Delta N_{\rm eff}$, from Planck'18 \cite{Planck:2018vyg} data were applied to constrain the parameter space. We further highlighted how future precision measurements of 
$\Delta N_{\rm eff}$ could serve as a low-energy probe of the UV dynamics underlying the solution to the PQ quality problem.  

More generally, the cosmological signatures of the 
anomalon fields crucially depend on $f_a$. We present here a brief summary of our findings in two relevant benchmark scenarios: 
\begin{itemize}
\item[\text{(I)}] High-quality PQ (post-inflationary PQ-breaking scenario). All the anomalons contribute to dark radiation. In order to evade the current bound from the Planck '18 analysis we must enforce the condition that they never reach thermal equilibrium (freeze-in), implying very small flavor gauge coupling $g_{f_R}\lesssim 10^{-9}$, as well as small mixings with active neutrinos. This imposes severe constraints on the parameter space of the model. The contribution to $\Delta N_{\rm eff}$ is below the sensitivity of future planned experimental proposals, with the exception of the region of the parameter space at the border between the thermalization and freeze-in regimes, where we expect a larger amount of dark radiation but a more precise computation would be needed.
\item[\text{(II)}] Accidental PQ (both pre-inflationary and post-inflationary PQ-breaking can be realized). The parameters of the model are less constrained and the phenomenology is richer. The mixings with active neutrinos are typically smaller compared to the scenario (I). This, and the defining condition $f_a>T_{\rm RH}$, more naturally forbid anomalon thermalization in the pre-inflationary scenario. In the post-inflationary scenario, a small flavor gauge coupling is needed (even though larger values than in the high-quality PQ scenario are admissible). 
We discussed two sub cases: (IIa) $f_a \sim 10^{13}\,\mathrm{GeV}$, 
where the 16 heavy anomalons have masses $m_\perp\sim 10\,\mathrm{TeV}$ and decay before BBN. The 8 light anomalons have masses 
$m_0\sim \mathrm{keV}$ and can contribute to a small fraction of dark matter in a region of the parameter space; (IIb) $f_a \sim 10^{10}\,\mathrm{GeV}$, where $m_\perp\sim 10\,\mathrm{keV}$ potentially contributing to dark matter, and $m_0\sim\,\mathrm{\mu eV}$ contributing to $\Delta N_{\rm eff}$ (typically well below the current bound).
\end{itemize}

While this study addresses several open questions related to the PQ quality problem and the phenomenology of anomalons, we conclude by identifying a significant direction for each of these issues which warrants further exploration:  
\begin{itemize}[leftmargin=0.5cm,itemsep=0cm]  
\item
Our investigation of PQ quality revealed two shortcomings of the PQ quality analysis and/or model. Their relevance also extends to the $\SO(10)$ model of Ref.~\cite{DiLuzio:2020qio}, and to horizontal-vertical models more broadly.
\par
First, the gauge couplings develop a Landau pole soon after the Pati-Salam breaking scale, a feature that calls into question the PQ quality analysis with a Planck-scale cutoff. Ideally, one would apply the model building lessons learned from \sect{sec:PS-model}, and the investigation of how successful elements of the presented model conspire, to propose a fully realistic and perturbatively calculable alternative. Such a model is expected to share many qualitative features, such as anomalons and the protection mechanism for higher-dimensional operators, with the model presented here.
\par
Second, a complete analysis of the PQ quality problem involves also quantum (i.e.~loop-induced) effects, a non-trivial but ultimately essential next step, cf.~\app{app:PSmodel-PQ-at-loop-level}. Extending the tree-level investigation would provide a more robust understanding of the model's viability in addressing the PQ quality problem. Ideally one could attempt to build an alternative model that is manifestly immune to quantum enhancements. 
\par
Both issues are relevant in fact for a wider class of models: the proximity of the Landau pole seems to be an important issue for every model with vertical-horizontal symmetry interplay, while quantum enhancements are potentially dangerous to any model of PQ quality in general. 
\item 
The computation of anomalon abundance via neutrino mixing in the early universe, in the regime $m_\Psi \lesssim m_\nu$ (parametrically favored by the solution to the PQ-quality problem), requires a treatment based on quantum kinetic equations. This is a computationally intensive task, particularly when considering multiple sterile states and flavor mixing among active neutrino states. While analytical approximations have been employed for sterile neutrino dark matter, a full numerical solution is necessary to accurately determine the anomalon production rates in this regime. Addressing these challenges will refine the constraints on the parameter space of this and similar vertical-horizontal models by providing a more robust determination of $\Delta N_{\rm eff}$.  
\end{itemize}

\noindent
We defer the investigation of these issues to future work, aiming to further advance our understanding of the origin of the PQ symmetry and its  implications for axion phenomenology.

%%%%%%%%%%%%%%%%%%%%%%%%%%%%%%%%%%%%%%%%%%%%%%
%%%%%%%%%%%%%%%%%%%%%%%%%%%%%%%%%%%%%%%%%%%%%%
\section*{Acknowledgments} 
We thank Maximilian Berbig, Enrico Nardi, Shaikh Saad and Luca Vecchi for useful discussions. 
The work of LDL, FM and VS is supported
by the European Union -- Next Generation EU and
by the Italian Ministry of University and Research (MUR) 
via the PRIN 2022 project n.~2022K4B58X -- AxionOrigins.  
GL is supported by the Generalitat Valenciana APOSTD / 2023 Grant No.~CIAPOS/2022/193. GL is grateful to the theory group of Frascati National Labs for their warm hospitality, during which part of this work was conducted. 
This article is also based upon work from COST Action COSMIC WISPers CA21106, supported by COST (European Cooperation in Science and Technology).

\appendix

%%%%%%%%%%%%%%%%%%%%%%%%%%%%%%
%%%%%%%%%%%%%%%%%%%%%%%%%%%%%%

\section{Low-energy axion couplings}
\label{sec:loweaxion}

In this Appendix we compute low-energy axion couplings to SM matter fields, following the strategy outlined in 
Ref.~\cite{DiLuzio:2020qio} (see also \cite{Ernst:2018bib,Bertolini:2020hjc}). 
This requires an extension of the analysis 
provided in 
\sect{sec:axionembedding}, including also electroweak VEVs 
for the identification of the physical axion field.  
\Table{tab:fieldsEW} summarizes the global (PQ) and gauge 
($B-L$ and $Y$) charges of the Pati-Salam 
sub-multiplets 
which host the axion as an angular component,  
thus requiring a complex scalar multiplet
with a $Q = T^3_L + Y = 0$ 
neutral component. Denoting the complex multiplet by $\phi$, 
as in \eq{eq:defphiax}, the list of fields includes 
$\phi \in \{ \chi, \Delta, \Phi_u, \Phi_d, \Sigma_u, 
\Sigma_d, \Sigma'_u, \Sigma'_d \}$. 

\begin{table}[hb]
$$\begin{array}{c|c|c|c|c|c|c|c|c}
& \chi 
& \Delta 
& \Phi_u 
& \Phi_d
& \Sigma_u
& \Sigma_d
& \Sigma'_u
& \Sigma'_d 
\\ \hline
\PQ & -1 & 2 & 2 & 2 & 2 & 2 & 2 & 2 \\
B-L & -1 & -2 & 0 & 0 & 0 & 0 & 0 & 0 \\
Y & 0 & 0 & -1/2 & 1/2 & -1/2 & 1/2 & -1/2 & 1/2 \\
\end{array}$$
\caption{Global (PQ) and local ($B-L$ and $Y$) 
charges of the Pati-Salam
sub-multiplets hosting the physical axion field.}
\label{tab:fieldsEW}
\end{table}

The low-energy PQ symmetry, with charge $q$, is a linear combination 
of the UV PQ charges and two broken Cartan generators, 
which can be chosen as $B-L$ and $Y$, namely 
\beq 
\label{eq:generalUVIRPQ2}
q = c_1 \PQ + c_2 (B-L) + c_3 Y \, .  
\eeq
It can be shown, analogously to the case discussed in \sect{sec:axionembedding}, 
that in order to match $\U(1)_{\PQ}$ anomalies in terms of UV and IR charges, 
$c_1 = 1$, which we henceforth assume in the following. 
Given the PQ current $J_\mu^{\rm PQ} = \sum_\phi q_{\phi} \sqrt{2} V_{\phi} \partial_\mu a_{\phi}$, 
the canonical axion field is defined as 
\beq 
\label{eq:axiondef2}
a = \frac{1}{V_a} \sum_\phi q_{\phi} V_{\phi} a_{\phi} \, , \qquad 
V_a^2 = \sum_\phi q^2_{\phi} V^2_{\phi} \, , 
\eeq
so that $J_\mu^{\rm PQ} = \sqrt{2} V_a \partial_\mu a$ and, 
compatibly with the Goldstone theorem, 
$\langle 0 | J_\mu^{\rm PQ} | a \rangle = i \sqrt{2} V_a p_\mu$.  
Under a PQ transformation, $a_{\phi} \to a_{\phi} 
+ \kappa \, q_{\phi} \sqrt{2} V_{\phi}$, 
the axion field transforms as $a \to a + \kappa \sqrt{2} V_a$. 
Inverting the orthogonal transformation in \eq{eq:axiondef2}, 
one readily 
obtains the 
projection of the angular modes on the axion field: 
\beq 
\label{eq:aproj2}
a_{\phi} \to q_{\phi} V_{\phi} \frac{a}{V_a} \, .  
\eeq 
To determine the $q$ charges we proceed as follows. 
First, we require the orthogonality between the axion current and the 
gauge currents $J_{B-L} = \sum_i (B-L)_{\phi} \sqrt{2} V_{\phi} \partial_\mu a_{\phi}$ 
and $J_Y = \sum_\phi Y_{\phi} \sqrt{2} V_{\phi} \partial_\mu a_{\phi}$ (to avoid kinetic mixings of the axion field 
with massive gauge bosons). This yields, respectively\footnote{Here, in analogy to the derivation in \sect{sec:axionembedding}, 
the VEVs denote norms in flavor space, 
i.e.~$V^2_\chi = \sum_{A} |\colorPQ{V}_{A}|^2$, 
$v_{\Phi_u}^2 = \sum_{A} |\vevv{u\Phi}{A}|^2$, etc.} 
\begin{align}
- q_{\chi} V^2_{\chi} 
-2 q_{\Delta} V^2_{\Delta} &= 0  \, ,  \\
- q_{\Phi_u} v_{\Phi_u}^2 
+ q_{\Phi_d} v_{\Phi_d}^2 
- q_{\Sigma_u} v_{\Sigma_u}^2 
+ q_{\Sigma_d} v_{\Sigma_d}^2 
- q_{\Sigma'_u} v_{\Sigma'_u}^2 
+ q_{\Sigma'_d} v_{\Sigma'_d}^2 
&= 0 \, .   
\end{align}
Second, by decomposing the invariants with non-trivial global re-phasings 
in the Pati-Salam scalar potential (cf.~\eq{eq:VC})
we obtain the  
following extra constraints on $q$ charges 
\begin{align}
\label{eq:EWPQchargesu}
q_{\Phi_u} &= q_{\Sigma_u} = q_{\Sigma'_u} \, , \\ 
\label{eq:EWPQchargesd}
q_{\Phi_d} &= q_{\Sigma_d} = q_{\Sigma'_d} \, . 
\end{align}
To close the system of linear equations, in order to extract the $q$ charges, 
it is necessary to include also the matching between UV and IR PQ charges 
in \eq{eq:generalUVIRPQ2} for all the scalar fields.   
Therefore we obtain
\begin{align}
c_2 &= - \frac{V_\chi^2 - 4 V_\Delta^2}{V_\chi^2 + 4 V_\Delta^2} \, , & 
c_3 &= 4 \frac{v^2_u - v^2_d}{v^2} \, , \\
q_{\chi} &= - \frac{8 V^2_{\Delta}}{V^2_{\chi} + 4V^2_{\Delta}} \, , &
q_{\Delta} &= \frac{4 V^2_{\chi}}{V^2_{\chi} + 4V^2_{\Delta}} \, , \\
q_{\Phi_u} &= q_{\Sigma_u} = q_{\Sigma'_u}
= 4 \frac{v^2_d}{v^2}  \, , & 
q_{\Phi_d} &= q_{\Sigma_d} = q_{\Sigma'_d}
= 4 \frac{v^2_u}{v^2}  \, , \\
\label{eq:V2expr}
V_a^2 &= 16 \( \frac{V^2_{\chi}V^2_{\Delta}}{V^2_{\chi} + 4 V^2_{\Delta}} 
+ \frac{v_u^2v_d^2}{v_{\rm EW}^2} \) \, ,
\end{align}
where we defined 
$v_u^2 = v_{\Phi_u}^2 + v_{\Sigma_u}^2 + v_{\Sigma'_u}^2$, 
$v_d^2 = v_{\Phi_d}^2 + v_{\Sigma_d}^2 + v_{\Sigma'_d}^2$ 
and  
$v_{\rm EW}^2 = v^2_u + v^2_d$, with $v_{\rm EW} = 246/\sqrt{2}\,\mathrm{GeV}$.

To compute low-energy axion couplings to SM charged fermions, 
we decompose the Pati-Salam Yukawa interaction $\bar Q_L Q_R \Phi \supset \bar q_L u_R \Phi_u + \bar q_L d_R \Phi_d$, 
and similarly for the Yukawa contribution of the $\Sigma$ and 
$\Sigma'$ fields. The axion is hence removed from the Yukawa Lagrangian via 
a family universal,  
axion-dependent transformation 
(suppressing generation indices)
\begin{align}
u &\to e^{-i \gamma_5 q_{\Phi_u} \frac{a}{2\sqrt{2}V_a}} u \, , \\
d &\to e^{-i \gamma_5 q_{\Phi_d} \frac{a}{2\sqrt{2}V_a}} d \, , \\
e &\to e^{-i \gamma_5 q_{\Phi_d} \frac{a}{2\sqrt{2}V_a}} e \, . 
\end{align}
Note that the same rotation also removes the axion from the 
Yukawa interaction involving $\Sigma$ and $\Sigma'$, 
due to the alignment of the (electroweak) PQ charges in 
\eqs{eq:EWPQchargesu}{eq:EWPQchargesd}. 

The above axial field redefinitions 
generate the anomalous terms 
\beq 
\delta \mathcal{L} = 
\frac{\alpha_s N}{4 \pi} \frac{a}{\sqrt{2}V_a} G \tilde G + 
\frac{\alpha E}{4 \pi} \frac{a}{\sqrt{2}V_a} F \tilde F \, ,
\eeq 
with the anomaly factors 
\begin{align}
N &= n_g T(3) \(2 \frac{q_{H_u^{10}}}{2} + 2 \frac{q_{H_d^{10}}}{2} \) = 6 \, , \\
E &= n_g \(q_{H_u^{10}} 3 (2/3)^2 + q_{H_d^{10}} 3 (-1/3)^2 
+ q_{H_d^{10}} (-1)^2 \) = 16 \, ,  
\end{align} 
using $n_g = 3$, $T(3) = 1/2$, etc.  
So, in particular, upon choosing the standard normalization of the 
axion-gluon coupling in terms of the Lagrangian term 
\beq 
\mathcal{L} \supset \frac{\alpha_s}{8 \pi} \frac{a}{f_a} G \tilde G \, ,
\eeq
the axion decay constant reads (using \eq{eq:V2expr})
\beq 
\label{eq:deffaV}
f_a = \frac{\sqrt{2} V_a}{2N} = \frac{1}{3} \sqrt{\frac{V^2_{\chi}V^2_{\Delta}}{V^2_{\chi} + 4 V^2_{\Delta}} 
+ \frac{v_u^2v_d^2}{v_{\rm EW}^2}} \, , 
\eeq
while $E/N = 8/3$, which enters the axion-photon coupling \cite{diCortona:2015ldu}
\beq 
\label{eq:axionph}
C_{a\gamma} =  E/N - 1.92(4) \, ,
\eeq 
defined in terms of the Lagrangian term 
\beq 
\mathcal{L} \supset \frac{\alpha_{\rm EM}}{8\pi f_a} C_{a\gamma} a F \tilde F \, . \eeq 
On the other hand, 
the variation of the fermion kinetic terms yields
\begin{align}
\delta(\bar u i \slashed{\partial} u) 
&= q_{\Phi_u} \frac{\partial_\mu a}{2 \sqrt{2}V_a} \bar u \gamma^\mu \gamma_5 u \, , \\
\delta(\bar d i \slashed{\partial} d) 
&= q_{\Phi_d} \frac{\partial_\mu a}{2 \sqrt{2}V_a} \bar d \gamma^\mu \gamma_5 d \, , \\
\delta(\bar e i \slashed{\partial} e) 
&= q_{\Phi_d} \frac{\partial_\mu a}{2 \sqrt{2}V_a} \bar e \gamma^\mu \gamma_5 e \, . 
\end{align}
Hence, defining the axion coupling to SM fermions via 
\beq 
\label{eq:defcf}
\mathcal{L} \supset c_f \frac{\partial_\mu a}{2 f_a} \bar f \gamma_\mu \gamma_5 f \, , 
\eeq
and replacing $f_a = \sqrt{2} V_a / (2N)$, we find the (family universal) couplings
\begin{align}
\label{eq:axioncu}
c_u &= 
\frac{q_{\Phi_u}}{2N} = \frac{v^2_d}{3v^2} 
\equiv \frac{1}{3} \cos^2\beta
\, , \\
\label{eq:axioncde}
c_d &= c_e =
\frac{q_{\Phi_d}}{2N} = \frac{v^2_u}{3v^2} 
\equiv \frac{1}{3} \sin^2\beta \, , 
\end{align}
where in the last step we introduced 
\beq 
\label{eq:deftanbeta}
\tan\beta \equiv \frac{v_u}{v_d} = 
\frac{\sqrt{
\sum_A |\vevv{u\Phi}{A}|^2 
+ \sum_A |\vevv{u\Sigma}{A}|^2 + 
\sum_A|\vevv{u\Sigma'}{A}|^2}}{\sqrt{
\sum_A |\vevv{d\Phi}{A}|^2 
+ \sum_A |\vevv{d\Sigma}{A}|^2 + 
\sum_A|\vevv{d\Sigma'}{A}|^2}
} \, .
\eeq

\section{In-depth analysis of the Pati-Salam  model}\label{app:PSmodel}

In this Appendix we provide further details regarding the Pati-Salam model discussed in the main part of the manuscript. It encompasses discussions on group-theoretical aspects, computational analysis of PQ quality within the scalar sector, insights into the Yukawa sector, and an examination of the perturbativity issue.

%%%%%%%%%%%%%%%%%%%%%%%%%%%%%%
%%%%%%%%%%%%%%%%%%%%%%%%%%%%%%
\subsection{Explicit VEV definitions \label{app:procedure-explicit}}

For any explicit computation in the Pati-Salam model, the most convenient way is through the use of tensor methods. We gather here all the related conventions, and identify the VEV directions for every irrep.

First, we write the index structure of the representations from \Table{tab:PSirrep}. The fermions, written as right-chiral Weyl fermions (but suppressing the Weyl index), have the index structure
\begin{align}
    (\bar{Q}_L)_{ai}\, ,&&
    (Q_R)^{ai'A}\, ,&&
    (\Psi_R)_A\, ,
\end{align}
while the scalars can be written as
\begin{align}
	\Phi^{ii'}{}_{A}\, ,&&
	\Sigma^{a}{}_{b}{}^{ii'}{}_{A}\, ,&&
	\Delta^{(ab)(i'j')(AB)}\, ,&&
	\chi^{ai'}{}_{A}\, ,&&
	\xi^{a}{}_{b}{}^{(i'j')} \, .
\end{align}
In the above notation, the indices $\{a,b,\ldots\}$ are used for $\SU(4)_{\rm PS}$, $\{i,j,\ldots\}$ for $\SU(2)_L$, $\{i',j',\ldots\}$ for $\SU(2)_R$, and $\{A,B,\ldots\}$ for $\SU(3)_{f_{R}}$, while parentheses indicate symmetrization of enclosed indices.

Expanding in component fields and using an index $I$ for the $3$ copies of $Q_{L}$ (not gauged) and $K$ for the $8$ generations of the anomalons, the fermions can be written as
\begin{align}
    (\bar{Q}_L)^{I}{}_{a i}&=
        \mymatrixs{ 
            \bar{u}_{L1}^{I} & \bar{d}_{L1}^{I} \\
            \bar{u}_{L2}^{I} & \bar{d}_{L2}^{I} \\
            \bar{u}_{L3}^{I} & \bar{d}_{L3}^{I} \\
            \bar{\nu}^{I}_{L} & \bar{e}^{I}_{L} \\
            } \, , 
        &
    (Q_{R})^{ai'A}&=
        \mymatrixs{
            u_{R1}^{A} & d_{R1}^{A}\\
            u_{R2}^{A} & d_{R2}^{A}\\
            u_{R3}^{A} & d_{R3}^{A}\\
            \nu_{R}^{A} & e_{R}^{A}\\
        } \, , &
    (\Psi_R)^{K}{}_{A} \, , 
    \label{eq:ansatz-fermions}
\end{align}
where $L$ and $R$ denote whether the field was originally left- or right-chiral, respectively. 

The vacuum alignments in scalar irreps are found to be along the directions
\begin{align}
	\langle \Phi^{ii'}{}_{A}\rangle &= \mymatrixs{0&1\\0&0\\}\otimes \vevv{u\Phi}{A}-
	\mymatrixs{0&0\\1&0\\}\otimes \vevv{d\Phi}{A} \, , \label{eq:vevs-begin}\\[4pt]
%%%
	\langle \Sigma^{a}{}_{b}{}^{ii'}{}_{A}\rangle &= \tfrac{1}{\sqrt{12}}\,\mathrm{diag}\,(1,1,1,-3)\otimes\left(\mymatrixs{0&1\\0&0\\}\otimes \vevv{u\Sigma}{A}-
	\mymatrixs{0&0\\1&0\\}\otimes \vevv{d\Sigma}{A}\right) \, ,\\[4pt]
%%%
    \langle \Sigma'^{a}{}_{b}{}^{ii'}{}_{A}\rangle &= \tfrac{1}{\sqrt{12}}\,\mathrm{diag}\,(1,1,1,-3)\otimes\left(\mymatrixs{0&1\\0&0\\}\otimes \vevv{u\Sigma'}{A}-
	\mymatrixs{0&0\\1&0\\}\otimes \vevv{d\Sigma'}{A}\right) \, ,\\[4pt]
%%%
	\langle \Delta^{abi'j' AB}\rangle &= 
		\mymatrixs{0&0&0&0\\0&0&0&0\\0&0&0&0\\0&0&0&1\\}
		\otimes\mymatrixs{1&0\\0&0\\}\otimes\colorPQ{Z}^{AB} \, ,
        \label{eq:vevs-delta}\\[4pt]
	\langle \chi^{ai'}{}_{A}\rangle&= (0,0,0,1)\otimes (1,0)\otimes \colorPQ{V}_{A} \, ,
        \label{eq:vevs-chi}\\[4pt]
	\langle \xi^{a}{}_{b}{}^{i'j'}\rangle &= \colorPQ{W}\;\tfrac{1}{\sqrt{12}}\,\mathrm{diag}\,(1,1,1,-3)\otimes 
    \tfrac{1}{\sqrt{2}}\mymatrixs{0 & 1\\ 1&0\\}\, ,\label{eq:vevs-end}
\end{align}
where the normalization of the VEVs is part our conventions.

The VEVs are labeled \colorEW{blue} for EW scale and \colorPQ{red} for PS scale, where  the benchmark values of the mass scales are given in \eq{eq:vev-scaling}. The EW VEVs are those
listed in \eq{eq:labels-EW-VEVs}. Since $\Delta$ is symmetric in the indices $(AB)$, there are only $6$ independent $Z$-type VEVs.

%%%%%%%%%%%%%%%%%%%%%%%%%%%%%%
%%%%%%%%%%%%%%%%%%%%%%%%%%%%%%
\subsection{Procedure for enumerating invariants \label{app:procedure-invariant-enumeration}}
This subsection provides some technical details on our procedure of searching for and enumerating invariants. 
It applies for invariants both in the scalar potential, which are relevant for assessing PQ quality in \sect{sec:PQquality} and further details in \app{app:PSmodel-PQ-at-tree-level}, as well as the Yukawa sector in \sect{sec:SMflavour}.

In terms of field-powers, a general invariant of dimension $d$ 
in the scalar potential will have the form
\begin{align}
    \Phi^{n_{1}}\Phi^{\ast n_{2}}\,
    \Sigma^{n_{3}}\Sigma^{\ast n_{4}}\,
    \Delta^{n_{5}}\Delta^{\ast n_{6}}\,
    \chi^{n_{7}}\chi^{\ast n_{8}}\,
    \xi^{n_{9}}\label{eq:invariant-general-form} \, , 
\end{align}
where $n_{i}\in\mathbb{Z}_{+}$ are non-negative integers, and $\sum_{i}n_{i}=d$. We search for all $d\leq 9$ invariants by systematically going through all admissible configurations of $n_{i}$. For each candidate configuration, one can perform a sequence of increasingly sophisticated tests:
\begin{enumerate}
    \item\label{item:layers-1}  \textbf{Invariance under center}. A necessary condition to form an invariant of a group $G$ is that the action of the center of $G$ acts trivially on the expression of \eq{eq:invariant-general-form}. The center of the group $\SU(4)_{\rm PS}\times\SU(2)_{L}\times\SU(2)_R\times\SU(3)_{f_R}$ is simply $\mathbb{Z}_{4}\times\mathbb{Z}_{2}\times\mathbb{Z}_{2}\times\mathbb{Z}_{3}$, so irreps have a discrete charge under every one of the factors. For $\SU(2)_L$, the condition can be summarized that the sum of $L$-isospins of the irreps in the product must be integer rather than half-integer, and an analogous condition holds for  $R$-isospin of $\SU(2)_R$.
    The other constraint comes from the charges under
    $\mathbb{Z}_{4}$ and $\mathbb{Z}_{3}$, which are listed for every irrep in \Table{tab:PSirrep}. 
    \item\label{item:layers-2}  \textbf{Tensor product contains singlet}. A stricter condition is to check that the tensor product of the representations in \eq{eq:invariant-general-form} contains at least one singlet. This can be straightforwardly computed using the public \textit{Mathematica} package \texttt{GroupMath}~\cite{Fonseca:2020vke}, or alternatively also \texttt{LieART}~\cite{Feger:2012bs,Feger:2019tvk}. Note that this is not necessarily a sufficient condition for an invariant to exist, since the multiplicity of each type of irrep is not taken into account. Namely, the tensor product keeps track of parts both symmetric and anti-symmetric in its factors, with the latter vanishing identically if we have only one copy of an irrep.
    \item\label{item:layers-3} \textbf{Symmetrized tensor product contains singlet}. 
    This operation discards all (partially) anti-symmetric parts from a tensor power of a representation, thus correctly handling the presence of a single copy of an irrep in an invariant. Each $n_{i}$ in \eq{eq:invariant-general-form} is thus taken as the symmetric tensor power of its corresponding representation, while the ordinary tensor product is then used between these factors. 
    The number of singlets in such a product in fact equals the number of independent invariants of that type in the Lagrangian; in particular, a non-zero number of singlets demonstrates the existence of the invariant. The method can be extend to having multiple copies of an irrep by treating each one as an independent irrep with respect to which the symmetrized tensor power is taken. To compute symmetrized tensor products, we use the public software \texttt{LiE}~\cite{Leeuwen1992LiE}.
\end{enumerate}

These criteria are necessary conditions of increasing strictness, with condition~\ref{item:layers-3} finally becoming also sufficient for the existence of an invariant. As soon as a configuration fails a test, it can be discarded as 
an invariant candidate without performing the subsequent computationally more intensive tests. 

If used all the way up to condition~\ref{item:layers-3}, the above procedure can be used to produce a list of configurations (field-powers) that yield an invariant. If implementing only up to condition~\ref{item:layers-1} or \ref{item:layers-2}, the procedure yields all potential candidates for an invariant. For scalar operators in Sect.~\ref{sec:PS-model}, \sect{sec:PQquality}, and \app{app:PSmodel-PQ-at-tree-level}, 
we evaluated configurations with $d\leq 4$ up to the final condition, and configurations with $5\leq d\leq 9$ up to condition~\ref{item:layers-2}.

The above procedure for finding invariants can be extended to 
include fermions. One simply includes a $\overline{Q_L}^{m_{1}}Q_{R}^{m_{2}}\Psi^{m_{3}}$ to the form of \eq{eq:invariant-general-form}, and then evaluates the same conditions \ref{item:layers-1}--\ref{item:layers-3}, Lorentz structure not withstanding. Taking the total fermionic power to be $2$, i.e.~$m_{1}+m_{2}+m_{3}=2$, and omitting the conjugates of fermionic irreps, as is relevant for the Yukawa sector in \sect{sec:SMflavour}, automatically also produces a Lorentz invariant.

We conclude this discussion by elaborating on condition~\ref{item:layers-1} in our concrete model, since it can be easily evaluated by hand. The vanishing of the four discrete charges of the center $\mathbb{Z}_{4}\times\mathbb{Z}_{2}\times\mathbb{Z}_{2}\times\mathbb{Z}_{3}$ for an expression of \eq{eq:invariant-general-form} to be an invariant yields the following conditions:
    \begin{align}
        2(n_{5}-n_{6})+(n_{7}-n_{8})&\equiv 0 \pmod{4},\label{eq:discrete-condition-Z4}\\
        (n_{1}-n_{2})+(n_{3}-n_{4})&\equiv 0 \pmod{2},\label{eq:discrete-condition-Z2L}\\
        (n_{1}-n_{2})+(n_{3}-n_{4})+(n_{7}-n_{8}) & \equiv 0 \pmod{2},\label{eq:discrete-condition-Z2R}\\
        -(n_{1}-n_{2})-(n_{3}-n_{4})-(n_{5}-n_{6})-(n_{7}-n_{8})&\equiv 0 \pmod{3}.\label{eq:discrete-condition-Z3}
    \end{align}
These conditions are the formal underpinning of the rules \ref{item:intuition1}--\ref{item:intuition4} for building invariants given in \sect{sec:PQquality}.

%%%%%%%%%%%%%%%%%%%%%%%%%%%%%%%%%%%%%%%%%%%%%%%%%%%%%%%%%%%%
\subsection{Aspects of PQ quality 
    \label{app:PSmodel-quality}
}

In this section, we elaborate on the uniqueness of the PQ symmetry in our Pati-Salam model, cf.~\app{app:number-of-U1s}, as well as on the question of its quality. We provide details on determining the dominant PQ-breaking vacuum contributions in a tree-level analysis in \app{app:PSmodel-PQ-at-tree-level}, and comment on how loop corrections may complicate the picture in \app{app:PSmodel-PQ-at-loop-level}.

%%%%%%%%%%%%%%%%%%%%%%%%%%%%%%%%%%%%%%%%%%%%%%%%%%%%%%%%%%%%
\subsubsection{Uniqueness of PQ: invisible axion and counting $\U(1)$s
    \label{app:number-of-U1s}
}

\def\PHASE{\varphi}

We show now that a single accidental $\U(1)$ emerges out of the scalar potential in the model (leading to a single invisible axion), as claimed in the main part of \sect{sec:PS-model}. 

It is most straightforward to derive this from $\mathcal{V}_{\mathbb{C}}$ in \eqref{eq:VC}. The model contains $4$ different complex scalar irreducible representations\footnote{
    In the subsequent analysis the multiplicity of each representation can be ignored, since a multiplicity greater than one also increases the number of invariant terms, which however enforce the same charge assignment for all irrep copies under any preserved $\U(1)$.
}, namely $\{\Phi,\Sigma,\Delta,\chi\}$ , and hence $4$ complex irrep phases relevant when assigning $\U(1)$ charges: $\{\PHASE_{\Phi},\PHASE_{\Sigma},\PHASE_{\Delta},\PHASE_{\chi}\}$.  We use this basis to describe the phase $\vec{\PHASE{}}$ of any scalar operator of the form given in \eq{eq:invariant-general-form}: 
\begin{align}
    \vec{\PHASE{}}=(n_{1}-n_{2},n_{3}-n_{4},n_{5}-n_{6},n_{7}-n_{8}).
\end{align}
Analogously, we define a vector of charge assignments as
\begin{align}
    \vec{q}&=([\Phi],[\Sigma],[\Delta],[\chi]),
\end{align}
where the square brackets denote the charge of the field inside them. An invariant with phase $\vec{\PHASE}$ preserves charge $\vec{q}$ if and only if $\vec{\PHASE}\cdot \vec{q}=0$.

The invariants in \eqref{eq:VC} all lie in the 3-dimensional subspace spanned by
\begin{align}
    \vec{\PHASE}_{1}&:=(1,-1,0,0), &
    \vec{\PHASE}_{2}&:=(0,-2,2,0), &
    \vec{\PHASE}_{3}&:=(0,0,1,2), 
\end{align}
leading to a 1-dimensional orthogonal complement --- the preserved PQ charge defined in \Table{tab:PSirrep}:
\begin{align}
    \vec{q}_{\text{PQ}}&:= (2,2,2,-1).
\end{align}
We could in principle stop the analysis here. There is however a subtlety we wish to clarify, which we focus on in the remainder of this subsection~\ref{app:number-of-U1s}. 

When considering contributions to the vacuum, only SM or EW singlets will acquire VEVs. Each VEV carries a well-defined $B-L$ charge, and since all VEV entries in our specific Pati-Salam irreps carry the same value of this charge, cf.~\Table{tab:BLnumbers}, it is useful to define a $\U(1)$ charge for irreps based on the $B-L$ of their VEV component: 
\begin{align}
	\vec{q}_{BL}& := (0,0,-2,-1).
\end{align}
An invariant with phase $\vec{\PHASE}$ can contribute to the vacuum with its VEVs if and only if $\vec{\PHASE}\cdot\vec{q}_{BL}=0$.

Notice that 
\begin{align}
    \vec{\PHASE}_{1}\cdot \vec{q}_{BL}&=0, &
    \vec{\PHASE}_{2}\cdot \vec{q}_{BL}&\neq 0, &
    \vec{\PHASE}_{3}\cdot \vec{q}_{BL}&\neq 0,
\end{align}
implying that only invariants proportional to phase $\vec{\PHASE}_{1}$ can contribute to the vacuum. In particular, $\Sigma^{*2}\Delta^{2}$ and $\Delta\chi^{2}\xi$ do not contribute to the vacuum. This implies that in the subspace of SM and EW singlets, there are $4-1=3$ preserved $U(1)$ charges, i.e.~$3$ charges for which $\vec{\PHASE}_{1}\cdot \vec{q}=0$. They are $\vec{q}_{\rm{PQ}}$, $\vec{q}_{BL}$ and $\vec{q}_{\rm{new}}$, where the last charge can be taken as
\begin{align}
    \vec{q}_{\rm{new}}:=(1,1,0,0).
\end{align}

Since $\vec{q}_{BL}$ is gauged on the space of singlets, the corresponding massless phase represents an unphysical would-be Goldstone boson. Conversely, the preservation of $\vec{q}_{\rm{PQ}}$ and $\vec{q}_{\rm{new}}$ leads to two physical nominally massless phases, and they correspond to a Goldstone (axion) and a pseudo-Goldstone boson, respectively, as we shall see.

The crucial point here is that the above analysis of massless modes is derived from analyzing the singlet subspace, and as such applies only to the tree-level spectrum. Since $\vec{q}_{\rm{new}}$ is only preserved on this subspace and is not an actual symmetry of the Lagrangian, the generic QFT expectation is that the corresponding phase is massless at tree-level only `accidentally', hence its classification as a pseudo-Goldstone mode, and it becomes massive due to loop corrections once non-singlet states become involved on internal lines of diagrams.

Note that the loop generation of mass for the phase corresponding to $\vec{q}_{\rm{new}}$ is crucial for obtaining a viable theory of the axion. Namely, at the PS-breaking scale only two complex irreps $\Delta$ and $\chi$ obtain VEVs, and hence only two charges are spontaneously broken: $\vec{q}_{\rm{PQ}}$ and $\vec{q}_{BL}$. Since $\vec{q}_{\rm{new}}$ does not involve either of the two irreps with high-scale VEVs, it is broken explicitly by the loop-generated mass term, or in its absence it is broken spontaneously only at the EW scale. The latter possibility would lead to a visible QCD axion of the WW-type (Weinberg-Wilczek), with the majority component coming from the Goldstone of the new charge (which is also anomalous under QCD) rather than the original PQ charge. Of course, a visible axion would be phenomenologically excluded. However, as we will show below, this is not the case here, since the mass of the would-be WW axion is lifted by explicit breaking at the loop level.

To confirm the loop-generation of this mass explicitly, we search for a 1-PI loop diagram generated in the effective potential from renormalizable couplings, which breaks $\vec{q}_{\rm{new}}$ yet contributes to the vacuum. 

Let's first focus on its outer legs, and specifically on its phase $\vec{\PHASE}$. It must correspond to an invariant that preserves $\vec{q}_{BL}$ (necessary for a non-vanishing vacuum contribution) as well as $\vec{q}_{\rm{PQ}}$ (since renormalizable couplings preserve PQ), hence $\vec{\PHASE}\cdot\vec{q}_{\rm{PQ}}=\vec{\PHASE}\cdot\vec{q}_{BL}=0$. If one amends to this the necessary conditions for an invariant (from the action of the gauge center) in \eqs{eq:discrete-condition-Z4}{eq:discrete-condition-Z3}, one can derive that the phase of the operator is necessarily of the form
\begin{align}
    \vec{\PHASE}&= (a,2b-a,-b,2b), \label{eq:phase-of-operator-breaking-qnew} 
\end{align}
where $a,b\in\mathbb{Z}$. Its breaking of $\vec{q}_{\rm{new}}$, i.e.~$\vec{\PHASE}\cdot \vec{q}_{\rm{new}}\neq 0$, implies $b\neq 0$. The simplest case $(a,b)=(2,1)$ suggests the candidate operator
\begin{align}
    \mathcal{O}&=\Phi^{2}\,\Delta^*\,\chi^2. \label{eq:operator-candidate}
\end{align}

\begin{figure}[htb]
	\centering
	\includegraphics[width=0.45\linewidth]{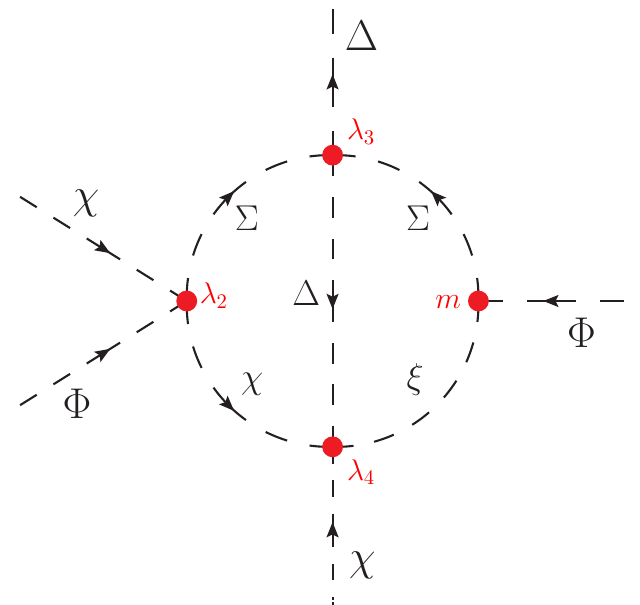}
	\hspace{0.05\linewidth}
	\includegraphics[width=0.45\linewidth]{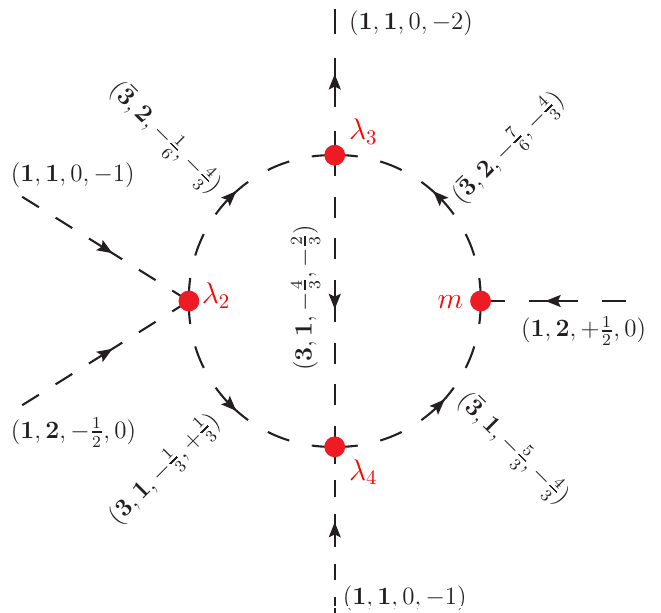}
	\caption{An example of a $2$-loop diagram constructed from renormalizable couplings contributing to the operator $\mathcal{O}$ of \eq{eq:operator-candidate} in the effective potential. The diagram is written with Pati-Salam fields (left) and contributes to the vacuum, as demonstrated by the subcontribution (right) written in the   
	$\SU(3)_C\times\SU(2)_L\times\U(1)_{Y}\times\U(1)_{B-L}$ language. In the right diagram, Clebsch coefficients in each vertex are ignored. \label{fig:2-loop-diag-newphase}}
\end{figure}
This operator explicitly breaks $\vec{q}_{\text{new}}$ and can indeed be generated at loop level, see \fig{fig:2-loop-diag-newphase}. 

We now verify the non-vanishing nature of this contribution explicitly. The dimensionful and dimensionless vertices in the diagram of \fig{fig:2-loop-diag-newphase} are labeled by $m$ and $\lambda_{2,3,4}$, respectively. While there are multiple independent invariants $\Delta\Delta^\ast \Sigma^* \Phi$ and $\Sigma^2\Delta^{*2}$, the respective couplings $\lambda_{2}$ and $\lambda_{3}$ refer to one particular index contraction for each. Using the index notation of \app{app:procedure-explicit}, we define the couplings as shown below:
\begin{align}
\begin{split}
\mathcal{L} \supset 
    &\quad
	\textcolor{red}{m}\;
		\Phi^{ii'}{}_{A}\,
		\Sigma^{*}{}_{a}{}^{b}{}_{ij'}{}^{A}\,
		\xi^{a}{}_{b}{}^{j'k'}\; 
		\epsilon_{i'k'}\\
	&+
    \textcolor{red}{\lambda_{2}}\;
		\chi^{ai'}{}_{A} \,
		\chi^{*}{}_{bj'}{}^{B} \,
		\Sigma^{*}{}_{a}{}^{b}{}_{ii'}{}^{A} \,
		\Phi^{ij'}{}_{B}\\
	&+
    \textcolor{red}{\lambda_{3}}\;
		\Sigma^{e}{}_{a}{}^{ii'}{}_{A}\,
		\Sigma^{f}{}_{b}{}^{jj'}{}_{B}\,
		\Delta^{*}{}_{eci'k'CE}\,
		\Delta^{*}{}_{fdj'l'DF}\;
		\epsilon^{abcd}\,
		\epsilon_{ij}\,
		\epsilon^{k'l'}\,
		\epsilon^{ACD}\,
		\epsilon^{BEF}\\
	&+
    \textcolor{red}{\lambda_{4}}\;
		\Delta^{aei'j'AB} \,
		\chi^{ck'}{}_{A} \,
		\chi^{dl'}{}_{B} \,
		\xi^{b}{}_{e}{}^{m'n'}\;
		\epsilon_{abcd}\,
		\epsilon_{i'k'}\,
		\epsilon_{j'm'}\,
		\epsilon_{l'n'}.
\end{split}
\end{align}

These terms lead to the following Feynman rules (symmetrizing when multiple legs with the same field are present), where the bracketing notation $\langle.\rangle$ and $[.]$ already anticipates which fields are taken as external-leg VEVs or internal legs in our $2$-loop diagram, respectively:
\begin{align}
\textcolor{red}{m}:& \quad
	[{\Sigma^\ast}]_{a}{}^{b}{}_{ij'}{}^{A}\;
	[\xi]^{c}{}_{d}{}^{k'l'}\;
	\langle \Phi^{ii'}{}_{A} \rangle \;
	\delta^{a}{}_{c}\,
	\delta_{b}{}^{d}\,
	\delta^{j'}{}_{l'}\,
	\epsilon_{i'k'},\\
\textcolor{red}{\lambda_{2}}:&\quad
	[\Sigma^*]_{a}{}^{b}{}_{ii'}{}^{A} \;
	[\chi^*]_{cj'}{}^{B} \;
	\langle \chi^{ai'}{}_{A} \rangle\;
	\langle \Phi^{ij'}{}_{B} \rangle\;
	\delta_{b}{}^{c}
	,\\
\textcolor{red}{\lambda_{3}}:&\quad
	[\Sigma ]^{e}{}_{a}{}^{ii'}{}_{A} \;
	[\Sigma ]^{f}{}_{b}{}^{jj'}{}_{B} \;
	[\Delta^*]_{gck'm'CE}
	\langle\Delta^*{}_{hdl'n'DF} \rangle \nonumber\\
&\qquad\quad\cdot
	\epsilon^{abcd} \,	
	\epsilon_{ij} \,
	\epsilon^{k'l'} \,
	\epsilon^{ACD} \,
	\epsilon^{BEF} \;
	\tfrac{1}{2}\left(
		\delta_{e}{}^{g} \,
		\delta_{f}{}^{h} \,
		\delta_{i'}{}^{m'}\,
		\delta_{j'}{}^{n'}
		+
		\delta_{f}{}^{g} \,
		\delta_{e}{}^{h} \,
		\delta_{j'}{}^{m'}\,
		\delta_{i'}{}^{n'}
	\right), \\
\textcolor{red}{\lambda_{4}}: &\quad 
	[\Delta]^{aei'j'AB} \;
	[\chi]^{ck'}{}_{C} \;
	[\xi]^{b}{}_{f}{}^{m'n'} \;
	\langle \chi^{dl'}{}_{B} \rangle \;
	\delta_{A}{}^{C} \,
	\delta_{e}{}^{f} \,
	\epsilon_{abcd} \,	
	\epsilon_{j'm'}\,
	\tfrac{1}{2}\left(
		\epsilon_{i'k'}\,
		\epsilon_{l'n'}
		-
		\epsilon_{i'l'}\,
		\epsilon_{k'n'}
	\right).
\end{align}
The Levi-civita tensors of appropriate dimension are denoted by $\epsilon$,
with the convention $\epsilon_{12}=-\epsilon^{12}=1$. Next, amputating the internal states, the rule for each vertex in the diagram is written as
\begin{align}
[\textcolor{red}{m}_{\Sigma^* \xi}]
	[\Sigma]^{a}{}_{b}{}^{ij'}{}_{A}
	[\xi^*]_{c}{}^{d}{}_{k'l'} 
	&:=
	\Big[
	\langle \Phi^{ii'}{}_{A} \rangle \;
	\delta^{a}{}_{c}\,
	\delta_{b}{}^{d}\,
	\delta^{j'}{}_{l'}\,
	\epsilon_{i'k'}
	\Big]_{\star_{m}}, \label{eq:vertex-rules-begin}\\
[\textcolor{red}{\lambda_{2}}_{\Sigma^* \chi^*}]
	[\Sigma]^{a}{}_{b}{}^{ii'}{}_{A}
	[\chi]^{cj'}{}_{B} 
	&:=
	\Big[
	\langle \chi^{ai'}{}_{A} \rangle\;
	\langle \Phi^{ij'}{}_{B} \rangle\;
	\delta_{b}{}^{c}	
	\Big]_{\star_{2}},\\
[\textcolor{red}{\lambda_{3}}_{\Sigma\Sigma\Delta^*}]
	[\Sigma^\ast]_{e}{}^{a}{}_{ii'}{}^{A} \;
	[\Sigma^\ast]_{f}{}^{b}{}_{jj'}{}^{B} \;
	[\Delta]^{gck'm'CE}
	&:=
	\Big[ 
	\langle\Delta^*{}_{fdj'l'DF} \rangle \;
	\delta_{e}{}^{g} \,
	\delta_{i'}{}^{m'} \,
	\epsilon^{abcd} \,
	\epsilon_{ij} \,
	\epsilon^{k'l'} \,
	\epsilon^{ACD} \,
	\epsilon^{BEF}
	\Big]_{\star_{3}},
	\\
[\textcolor{red}{\lambda_{4}}_{\Delta\chi\xi}]
	[\Delta^*]_{aei'j'AB} \;
	[\chi^*]_{ck'}{}^{C} \;
	[\xi^*]_{b}{}^{f}{}_{m'n'}
	&:=
	\Big[
	\langle \chi^{dl'}{}_{B} \rangle \;
	\delta_{A}{}^{C} \,
	\delta_{e}{}^{f} \,
	\epsilon_{abcd} \,
	\epsilon_{i'k'}\,
	\epsilon_{j'm'}\,
	\epsilon_{l'n'}
	\Big]_{\star_{4}}, \label{eq:vertex-rules-end}
\end{align}
where the $\star$-conditions enforce that contractions along internal lines will proceed according to the irrep being exchanged: 
\begin{align}
	\star_{m}&= \{ (ab)_{\dagger\mathrm{tr}},(cd)_{\dagger\mathrm{tr}}, (k'l') \},\\
	\star_{2}&= \{ (ab)_{\dagger\mathrm{tr}}	\},\\
	\star_{3}&= \{
		(ea)_{\dagger\mathrm{tr}},
		(fb)_{\dagger\mathrm{tr}},
		(gc),
		(k'm'),
		(CE)
		\}, \\
	\star_{4}&= \{
		(ae),
		(i'j'),
		(AB),
		(bf)_{\dagger\mathrm{tr}},
		(m'n')
		\}.
\end{align}
Parentheses without a subscript denote symmetrization of enclosed indices, while the subscript $\dagger\mathrm{tr}$ implies taking a hermitian traceless combination of the enclosed indices. Furthermore, the vertex labels encompass the entire expression on the left-hand side of \eqs{eq:vertex-rules-begin}{eq:vertex-rules-end}, keeping track of which fields are taken in internal lines and remain uncontracted (the subscript under the coefficient label), as well as to which internal line the free indices belong to; an individual vertex labels thus comprises of multiple bracketed parts.

The easiest way to confirm a non-vanishing contribution is to contract the couplings 
along propagators in the unbroken phase, such that the structure from vertices factorizes from the loop diagram. Using the VEV directions specified in \app{app:procedure-explicit}, the properly contracted vertex structure $\lambda_{\mathcal{O}}$ explicitly comes out as
\begin{align}
\lambda_{\mathcal{O}}=\quad&
    [\textcolor{red}{m}_{\Sigma^* \xi}]
	[\Sigma]^{a}{}_{b}{}^{ii'}{}_{A}
	[\xi^*]_{c}{}^{d}{}_{j'k'} \nonumber\\
&[\textcolor{red}{\lambda_{2}}_{\Sigma^* \chi^*}]
	[\Sigma]^{f}{}_{g}{}^{jm'}{}_{C}
	[\chi]^{el'}{}_{B} \nonumber\\
&[\textcolor{red}{\lambda_{3}}_{\Sigma\Sigma\Delta^*}]
	[\Sigma^\ast]_{a}{}^{b}{}_{ii'}{}^{A} \;
	[\Sigma^\ast]_{f}{}^{g}{}_{jm'}{}^{C} \;
	[\Delta]^{xyn'o'DE} \nonumber\\
&[\textcolor{red}{\lambda_{4}}_{\Delta\chi\xi}]
	[\Delta^*]_{xyn'o'DE} \;
	[\chi^*]_{el'}{}^{B} \;
	[\xi^*]_{d}{}^{c}{}_{p'q'} \nonumber\\
&\epsilon^{j'p'}\,\epsilon^{k'q'} \\
=\quad& -\tfrac{57}{64}\;
		\vevv{d\Phi}{A}\,
		\vevv{u\Phi}{B}\,
		\vevV{C}\,
		\vevV{D}\,
		\vevZc{EF}\;
		\epsilon^{ACE}\,
		\epsilon^{BDF}. \label{eq:two-loop-operator-result}
\end{align}
It is thus indeed a non-vanishing contribution to the explicit breaking of $\vec{q}_{\text{new}}$. The propagators involve PS-scale masses; based on dimensional analysis, the contribution to the mass $m_{\text{new}}$ of the phase associated with $\vec{q}_{\text{new}}$ is thus expected to be roughly
	\begin{align}
		m_{\text{new}}^{4}&\sim \left(\tfrac{1}{16\pi^2}\right)^{2}\,
        (\lambda_{2}\lambda_{3}\lambda_{4}\, m v^2 V)
			\sim \left(\tfrac{1}{16\pi^2}\right)^{2}\,(v^2 V^2),
    \end{align}
and thus
    \begin{align}
		m_{\text{new}}&\sim \frac{1}{4\pi}\sqrt{vV} \sim 25\,\mathrm{TeV}
	\end{align}
for the high-quality scenario.

Lastly, we make the curious observation that an explicit breaking of $\vec{q}_{\rm{new}}$ does not manifest at 1-loop order, so a 2-loop diagram as in \fig{fig:2-loop-diag-newphase} is indeed necessary. This can be seen as follows. In a 1-loop diagram, all the fields in the loop must be of the same Lorentz type, namely gauge bosons, fermions or scalars. A construction with a gauge boson loop cannot provide a non-trivial phase, since gauge couplings preserve all $\U(1)$ charges associated with phase redefinitions of irreps. In a construction with fermions, the renormalizable Yukawa couplings from \eq{eq:Yukawa-sector} do not involve the field $\chi$, while any operator breaking $\vec{q}_{\text{new}}$ would involve such a field due to the 4th component of $\vec{\PHASE}$ in \eq{eq:phase-of-operator-breaking-qnew} required to be non-vanishing (since $b\neq 0$). Finally, a construction with a single scalar loop is not possible, since the use of the crucial invariants $\Sigma^{*2}\Delta^{2}$ and $\Delta\chi^2\xi$ in such a loop requires setting two of their four legs to the vacuum, which is  already sufficient for either of them to vanish. This last observation can be checked either explicitly, cf.~\app{app:procedure-explicit} for VEV directions, or by encountering incompatible $B-L$ charges when attempting to combine irrep parts in the language of the $\SU(3)_{c}\times\SU(2)_L\times\U(1)_{Y}\times\U(1)_{B-L}$ subgroup.

%%%%%%%%%%%%%%%%%%%%%%%%%%%%%%%%%%%%%%%%%%%%%%%%%%%%%%%%%%%%
\subsubsection{PQ quality at tree level --- dominant vacuum contributions
    \label{app:PSmodel-PQ-at-tree-level}
}

The operators that give dominant (tree-level) contributions to the axion potential are listed in \eq{eq:operators-dominant-quality-contribution} of \sect{sec:PQquality}. We provide here further details on how this result is obtained.

To compile a list of candidates, we conceptually follow steps \ref{list:dominant-contributions-begin}--\ref{list:dominant-contributions-end} from \sect{sec:PQquality}: enumerate all invariant candidates up to $d=9$ (procedure of \app{app:procedure-invariant-enumeration} up to condition~\ref{item:layers-2}),
identify those that break PQ symmetry (based on the charges in \Table{tab:PSirrep}), and retain only those that can contribute to the vacuum (based on the vanishing $B-L$ charge upon VEV insertion into every field). Every irrep acquires a single type of VEV (PS- or EW-breaking), whose $B-L$ charge is given in \Table{tab:BLnumbers}.

\begin{table}[htb]
    \centering
    \begin{tabular}{lrrrrr}
         \toprule
        irrep& $\Phi$ & $\Sigma$ & $\Delta$ & $\chi$ & $\xi$ \\
        \midrule
        $B-L$ of VEV & $0$ & $0$ & $-2$ & $-1$ & $0$\\ 
        \bottomrule
    \end{tabular}
    \caption{
        The $B-L$ charges of the VEVs in every irrep. 
        \label{tab:BLnumbers}
    }
\end{table}

The process just described results in $127$ operator candidates up to $d\leq 9$ that possibly contribute to the vacuum, with the first ones arising at $d=6$. They are listed in \Table{tab:operators-for-axion-potential} based on dimension $d$, and ordered by the associated scale of their vacuum contribution from largest to smallest given the VEV hierarchy $v\ll V$. 

Since $\Phi$ and $\Sigma$ in invariants are often interchangeable, we can compactly write an associated set of operators by using an index $k$, which takes all integer values for which the powers of the fields are non-negative. In cases where the same field is written in two separate factors, we take $k$ for which both are positive. For example, in $\Phi^{4-k}\Sigma^{k}\Sigma^2$ the index should be interpreted to take values $k\in\{0,1,2,3,4\}$, while $k=-1$ is forbidden due to the first $\Sigma$-factor. The number of values the index $k$ takes is denoted in \Table{tab:operators-for-axion-potential} by $\#$.

Note that \Table{tab:operators-for-axion-potential} merely compiles a list of candidates: it is complete in the sense that no actual contribution is missing from the list, but some listed entries may in fact vanish. The reason is that we tested only the necessary condition that an all-VEV term has a vanishing $B-L$ charge, while the contribution may ultimately still vanish, as we shall see below. Ultimately the presence of a contribution is confirmed only by an explicit index contraction, whereby multiple\footnote{
    The number of independent invariants with fixed powers of fields can be larger than $10^5$ for some $d=9$ cases, as is revealed by using the counting method of condition~\ref{item:layers-3} in \app{app:procedure-invariant-enumeration}. The large number of contractions should not be surprising, since every irrep carries many indices, and the number of indices (and thus contractions) increases exponentially with operator dimension.
    Due to computational intensity, we did not check the invariant count for all entries listed in \Table{tab:operators-for-axion-potential}.
} 
independent contractions of any given invariant are usually possible (with fixed powers $n_{i}$ of the involved fields, cg.~\eq{eq:invariant-general-form}. 

\def\SKIP{4pt}
\begin{table}[t!]
    \begin{center}
\begin{minipage}[t]{0.4\linewidth}
    \vspace{0pt}
%%% dimension 6
    \begin{tabular}{lll}
        \toprule
        $\mathcal{O}$ ($d=6$)& $\langle\mathcal{O}\rangle$&$\#$\\
        \midrule
        $\Phi^{6}$ & $v^6\NO$&$1$\\[\SKIP]
        $\Phi^{4-k}\Sigma^{k}\Sigma^2$ & $v^6\YES$&$5$\\
       \bottomrule
%    \end{tabular}
%%% dimension 7
%    \vskip 0.2cm
%    \begin{tabular}{lll}
\addlinespace[2pt]
        \toprule
        $\mathcal{O}$ ($d=7$)& $\langle\mathcal{O}\rangle$&$\#$\\
        \midrule
        $\Phi^{4-k}\Sigma^{k}\Delta\chi^{*2}$ & $v^4 V^{3}\YES$&$5$\\[\SKIP]
        $\Phi^{5-k}\Sigma^{k}\Sigma\xi$ & $v^6 V$&$6$\\
        \bottomrule
%    \end{tabular}
%%% dimension 8
%   \vskip 0.2cm
%    \begin{tabular}{lll}
\addlinespace[2pt]
\toprule
        $\mathcal{O}$ ($d=8$)& $\langle\mathcal{O}\rangle$&$\#$\\
        \midrule
        $\Phi^{2-k}\Sigma^{k}\Delta^2\chi^{*4}$&$v^2 V^6\YES$&$3$\\[\SKIP]
        $\Phi^{4-k}\Sigma^{k}\Delta\chi^{*2}\xi$&$v^4 V^4$&$5$\\[\SKIP]
        $\Phi^{6-k}\Sigma^{k}\Delta\Delta^*$&$v^6 V^2$&$7$\\[\SKIP]
        $\Phi^{6-k}\Sigma^{k}\chi\chi^*$&$v^6 V^2$&$7$\\[\SKIP]
        $\Phi^{6-k}\Sigma^{k}\xi^2$&$v^6 V^2$&$7$\\[\SKIP]
        $\Phi^{4-k}\Sigma^{k}\Phi\Phi^* \Sigma^{2}$&$v^8$&$5$\\[\SKIP]
        $\Phi^{6-k}\Sigma^{k}\Sigma\Sigma^*$&$v^8$&$7$\\[\SKIP]
        $\Phi^{7}\Phi^*$&$v^8$&$1$\\[\SKIP]
        $\Phi\Sigma^{\ast 7}$&$v^{8}$&$1$\\
        \bottomrule
    \end{tabular}
\end{minipage}
\hspace{0.5cm}
\begin{minipage}[t]{0.4\linewidth}
    \vspace{0pt}
%%% dimension 9
    \begin{tabular}{lll}
        \toprule
        $\mathcal{O}$ ($d=9$)& $\langle\mathcal{O}\rangle$&$\#$\\
        \midrule
        $\Delta^3\chi^{*6}$&$V^9\NO$&$1$\\[\SKIP]
        $\Phi^{2-k}\Sigma^{k}\Delta^2\chi^{*4}\xi$&$v^2 V^7$&$3$\\[\SKIP]
        $\Phi^{4-k}\Sigma^{k}\Delta^2 \Delta^\ast \chi^{\ast 2}$&$v^{4}V^{5}$&$5$\\[\SKIP]
        $\Phi^{4-k}\Sigma^{k}\Delta \chi^{\ast 2}\xi^{2}$&$v^{4}V^{5}$&$5$\\[\SKIP]
        $\Phi^{4-k}\Sigma^{k}\Delta \chi \chi^{\ast 3}$&$v^{4}V^{5}$&$5$\\[\SKIP]
        $\Phi^{6-k}\Sigma^{k}\xi^{3}$&$v^{6}V^{3}$&$7$\\[\SKIP]
        $\Phi^{6-k}\Sigma^{k}\Delta\Delta^\ast\xi$&$v^{6}V^{3}$&$7$\\[\SKIP]
        $\Phi^{6-k}\Sigma^{k}\chi\chi^\ast\xi$&$v^{6}V^{3}$&$7$\\[\SKIP]
        $\Phi^{4-k}\Sigma^{k}\Phi\Phi^{\ast}\Delta\chi^{\ast 2}$&$v^{6}V^{3}$&$5$\\[\SKIP]
        $\Phi^{5-k}\Sigma^{k}\Sigma^\ast\Delta\chi^{\ast 2}$&$v^{6}V^{3}$&$6$\\[\SKIP]
        $\Phi^{\ast}\Sigma^{5}\Delta\chi^{\ast 2}$&$v^{6}V^{3}$&$1$\\[\SKIP]
        $\Phi^{5-k}\Sigma^{k}\Phi\Phi^{\ast}\Sigma\xi$&$v^{8}V$&$6$\\[\SKIP]
        $\Phi^{7-k}\Sigma^{k}\Sigma^\ast\xi$&$v^{8}V$&$8$\\[\SKIP]
        $\Phi\Sigma^{\ast 7}\xi$&$v^{8}V$&$1$\\
        \bottomrule
    \end{tabular}
\end{minipage}
%%%%%%%%%%%%%%%%%%%%%%%%%%%%%
    \end{center}
    \caption{Non-renormalizable operators $\mathcal{O}$ of dimensions $d\leq 9$ that may contribute to the axion potential: they violate PQ (their PQ-charge is $\pm12$) and have a vanishing $B-L$ charge upon VEV insertion. The vacuum contribution is denoted by $\langle\mathcal{O}\rangle$, while $\#$ specifies the number of operators, given that the index $k$ runs over all integer possibilities, such that all exponents are non-negative. For the dominant contributions, we indicate with an arrow whether the vacuum contribution vanishes or not
    based on explicit calculation. 
        \label{tab:operators-for-axion-potential}
    }
    \end{table}

A search through all contractions is not easily amenable to automation, so we  
identify the dominant candidates in \Table{tab:operators-for-axion-potential} and check them explicitly with the tensor methods from \app{app:procedure-explicit}.  The vacuum contributions necessarily have the form 
$v^{m} V^{n} \LUV^{4-m-n}$ for even $m$, see \sect{sec:PQquality} or \app{app:procedure-invariant-enumeration}, so we need to find the contribution with the smallest $n$ for each of $m\in\{0,2,4,6\}$. We are satisfied with one non-vanishing contraction, or alternatively we need to show all contractions of an invariant type vanish. The analysis is as follows:
\begin{enumerate}
    \item \label{item:V9} For $m=0$ we have $n=9$: the dominant candidate is $V^{9}$, which can come only from the operator $\Delta^{3}\chi^{\ast 6}$. This contribution in fact vanishes on the vacuum, essentially due to anti-symmetrization in flavor contraction while only having one copy of $\Delta$ and $\chi$ available. The argument applies to all $100$ independent contractions of this type of invariant, where the counting was performed using condition~\ref{item:layers-3} from \app{app:procedure-invariant-enumeration}. To demonstrate this somewhat remarkable result, notice that there is only one SM-singlet in the Pati-Salam part of irreps $\Delta$ and $\chi$, i.e.~there is one flavor $\6$-plet VEV in $\Delta$ and one flavor $\3$-plet VEV in $\chi^{\ast}$. Using \texttt{LiE} for only the $\SU(3)_{f_R}$ factor, the number of singlets $\1$ in the tensor product $\6^{\otimes_{s} 3}\otimes \3^{\otimes_{s} 6}$ is zero, where an exponent $\otimes_{s} n$ denotes the $n$-th symmetrized tensor power. This argument applies regardless of the choice of index contraction in $\Delta^{3}\chi^{\ast 6}$.
    \par
    It is instructive to demonstrate explicitly how this argument works. Given the VEV structure from Eqs.~\eqref{eq:vevs-delta} and \eqref{eq:vevs-chi}, the $\SU(4)_{\rm PS}$ indices must take the value $4$ and $\SU(2)_R$ indices must take the value $1$, and thus a contribution must necessarily have the VEV form
        \begin{align}
            \colorPQ{Z}^{AB}\;
            \colorPQ{Z}^{CD}\;
            \colorPQ{Z}^{EF}\;
            \colorPQ{V}^{\ast G}\;
            \colorPQ{V}^{\ast H}\;
            \colorPQ{V}^{\ast I}\;
            \colorPQ{V}^{\ast J}\;
            \colorPQ{V}^{\ast K}\;
            \colorPQ{V}^{\ast L},
            \label{eq:vevs-V9}
        \end{align}
        where the $12$ flavor indices need to be contracted with an $\SU(3)$-invariant tensor, for example four Levi-Civita tensors.
        Such a contraction will always contain sufficient anti-symmetry
        in index exchange contraction so that \eq{eq:vevs-V9} vanishes. 
    \item For $m=2$, we have $n=6$: the contribution $v^{2}V^{6}$ can come only from operators of the type $\Phi^{2-k}\Sigma^{k}\Delta^{2}\chi^{\ast 4}$.
    Note that an anti-symmetric contraction of $\chi^{\ast 2}$ in flavor vanishes on the vacuum, i.e.,
    \begin{align}
        (\chi^{\ast})_{ii'}{}^{A}\;
        (\chi^{\ast})_{jj'}{}^{B}\;
        \epsilon_{ABC}
        &=0, \label{eq:vacuum-chichi}
    \end{align}
    hence it must be avoided with an elaborate contraction in $\SU(3)_{f_R}$ indices. This is indeed possible, since there is one singlet $\1$ in the tensor product $\overline{\3}^{\otimes 2}\otimes \6^{\otimes_{s} 2}\otimes\3^{\otimes_{s} 4}$, with the exponent $\otimes 2$ denoting the (ordinary) 2nd tensor power. Focusing on $k=0$, we find the following example of a non-vanishing contraction:
        \begin{align}
            \begin{split}
            &\left\langle
                \Phi^{ii'}{}_{A}\;
                \Phi^{ij'}{}_{B}\;
                \Delta^{abk'l'CD}\;
                \Delta^{cdm'n'EF}\;
                (\chi^\ast)_{ak'}{}^{G}\;
                (\chi^\ast)_{bl'}{}^{H}\;
                (\chi^\ast)_{cm'}{}^{I}\;
                (\chi^\ast)_{dn'}{}^{J}\;
                \epsilon_{ij}\,
                \epsilon_{i'j'}
            \right.\\
            &\quad \left.
                \epsilon_{KCG}\,
                \epsilon_{LDH}\,
                \epsilon_{MEI}\,
                \epsilon_{NFJ}\,
                \epsilon^{AKN}\,
                \epsilon^{BLM}\,
            \right\rangle \\
            &=
            2\,
            \left(
                \colorPQ{V}^{\ast A}\; 
                \vevv{u\Phi}{A}
            \right)
            \left(
                \colorPQ{V}^{\ast B}\; 
                \vevv{d\Phi}{B}
            \right)
            \left(
                \colorPQ{Z}^{CD}\;
                \colorPQ{Z}^{EF}\;
                \colorPQ{V}^{\ast G}\;
                \colorPQ{V}^{\ast H}\;
                \epsilon_{CEG}\,
                \epsilon_{DFH}
            \right).
            \end{split}
        \end{align} 
    \item For $m=4$ we get $n=3$: the contribution $v^{4}V^{3}$ comes from operators of the type $\Phi^{4-k} \Sigma^{k}  \Delta \chi^{\ast 2}$,
        which for $k=0$ has a possible contraction
        \begin{align}
            \begin{split}
                &\left\langle
                    \Phi^{ii'}{}_{A}\;
                    \Phi^{jj'}{}_{B}\;
                    \Phi^{kk'}{}_{C}\;
                    \Phi^{ll'}{}_{D}\;
                    \Delta^{abm'n'AB}\;
                    (\chi^{\ast})_{am'}{}^{C}\;
                    (\chi^{\ast})_{bn'}{}^{D}\;
                    \epsilon_{ij}\,
                    \epsilon_{kl}\,
                    \epsilon_{i'j'}\,
                    \epsilon_{k'l'}
                \right\rangle \\
                &
                =
                4
                \left(
                    \colorPQ{Z}^{AB}\;
                    \vevv{u\Phi}{A}\;
                    \vevv{d\Phi}{B}
                \right)
                \left(
                    \colorPQ{V}^{\ast C}\; 
                    \vevv{u\Phi}{C}
                \right)
                \left(
                    \colorPQ{V}^{\ast D}\; 
                    \vevv{d\Phi}{D}
                \right).
            \end{split}
        \end{align}       
    \item For $m=6$ we get $n=0$: the contribution $v^{6}$ comes from $\Phi^{6-k} \Sigma^{k}$, which for $k=2$ gives a possible contraction
        \begin{align}
            \begin{split}
                &\left\langle
                    \Sigma^{a}{}_{b}{}^{ii'}{}_{A}\;
                    \Phi^{jj'}{}_{B}\;
                    \Phi^{kk'}{}_{C}\;
                    \Sigma^{b}{}_{a}{}^{ll'}{}_{D}\;
                    \Phi^{mm'}{}_{E}\;
                    \Phi^{nn'}{}_{F}\;
                    \epsilon^{ABC}\,
                    \epsilon^{DEF}\,
                    \epsilon_{il}\,
                    \epsilon_{i'l'}\,
                    \epsilon_{jm}\,
                    \epsilon_{j'm'}\,
                    \epsilon_{kn}\,
                    \epsilon_{k'n'}\,
                \right\rangle \\
                &
                = -4 \;
                \left(
                    \vevv{u\Sigma}{A}\;
                    \vevv{u\Phi}{B}\;
                    \vevv{d\Phi}{C}\;
                    \epsilon^{ABC}
                    \right)
                    \left(
                    \vevv{d\Sigma}{D}\;
                    \vevv{u\Phi}{E}\;
                    \vevv{d\Phi}{F}\;
                    \epsilon^{DEF}
                \right).\\
    \end{split}
\end{align}
    It is worth mentioning that for $k=0$ the vacuum contribution must identically vanish: the EW VEVs of $\Phi$ come in two flavor $\overline{\3}$-plets, so a vacuum contribution would be constructed from a linear combination of $\overline{\3}^{\otimes_s l}\otimes\overline{\3}^{\otimes_{s} (6-l)}$ for integer $l$ between $0$ and $6$, none of which contain a $\1$. It is instructive to extend such an analysis to general $k$ as follows: $\Sigma$ also contains two flavor $\overline{\3}$-plet EW VEVs, so an all-VEV term would come from 
        \begin{align}
            \overline{3}^{\otimes_{s} l_{1}} \otimes
            \overline{3}^{\otimes_{s} l_{2}} \otimes
            \overline{3}^{\otimes_{s} l_{3}} \otimes
            \overline{3}^{\otimes_{s} l_{4}},
        \end{align}
     where $\sum_{i=1}^{4} l_i=6$. Up to permutations in $(l_{1},l_{2},l_{3},l_{4})$, only $(2,2,2,0)$ and $(2,2,1,1)$ contain a $\1$. Two of the $l_{i}$ must refer to VEVs in $\Phi$ and their sum must be $k$. Therefore we conclude that $\Phi^{6-k} \Sigma^{k}$ can yield  
     vacuum contributions only for $k=2,3,4$, contraction of PS indices notwithstanding. Note that $k=1$ does not even give an invariant, since one $\Sigma$ alone cannot form a singlet in the $\SU(4)_{\rm PS}$ factor.
\end{enumerate}
Given that the case $m=6$ already has no involvement of the high-scale VEV $V$, i.e.~it gives $n=0$, all contributions with $m>6$ will necessarily be subdominant and we can terminate the search. We indicate the explicit results on (non)-vanishing vacuum contributions in \Table{tab:operators-for-axion-potential} with an arrow.

Since the contribution $V^{9}$ is absent, one can wonder what is the smallest $n$ for which $V^{n}$ is present. It turns out it is $V^{11}$, as argued below:
\begin{itemize}
\item 
    The presence of only the Pati-Salam-breaking VEVs indicates an operator of the type $\Delta^{n_{5}}\Delta^{\ast n_6}\chi^{n_{7}}\chi^{\ast n_8}\xi^{n_9}$, i.e., $n_1=n_2=n_3=n_4=0$ in the ansatz of \eq{eq:invariant-general-form}. The constraints of Eq.~\eqref{eq:discrete-condition-Z4} and \eqref{eq:discrete-condition-Z3} then imply an operator of the type 
        \begin{align}
            (\Delta^{3} \chi^{\ast 6})^{n_{1}'} (\Delta\Delta^{\ast})^{n_{2}'} (\chi\chi^\ast)^{n_{3}'}\xi^{n_{4}'} \label{eq:vacuum-Vn-1}
        \end{align}
    (or its conjugate), where the fields were joined into convenient combinations and \hbox{$\{n_{1}',n_{2}',n_{3}',n_{4}'\}$} are non-negative integers. 
\item Analogously to the analysis for $V^{9}$ in item~\ref{item:V9}, each field acquires a VEV in a single Pati-Salam direction, and the contribution to the vacuum depends on the successful contraction of $\SU(3)_{f_R}$ indices. A contribution to the vacuum for an operator in \eq{eq:vacuum-Vn-1} can thus occur only when 
    \begin{align}
        \1&\subset 
        \6^{\otimes_{s} 3n_{1}'+n_{2}'}\otimes 
        \overline{\6}^{\otimes_{s} n_{2}'}\otimes 
        \overline{\3}^{\otimes_{s} n_{3}'} \otimes
        \3^{\otimes_{s} 6n_{1}'+n_{3}'}\otimes
        \1^{\otimes_{s} n_{4}}.
    \end{align}
    The minimal case will have $n_{4}'=0$, and the operator violates PQ if $n_{1}'\geq 1$. Using \texttt{LiE}, the minimal case when a singlet is present is $(n_{1}',n_{2}',n_{3}')=(1,1,0)$ for a contribution $V^{11}$, while no singlets are present for $(1,0,0)$ or $(1,0,1)$ (contributions $V^{9}$ and $V^{11}$, respectively).
\item We can confirm the presence of the $V^{11}$ contribution by explicit index contraction in the $\Delta^4\Delta^\ast\chi^{\ast 6}$ invariant:
    \begin{align}
    \begin{split}
        &
        \left\langle
        \Delta^{abi'j'AB}\;
        \Delta^{cdk'l'CD}\;
        \Delta^{efm'n'EF}\;
        \Delta^{gho'p'GH}\;
        (\Delta^\ast)_{gho'p'IJ}
        \right.\\
        &\qquad
        \left.
        (\chi^{\ast})_{ai'}{}^{K}\;
        (\chi^{\ast})_{bj'}{}^{L}\;
        (\chi^{\ast})_{ck'}{}^{M}\;
        (\chi^{\ast})_{dl'}{}^{N}\;
        (\chi^{\ast})_{em'}{}^{O}\;
        (\chi^{\ast})_{fn'}{}^{P}\;
        \right.\\
        &
        \qquad
        \left.
        \epsilon_{AKQ}\,
        \epsilon_{BLR}\,
        \epsilon_{CMS}\,
        \epsilon_{DNT}\,
        \epsilon_{EGO}\,
        \epsilon_{FHP}\,
        \epsilon^{QSI}\,
        \epsilon^{RTJ}
        \right\rangle \\
        &=
        \left(
        \colorPQ{Z}^{\ast}_{IJ}
        \colorPQ{V}^{\ast I}
        \colorPQ{V}^{\ast J}
        \right)
        \left(
        \colorPQ{Z}^{AB}
        \colorPQ{Z}^{CD}
        \colorPQ{V}^{\ast K}
        \colorPQ{V}^{\ast L}\,
        \epsilon_{ACK}\,
        \epsilon_{BDL}
        \right)^{2}.
    \end{split}
    \end{align}
\end{itemize}

%%%%%%%%%%%%%%%%%%%%%%%%%%%%%%
\subsubsection{PQ quality at the quantum level}
\label{app:PSmodel-PQ-at-loop-level}

The quality of Peccei-Quinn symmetry was discussed in \sect{sec:PQquality}
at tree-level of the EFT. Quantum corrections, however, can potentially spoil this neat picture by enhancing the contribution from PQ-violating non-renormalizable operators, a phenomenon which is usually not addressed in the literature. We discuss in this appendix the general circumstances under which this quantum degradation of PQ symmetry can occur, and outline the enormous challenge of explicitly analyzing this effect.

First, we elaborate on how a ``quantum enhancement'' can generally arise in an EFT. The calculation in EFTs is organized in powers of a UV cutoff scale $\LUV$, see e.g.~\cite{Manohar:2018aog}. In $4$ space-time dimensions and for an operator of mass dimension $d$, tree-level power counting requires its coefficient to be suppressed by the factor $(\LUV)^{4-d}$. If this same operator can be constructed at some loop level from a set of operators with dimensions $d_{i}$, then by EFT power counting the coefficient of the correction is $(\LUV)^{4-d'}$,
where
\begin{align}
    4-d' = \sum_{i} (4-d_i).
\end{align}
It can happen that $d'<d$, i.e.~the suppression from the loop 
is smaller than the ``naive'' tree-level one, leading to a quantum enhancement.

\begin{figure}[t!]
    \centering
    \includegraphics[width=0.35\linewidth]{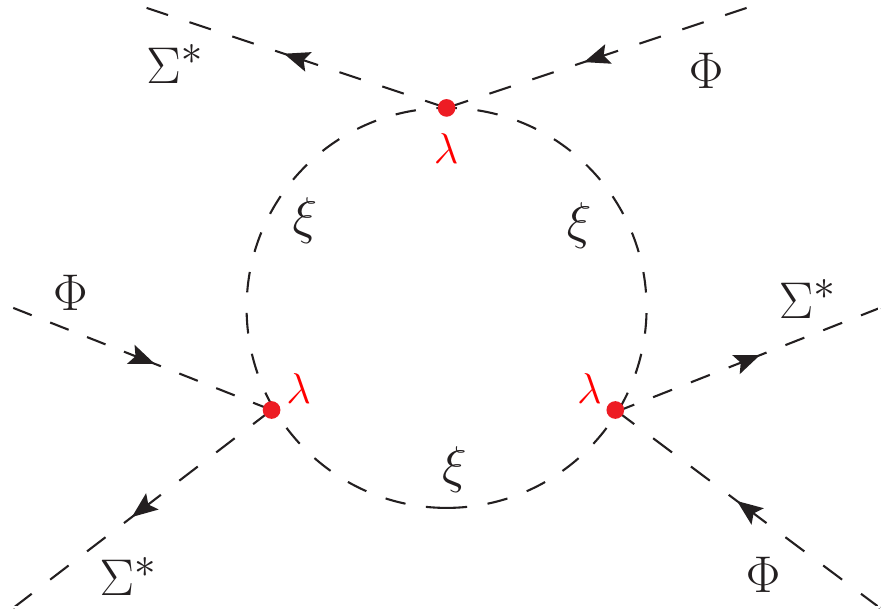}
    \hspace{0.1\linewidth}
    \includegraphics[width=0.35\linewidth]{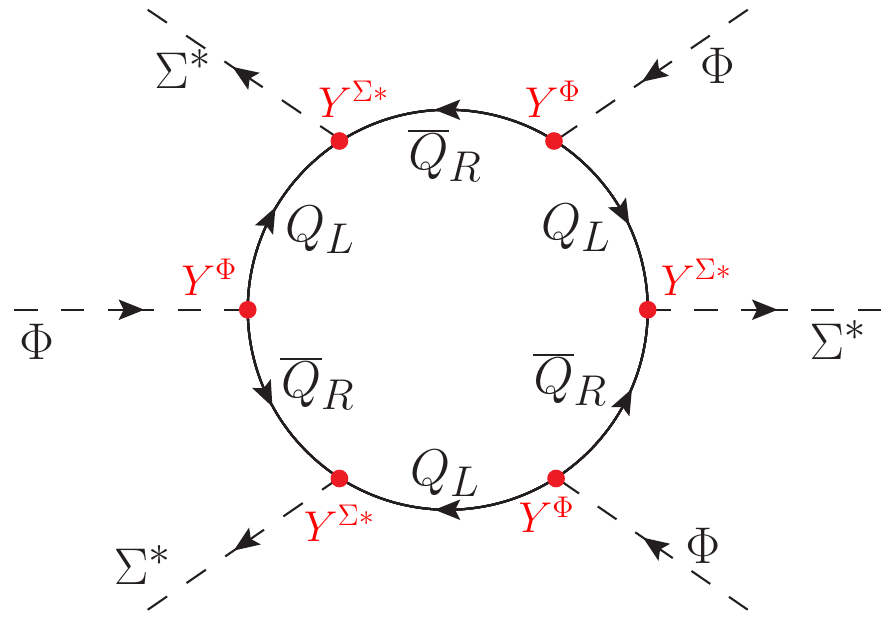}
    \caption{
        Examples of 1-PI one-loop diagrams that generate the quantum correction to the operator $\Phi^{3}\Sigma^{3 *}$ and induce a parametric enhancement to its vacuum contribution. Couplings are shown in \textcolor{red}{red} for visual clarity.
        \label{fig:loop-enhancement}
    }
\end{figure}

To give a concrete example in our model, consider the non-renormalizable operator $\mathcal{O}=\Phi^{3}\Sigma^{3 *}$. In the EFT at tree level, it generates a Lagrangian term
\begin{align}
    \mathcal{L}_{\mathcal{O}}&=\frac{C}{\LUV^2}\,\Phi^3 \Sigma^{3*} + \text{h.c.} \, ,
\end{align}
where we refrained from explicitly writing gauge indices. The same operator $\mathcal{O}$ may also be generated from $1$-PI diagrams. We show two one-loop examples in \fig{fig:loop-enhancement}: 
\begin{enumerate}[label=(\textit{\roman*}),itemsep=0cm,leftmargin=0.7cm]
    \item \label{item:diagram-scalar} 
        The left diagram in the figure is constructed from repeated use of the renormalizable scalar interaction $\Phi\Sigma^*\xi^2$ from \eq{eq:VC}, with the associated dimensionless coupling denoted by $\lambda$.
    \item \label{item:diagram-fermion} 
        The right diagram in the figure is constructed from renormalizable Yukawa interactions of Eq.~\eqref{eq:Yukawa-sector}. It makes use of couplings $\parY{\Phi}{}$ and $\parY{\Sigma}{}$, where $Q_{L}$-family indices have been suppressed.
\end{enumerate}

Since all the couplings in \fig{fig:loop-enhancement} are renormalizable (and dimensionless), the diagrams are not associated to the EFT cutoff scale $\LUV$ in any way; instead, the only scale involved other than the VEVs of $\Phi$ and $\Sigma$ are their masses of scale $V$ (the exception being the specific combination of entries corresponding to the SM Higgs).
On dimensional grounds, the contributions to the vacuum from the operator $\mathcal{O}$ are schematically the following (ignoring any symmetry factors from diagrams): 
\begin{align}
    \langle\mathcal{L}_{\mathcal{O}}+\delta\mathcal{L}_{\mathcal{O}}\rangle &\sim
        C\,\frac{v^{6}}{\LUV^2} + \frac{\lambda^{3}}{16\pi^{2}}\,\frac{v^6}{V^{2}}+ \frac{(\parY{\Phi}{}\parY{\Sigma}{} {}^{*})^{3}}{16\pi^{2}}\,\frac{v^6}{V^{2}}+\text{h.c.}+\ldots \, ,
\end{align}
where the 2nd and 3rd terms correspond to the quantum corrections, with a parametric enhancement of $(\LUV/V)^2$ alongside the $(16\pi^{2})^{-1}$ loop suppression. In other words, the tree-level coefficient comes with the power $\LUV^{-2}$, while the 1-loop correction comes with the power $\LUV^{0}$. The ellipsis at the end of the equation signifies loop corrections in $\mathcal{L}_{\mathcal{O}}$ from other possible loop diagrams.

In general, a shift in power counting $\LUV^{4-d}\mapsto \LUV^{4-d'}$ from an $l$-loop contribution induces an enhancement by a factor $(16\pi^{2})^{-l} (\LUV/V)^{d-d'}$ (ignoring the dimensionless couplings), where $V$ represents the mass scale of the involved fields in the loop.

We now investigate the circumstances under which a PQ-breaking operator can undergo a quantum enhancement that is potentially dangerous for PQ quality.
The following straightforward observations can be made: 
\begin{itemize}[itemsep=0cm,leftmargin=0.7cm]
    \item 
        The operators that need to be analyzed for possible enhancements
        are those that can contribute to the axion potential, i.e.~operators in the scalar potential that are PQ-violating and which contribute to the vacuum. We refer to them in the subsequent discussion as \EMPH{PQ-dangerous}. The list of such operators in our model for $d\leq 9$ is given in \Table{tab:operators-for-axion-potential}.
        Note that the example operator $\mathcal{O}$ from Figure~\ref{fig:loop-enhancement} is not in this category (it preserves PQ). 
    \item 
        A loop contribution to a \EMPH{PQ-dangerous} operator will include at least one PQ-violating vertex. This coupling can be of any type and does not necessarily have to come from a \EMPH{PQ-dangerous} operator itself --- it can also involve fermions, or it can be scalar-only but not contribute to the vacuum (and hence for our model not listed in \Table{tab:operators-for-axion-potential}), since some/all of their fields can be carried on internal legs of the loop diagram. We refer to such a PQ-violating building block operator as a \EMPH{PQ-source}.
    \item 
        Suppose $d_{0}$ is the smallest dimension of all \EMPH{PQ-source} operators. For a \EMPH{PQ-dangerous} operator of dimension $d$, the 
        cutoff scale power counting can be enhanced at most as $\LUV^{4-d}\mapsto \LUV^{4-d_{0}}$, corresponding to a case where all other couplings in the loop diagram are renormalizable (and hence not having any associated powers of $\LUV$).
        \par
        For a theory with a single scale $V$, the parametric enhancement can thus be at most $(\LUV/V)^{d-d_{0}}$. Thus only \EMPH{PQ-dangerous} operators of dimension $d>d_{0}$ need to be analyzed for true parametric enhancement; quantum enhancements dangerous to PQ quality are thus possible in single scale models only when the lowest dimensional \EMPH{PQ-source} operator is of strictly lower dimension than any \EMPH{PQ-dangerous} operator. For a given \EMPH{PQ-dangerous} operator of dimensions $d$, one needs to consider only loop diagrams constructed from couplings (any couplings, not just those in the scalar potential) of dimension smaller than $d$ and up to loop-order $l$ roughly satisfying
        \begin{align}
            (16\pi^{2})^{l} < (\LUV/V)^{d-d_{0}},
        \end{align}
        lest loop suppression overwhelms any possible enhancement. 
        \par
        The analysis is more complicated if multiple scales are present in the model, such as in our case. In fact, realistic models will always involve at least two scales, namely the EW symmetry breaking scale $v$ and a PQ-breaking scale $V$. The two scales cannot coincide --- the Weinberg-Wilczek axion is experimentally excluded. The multiscale case is analyzed on the example of our model after we finish with the general remarks.
    \item 
        Diagrams need to be considered at the level of contracting all gauge indices and putting the outer legs onto the vacuum, similar to our tree-level analysis in \app{app:PSmodel-PQ-at-tree-level}.
        A diagrammatic analysis that assigns only Pati-Salam irreps to each line, e.g.~as done in \fig{fig:loop-enhancement}, can be useful to identify candidate contribution for enhancement, but index contractions can still conspire for the contribution to vanish. 
    \item 
        At one loop, diagram topologies with a scalar or fermion loop can contribute, cf.~the examples of \fig{fig:loop-enhancement}, while a gauge loop does not violate PQ. The scalar/fermion loop can have an arbitrary number of vertices lying on it. 
    \item 
        An enhancement can be small enough in magnitude that it does not spoil or even degrade PQ quality. Although there is no general argument for the upper limit on $d$ of \EMPH{PQ-dangerous} operators to analyze, those of larger $d$ come with more powers of $\LUV$-suppression, and consequently the enhancement must be correspondingly larger to render the contribution PQ-spoiling. Heuristically we thus expect dangerous enhancements to be more likely for lower-dimensional operators.
\end{itemize}

\noindent
We extend the principles discussed above to our Pati-Salam model, which involves multiple scales $v$ and $V$:
\begin{itemize}[itemsep=0cm,leftmargin=0.7cm]
    \item 
        The lowest dimensional PQ-source operators are of dimension $d_{0}=6$. These include the \EMPH{PQ-dangerous} operators of dimension $6$ in \Table{tab:operators-for-axion-potential}. There are no dimension $5$ \EMPH{PQ-sources}; in the fermion sector, this was achieved by 
        assigning a suitable PQ charge to anomalons $\Psi_R$, cf.~\Table{tab:PSirrep}. 
    \item 
        With $d_{0}=6$, any \EMPH{PQ-dangerous} operator of dimension $d$ can be enhanced at most up to $\LUV^{-2}$ in cutoff power counting. The enhancement factor can now come, however, in powers of $(\Lambda/V)$ and $(\Lambda/v)$, depending on the fields running in the loops. Furthermore, the value at which an enhancement factor spoils quality no longer depends only on $d$, but on the powers of $v$ and $V$ with which the \EMPH{PQ-dangerous} operator contributes to the vacuum --- information provided by the right panel of \fig{fig:plotfa}. 
        \par
        As an example, consider $d=11$ \EMPH{PQ-dangerous} operators contributing to the vacuum as $V^{11}/\LUV^{7}$. Based on \fig{fig:plotfa}, an enhancement of $(\LUV/V)^{1}$ is still admissible, while $(\LUV/V)^{2}$ degrades quality. On the other hand, a $d=11$ operator contributing as $v^{2}V^{9}/\LUV^{7}$ can still suffer an enhancement $(\LUV/V)^{3}$ to be admissible.
        \par
        One can estimate for any given \EMPH{PQ-dangerous} operator an upper limit on the number of loops $l$ that need to be considered in constructing its diagrams in order to determine possible PQ spoilage. Suppose the \EMPH{PQ-dangerous} operator is of dimension $d$ and its contribution to the vacuum is of the form $v^{n}V^{d-n}/\LUV^{(4-d)}$. 
        For the worst case scenario of maximal enhancement, i.e.~$(\LUV/V)^{d-6}$, we can compare the loop contribution to the right-hand side of \eq{eq:PQquality}:
        \begin{align}
            \left(\frac{1}{16\pi^2}\right)^{l} \frac{v^{n}V^{6-n}}{\LUV^{2}}
            \lesssim 10^{-10}\,\chi_{\rm QCD}^{4}, \label{eq:loop-spoilage}
        \end{align}
        We assumed in this estimate there is always a heavy field in the loop, i.e.~$\LUV/V$ rather than $\LUV/v$ enhancements, no symmetry factors in diagrams, and a tree-level EFT expansion with natural dimensionless coefficients in front of operators. Under these assumptions the dependence on $d$ has dropped out. Taking $V=10^{9}\,\mathrm{GeV}$ in Eq.~\eqref{eq:loop-spoilage} consistent with the high PQ quality scenario, the operators with powers $v^0$, $v^{2}$ and $v^4$ in the EW VEV need to be checked for enhancements only up to and including $13$, $7$, and $1$ loop(s), respectively, while operators yielding an EW VEV power $v^{n}$ for $n\geq 6$ will always be safe. Barring any other advantageous structural features of the model, such an analysis is clearly an enormously difficult task.
\end{itemize}

Analogous considerations of quantum enhancement apply to any model of PQ quality.
We conclude by stating that while degradation of PQ quality from quantum corrections in our model is in principle possible, we are not aware of any specific dangerous contribution to this effect at the time of writing.

%%%%%%%%%%%%%%%%%%%%%%%%%%%%%%
%%%%%%%%%%%%%%%%%%%%%%%%%%%%%%
%%%%%%%%%%%%%%%%%%%%%%%%%%%%%%
\subsection{Computational framework for the Yukawa sector} 
\label{app:PSmodel-Yukawa-computational-details}

This section presents the technical specifics of the Yukawa sector contributions, as discussed in \sect{sec:SMflavour}, following the indexing conventions outlined in \app{app:procedure-explicit}. The content is systematically structured by topic for coherence:

\begin{enumerate}[label=(\arabic*),leftmargin=0.5cm,ref=\arabic*]
    \item \textit{Renormalizable operators:} \\[8pt]
        We explicitly express with indices some of the schematically written invariants from the main text, starting with the invariants from the renormalizable Yukawa sector in \eq{eq:Yukawa-sector}. The invariants are obtained with contractions
            \begin{align}
                \epsilon_{i'j'}\;(\bar{Q}_L)^{I}{}_{ai}\,(Q_{R})^{ai'A}\,(\Phi^{\alpha})^{ij'}{}_{A} \, , \label{eq:contraction-Y1}\\
                \epsilon_{i'j'}\;(\bar{Q}_L)^{I}{}_{ai}\,(Q_{R})^{bi'A}\,(\Sigma^{\alpha})^{a}{}_{b}{}^{ij'}{}_{A} \, , \label{eq:contraction-Y2}\\
                (Q_{R})^{ai'A} \, (Q_{R})^{bj'B} \,(\Delta^\ast)_{abi'j'AB} \, ,
                \label{eq:contraction-Y3}
            \end{align}
        where $I$ and $\alpha$ are the multiplicity indices and remain uncontracted. We assume one copy for $\Phi$ and two copies $\Sigma,\Sigma'$. We denote by $\epsilon$ the anti-symmetric Levi-Civita tensor (of any rank). The results of  \eqs{eq:mass-mu}{eq:mass-me} and \eqref{eq:Nblock-LR} follow from the above contractions, given the VEV definitions of 
        \app{app:procedure-explicit}.
    \item \textit{Leading non-renormalizable operators:} \\[8pt]
        The operators for leading non-renormalizable contributions have been determined by listing the invariants using the methods of \app{app:procedure-invariant-enumeration} and sorting by size of contribution. Those with leading contributions (unique in the entire range of $V$ given by \eq{eq:vev-scaling}) are listed with non-tilde coefficients in \Table{tab:Yukawa-nonrenormalizable}. Suppressing gauge indices, we write their Lagrangian terms as 
        \begin{align}
                -\mathcal{L}_{\Psi}
                &\supset 
                \sum_{I=1}^{3}\sum_{K=1}^{8} \frac{\parA_{IK}}{\LUV}\;(\overline{Q}_{L})^{I}(\Psi_R)^{K}\Phi\chi \nonumber\\
            %%%%%%%%%%%%
                &\quad + \sum_{I=1}^{3}\sum_{K=1}^{8} \frac{2\sqrt{3}\,\parB_{IK}}{\LUV}\,(\overline{Q}_{L})^{I}(\Psi_R)^{K}\Sigma\chi
                + \sum_{I=1}^{3}\sum_{K=1}^{8} \frac{2\sqrt{3}\,\parBp_{IK}}{\LUV}\,(\overline{Q}_{L})^{I}(\Psi_R)^{K}\Sigma'\chi \nonumber \\
            %%%%%%%%%%%%
                &\quad + \sum_{K=1}^{8} \frac{\parC_{K}}{\LUV}\;Q_R (\Psi_R)^{K}\, \Delta^* \chi + \sum_{K,N=1}^{8} \frac{1}{2}\frac{\parD_{KN}}{\LUV^2}\;(\Psi_R)^K(\Psi_R)^N\,\Delta^*\chi^2 
                + \text{h.c.} .
                \label{eq:Yukawa-nonrenormalizable-dominant}
        \end{align}
        We see from \Table{tab:Yukawa-nonrenormalizable} that each invariant has a unique gauge contraction; explicitly we take
        \begin{align}
            \mathcal{A}_{IK}:&&
                (\bar{Q}_L)^{I}{}_{ai}\,(\Psi_{R})^{K}{}_{A}
                \,\Phi^{ii'}{}_{B}\,\chi^{aj'}{}_{C}\;\epsilon^{ABC}\,\epsilon_{i'j'} \, ,\\[4pt]
            \mathcal{B}_{IK}:&&
                (\bar{Q}_L)^{I}{}_{ai}\,(\Psi_{R})^{K}{}_{A}
                \,\Sigma^{a}{}_{b}{}^{ii'}{}_{B}\,\chi^{bj'}{}_{C}\;\epsilon^{ABC}\,\epsilon_{i'j'} \, ,\\[4pt]
            \mathcal{C}_{K}:&&
                (Q_{R})^{ai'D} \, (\Psi_R)^{K}{}_{A}\, 
                (\Delta^\ast)_{abi'j'BD}\,\chi^{bj'}{}_{C}\;
                \epsilon^{ABC} \, ,
                \\[4pt]
            \mathcal{D}_{KN}:&&
                (\Psi_{R})^{K}{}_{A}\,(\Psi_{R})^{N}{}_{D}\,
                (\Delta^{\ast})_{abi'j'BE} \,
                \chi^{ai'}{}_{C} \, \chi^{bj'}{}_{F}\;
                \epsilon^{ABC}\,\epsilon^{DEF} \, ,
        \end{align}
        and the $\parB'$-term is obtained from the $\parB$-term by replacing $\Sigma \to \Sigma'$.
    \item \textit{Subleading non-renormalizable operators:} \\[8pt]
        We are interested in those subleading operators that involve the $\Psi_{0}$ anomalon modes defined in \eq{eq:Psi0-definition}, since these states have been left out by the leading operators. For an operator consisting of a given set of fields, there may now be too many independent contractions of gauge indices to list. This problem is mitigated by considering only independent contractions in $\SU(3)_{f_{R}}$ flavor indices, in other words the independent contributions to the neutrino-anomalon mass matrix.
        \par
        The strategy to find the most important subleading contributions is as follows:
            \begin{enumerate}[label=(\theenumi.\arabic*),leftmargin=0.5cm]
                \item We make use of the list of all operators up to $d=9$ involving a given pair of fermion fields. This has already been compiled for the purposes of determining leading operators. The list is ordered by the parametric size of the contribution to the mass for largest to smallest.
                \item We retain only those operators that can in principle contribute to the neutrino-anomalon masses based on Pati-Salam quantum numbers. Namely, in invariants all terms must have a vanishing total quantum number under any of the Pati-Salam diagonal generators, and this condition is checked for the term involving neutrinos/anomalons and VEVs.    
                For irreps $\Phi,\Sigma,\Sigma'$ with two EW VEVs, all possible combinations are considered in the search for a valid term.
                \item For each operator in the curated list we compute the number of different flavor contractions between VEVs --- the same method as already used in \app{app:PSmodel-PQ-at-tree-level} --- and then write down these contractions explicitly.
                \item For each contraction we check wether the contribution couples also to $\Psi_{0}$. This is trivially seen when the flavor index $A$ of the anomalon $(\Psi_R)_{A}$ contracts with the $\chi$-VEV $\vevV{B}$ via the Levi-Civita tensor $\epsilon^{ABC}$.
            \end{enumerate}
        Carrying out the above strategy results in \Table{tab:Yukawa-subleading-contributions}. For each type of contribution involving anomalons we list the subleading operators $\mathcal{O}$ that can contribute and all independent VEV contractions in flavor. The anomalon flavor indices for each block type, colored \OUTER{teal}, are the following: $\OUTER{A}$ for $L\Psi$, $\OUTER{B}$ for $R\Psi$ and both $\OUTER{A}$ and $\OUTER{B}$ for $\Psi\Psi$. The contributions involving $\Psi_{0}$ (when the aforementioned index does not connect to a $\vevV{}$-VEV via $\epsilon$) have a designated label, which is used for the coefficient of these relevant contributions in \Table{tab:Yukawa-nonrenormalizable} and \eqs{eq:Nblock-LL}{eq:Nblock-AA}. The listing of operators in the table is truncated, so that all largest subleading contributions up to the size when relevant contributions are first encountered are included. We took care to do this for all possible values of $V$ in \eq{eq:vev-scaling}, hence both the $v^{2}/\LUV$ and $V^{5}/\LUV^{4}$ contributions are present in the $\Psi\Psi$ block. We confirm the presence of each relevant contribution by specifying the corresponding index contraction explicitly (an example, there may be multiple yielding the same contribution):
            \begin{align}
	            \parAt : &&
		              (\bar{Q}_L)^{I}{}_{ai}\,
		              (\Psi_{R})^{K}{}_{A}\,
		              \Delta^{abi'j'AB}\,
		              (\Phi^\ast)_{ji'}{}^{C}\,
		              (\chi^*)_{bj'}{}^{D}\;
		              \epsilon_{BCD}\,\epsilon^{ij},\\[4pt]
            %%%%
	            \parBt : &&
		              (\bar{Q}_L)^{I}{}_{ai}\,
		              (\Psi_{R})^{K}{}_{A}\,
		              \Delta^{cbi'j'AB}\,
		              (\Sigma^\ast)_{c}{}^{a}{}_{ji'}{}^{C}\,
		              (\chi^*)_{bj'}{}^{D}\;
		              \epsilon_{BCD}\,\epsilon^{ij},\\[4pt]
            %%%%
                \parCt : &&
                    (Q_{R})^{ai'A} \,
                    (\Psi_{R})^{N}{}_{B}\,
                    \Delta^{cdk'l'EF} \,
                    \Delta^{*}_{cdk'l'ED}\,
                    \Delta^{*}_{abi'j'AC}\,
                    \chi^{bj'}{}_{F}\;
                    \epsilon^{BCD},\\
            %%%%
                \parDt :&&
                    (\Psi_{R})^{K}{}_{A}\,(\Psi_{R})^{N}{}_{B}\,
                    \Phi^{\ast}{}_{ii'}{}^{A}\,
                    \Phi^{\ast}{}_{jj'}{}^{B}\;
                    \epsilon^{ij}\,\epsilon^{i'j'},\\[4pt]
            %%%% 
                \parEt:&&
                    (\Psi_{R})^{K}{}_{A}\,(\Psi_{R})^{N}{}_{B}\,
                    \Sigma^{\ast}{}_{a}{}^{b}{}_{ii'}{}^{A}\,
                    \Sigma^{\ast}{}_{b}{}^{a}{}_{jj'}{}^{B}\;
                    \epsilon^{ij}\,\epsilon^{i'j'},\\[4pt]
            %%%%
                \parFt : &&
		              (\Psi_{R})^{K}{}_{A}\,
		              (\Psi_{R})^{N}{}_{B}\,		
		              \Delta^{cdk'l'GH}\,
		              \Delta^{\ast}_{cdk'l'CE}\,
		              \Delta^{\ast}_{abi'j'DF}\,
		              \chi^{ai'}{}_{G}\,
		              \chi^{bj'}{}_{H}\;
		              \epsilon^{ACD}\,\epsilon^{BEF},\\[4pt]
            %%%%
                \parGt : &&
		              (\Psi_{R})^{K}{}_{A}\,
		              (\Psi_{R})^{N}{}_{B}\,		
		              \Delta^{cdk'l'GH}\,
		              \Delta^{\ast}_{cdk'l'GD}\,
		              \Delta^{\ast}_{abi'j'CE}\,
		              \chi^{ai'}{}_{H}\,
		              \chi^{bj'}{}_{F}\;
		              \epsilon^{DEF}\,\epsilon^{ABC}.
            \end{align}
%%%% table        
        \def\SSKIP{2pt}
        \def\MSKIP{6pt}
        \def\BSKIP{12pt}
        \renewcommand{\topfraction}{1}
        \renewcommand{\textfraction}{.00}
        \begin{table}[t!]
            \centering
            \scalebox{0.99}{
            \begin{tabular}{lllclc}
            \toprule
                 type&$\mathcal{O}$&$\sim\langle\mathcal{O}\rangle$&$\#_{f_{R}}$&$\SU(3)_{f_{R}}$ VEV contractions&label\\
            \midrule
            %%%%
                 $(L\Psi)_{\OUTER{A}}$&
                    $\bar{Q}_L\Psi_R\Phi \chi \xi $&
                    $ vV^2/\LUV^2 $&
                    $1$&
                    $\epsilon^{\OUTER{A}BC}\,\vevv{u\Phi}{B}\vevW\vevV{C}$&
                    ---\\[\MSKIP]
            %%%%
                 &
                    $\bar{Q}_L\Psi_R\Sigma \chi \xi $&
                    $ vV^2/\LUV^2 $&
                    $1$&
                    $\epsilon^{\OUTER{A}BC}\,\vevv{u\Sigma}{B}\vevW\vevV{C}$&
                    ---\\[\MSKIP]
            %%%%
                 & 
                    $\bar{Q}_L\Psi_R\Phi^* \Delta \chi^* $&
                    $ v V^2/\LUV^2 $&
                    $1$&
                    $\vevZ{\OUTER{A}B}\,\epsilon_{BCD}\,\vevvc{d\Phi}{C}\vevVc{D}$&
                    $\parAt$\\[\MSKIP]
            %%%%
                 &
                    $\bar{Q}_L\Psi_R\Sigma^* \Delta \chi^* $&
                    $ v V^2/\LUV^2 $&
                    $1$&
                    $\vevZ{\OUTER{A}B}\,\epsilon_{BCD}\,\vevvc{d\Sigma}{C}\vevVc{D}$&
                    $\parBt$\\[\BSKIP]
            %%%%%%%%%%%%%%%%%% R Psi
                 $(R\Psi)^{A}{}_{\OUTER{B}}$&
                    $Q_R\Psi_R\Delta^*\chi \xi $&
                    $V^3/\LUV^2 $&
                    $1$&
                    $\epsilon^{\OUTER{B}CD}\,\vevZc{AC}\vevV{D}\vevW$&
                    ---\\[\MSKIP]
            %%%%
                &
                    $Q_R\Psi_R\Delta \Delta^{*2}\chi $&
                    $ V^4/\LUV^3 $&
                    $3$&
                    $\epsilon^{\OUTER{B}CD}\,\vevZ{EF}\vevZc{EF}\vevZc{AC}\vevV{D}$&
                    ---\\[\SSKIP]
                &&&&
                    $\epsilon^{\OUTER{B}CD}\,\vevZ{EF}\vevZc{EA}\vevZc{FC}\vevV{D}$&
                    ---\\[\SSKIP]
                &&&&
                    $\epsilon^{\OUTER{B}CD}\,\vevZ{EF}\vevZc{ED}\vevZc{AC}\vevV{F}$&
                    $\parCt$\\[\MSKIP]
            %%%%
                 &
                    $Q_R\Psi_R\Delta^*\chi^2 \chi^*$&
                    $ V^4/\LUV^3 $&
                    $2$&
                    $\epsilon^{\OUTER{B}CD}\,\vevZc{AC}\vevV{D}\vevV{E}\vevVc{E}$&
                    ---\\[\SSKIP]
                &&&&
                    $\epsilon^{\OUTER{B}CD}\,\vevZc{DE}\vevV{A}\vevV{C}\vevVc{E}$&
                    ---\\[\MSKIP]
            %%%%
                 &
                    $Q_R\Psi_R\Delta^*\chi \xi^2 $&
                    $ V^4/\LUV^3 $&
                    $1$&
                    $\epsilon^{\OUTER{B}CD}\,\vevZc{AC}\vevV{D}\vevW^{2}$&
                    ---\\[\BSKIP]
            %%%%%%%%%%%%%%%%%% PsiPsi
                $(\Psi\Psi)_{\OUTER{A}\OUTER{B}}$&
                    $\Psi_{R}\Psi_{R}\Phi^{*2} $&
                    $ v^{2}/\LUV $&
                    $1$&
                    $\vevvc{u\Phi}{\OUTER{A}}
                        \vevvc{d\Phi}{\OUTER{B}} +
                        [\OUTER{A}\leftrightarrow\OUTER{B}]$&
                    $\parDt$\\[\MSKIP]
            %%%%
                &
                    $\Psi_{R}\Psi_{R}\Sigma^{*2} $&
                    $ v^{2}/\LUV $&
                    $1$&
                    $\vevvc{u\Sigma}{\OUTER{A}}
                        \vevvc{d\Sigma}{\OUTER{B}} +
                        [\OUTER{A}\leftrightarrow\OUTER{B}]$&
                    $\parEt$\\[\MSKIP]
            %%%%        
                &
                    $\Psi_{R}\Psi_{R}\Delta^*\chi^2\xi $&
                    $ V^4/\LUV^3 $&
                    $1$&
                    $\epsilon^{\OUTER{A}CD}\epsilon^{\OUTER{B}EF}\,\vevZc{CE}\vevV{D}\vevV{F}\vevW$&
                    ---\\[\MSKIP]
            %%%%
                &
                    $\Psi_{R}\Psi_{R}\Delta \Delta^{*2}\chi^2 $&
                    $ V^5/\LUV^4 $ &
                    $5$&
                    $\epsilon^{\OUTER{A}CD}\epsilon^{\OUTER{B}EF}\,\vevZ{GH}\vevZc{GH}\vevZc{CE}\vevV{D}\vevV{F}$&
                    ---\\[\SSKIP]
                &&&&
                    $\epsilon^{\OUTER{A}CD}\epsilon^{\OUTER{B}EF} \,\vevZ{GH}\vevZc{GC}\vevZc{HE}\vevV{D}\vevV{F}$&
                    ---\\[\SSKIP]
                 &&&&
                    $\epsilon^{\OUTER{A}CD}\epsilon^{\OUTER{B}EF}\,\vevZ{GH}\vevZc{CE}\vevZc{DF}\vevV{G}\vevV{H}$&
                    $\parFt$\\[\SSKIP]
                 &&&&
                    $\epsilon^{\OUTER{A}CD}\epsilon^{\OUTER{B}EF}\,\vevZ{GH}\vevZc{GC}\vevZc{DE}\vevV{H}\vevV{F}$&
                    ---\\[\SSKIP]
                 &&&&
                    $\epsilon^{\OUTER{AB}C}\epsilon^{DEF}\,\vevZ{GH}\vevZc{GD}\vevZc{CE}\vevV{H}\vevV{F}$&
                    $\parGt$\\[\MSKIP]
            %%%%
                 &
                 $\Psi_{R}\Psi_{R}\Delta^*\chi^3\chi^*$&
                 $ V^5/\LUV^4 $&
                 $1$&
                 $\epsilon^{\OUTER{A}CD}\epsilon^{\OUTER{B}EF}\,\vevZc{CE}\vevV{D}\vevV{F}\,\vevV{G}\vevVc{G}$&
                 ---\\[\MSKIP]
            %%%%
                 &
                 $\Psi_{R}\Psi_{R}\Delta^*\chi^2\xi^2 $&
                 $ V^5/\LUV^4 $&
                 $1$&
                 $\epsilon^{\OUTER{A}CD}\epsilon^{\OUTER{B}EF}\,\vevZc{CE}\vevV{D}\vevV{F}\vevW^{2}$&
                 ---\\
            \bottomrule
            \end{tabular}
            }
            \caption{The list of subleading contributions from largest to smallest for all blocks in \eqref{eq:mass-matrix-LRA} involving anomalons: $L\Psi$, $R\Psi$, $\Psi\Psi$ involving anomalons we list the subleading operators $\mathcal{O}$ that can contribute. The operator (in terms of field powers) is denoted by $\mathcal{O}$, its VEV scaling by $\sim\langle\mathcal{O}\rangle$ and the number of independent contributions to the mass (number of independent flavor contractions) by $\#_{f_{R}}$, which are then listed explicitly in different rows. Contributions that involve $\Psi_{0}$ anomalon states from \eq{eq:Psi0-definition} have a label next to them, and are included as relevant contributions into \Table{tab:Yukawa-nonrenormalizable} and \eqs{eq:Nblock-LL}{eq:Nblock-AA}. The flavor indices of anomalons in the first column and the associated free indices in the contraction column are colored \OUTER{teal} as visual emphasis due to their structural importance, see main text; the standard color coding of \colorEW{blue} for EW VEVs and  \colorPQ{red} for the PQ-breaking Pati-Salam VEVs is also used. 
                \label{tab:Yukawa-subleading-contributions}
                }
        \end{table}
    \item \textit{Redundancy removal:} \\[8pt]
        We address here the redundancy in describing the Yukawa sector of our Pati-Salam model. The SM fermions are packed into two types of representations, namely $Q_L$ and $Q_R$, thus allowing for a global flavor transformation $\mathrm{U}(3)_{Q_L}\times\mathrm{U}(3)_{Q_R}$, transforming the multiplicity index $I$ and the gauged family index $A$ with the two respective factors. The second factor essentially contains the $\SU(3)_{f_R}$ gauge transformations. 
        \par
        An $\mathrm{SU}(3)$ transformation acting on a set of complex vectors $\vec{v}_{i}=(x_i,y_i,z_i)$ can bring one into the form $\vec{v}_{1}=(0,0,z_{1})$ and a subsequent $\mathrm{SU}(2)$ rotation in the first two components can bring a second vector to the form $\vec{v}_{2}=(0,y_{2},z_{2})$ without disrupting the form of $\vec{v}_{1}$. The VEV component $z_{1},y_{2}\in\mathbb{R}$ and $z_{2}\in\mathbb{C}$ in accordance with using the $8$ real degrees of freedom of $\SU(3)$ to remove the same number in $\vec{v}_{1}$ and $\vec{v}_{2}$. 
        \par
        The rotation $\mathrm{U}(3)_{Q_L}$ can be used on the parameters of the renormalizable Yukawa potential, where we have a set of $3$ vectors $\{\parY{\Phi}{I},\parY{\Sigma}{I},\parY{\Sigma'}{I}\}$, while the $\mathrm{U}(3)_{Q_R}$ rotation can be used on the set of the $6$ vectors of EW VEVs listed in \eq{eq:labels-EW-VEVs}. In particular, we find the removal of redundancies via the choice of \eq{eq:redundancy-removal-choice} the most convenient, especially when addressing the realistic nature of the Yukawa sector in \sect{sec:PSmodel-Yukawa-realistic}.
    \item \textit{Diagonalization of a hierarchical mass matrix:} \\[8pt]
        To obtain masses and mixings, the matrix of \eq{eq:mass-LRA-block4} is diagonalized via Takagi decomposition. A general complex symmetric matrix $\mathbf{M}$ can be decomposed as
        \begin{align}
            \mathbf{M}&=\mathbf{U}\mathbf{D}\mathbf{U}^{T} \, ,
        \end{align}
        where $\mathbf{D}$ is diagonal with non-negative real eigenvalues $m_{i}\equiv D_{ii}$, and $\mathbf{U}$ is a unitary matrix. The decomposition can be performed numerically, but an analytic approximation is also possible due to the hierarchical nature of our mass matrix. We elaborate on the general methods involving the analytic approach:
        \begin{enumerate}[leftmargin=0.5cm,label=(\theenumi\alph*)]
            \item 
            Consider a real matrix $\mathbf{M}$ to avoid unnecessarily complicated expressions; $\mathbf{U}$ then becomes real and orthogonal. 
            Assuming parametrically small off-diagonal entries in $\mathbf{M}$, and consistently expanding order-by-order in these small parameters, one can derive the following approximations for masses and mixing angles at next-to-leading order:  
            \begin{align}
	               m_{i}&\approx M_{ii} + \sum_{k\neq i} \frac{M_{ik}^{2}}{M_{ii}-M_{kk}}
                    \, , &
                \quad
	               U_{ij}&\approx \frac{M_{ij}}{M_{jj}-M_{ii}} -
		          \sum_{k\neq i,j} \frac{M_{ik}M_{jk}}{(M_{jj}-M_{ii})(M_{kk}-M_{jj})} \, ,
                \label{eq:mass-angle-approximation}
            \end{align}
            where $M_{ij}$ are the components of $\mathbf{M}$, and $U_{ij}$ are off-diagonal elements of $\mathbf{U}$ for $i<j$ (note that the expression is not $i\leftrightarrow j$ symmetric). Orthogonality of $\mathbf{U}$ implies that specifying the upper-diagonal part uniquely determines the entire matrix, which can be reconstructed as follows:
            \begin{align}
	           (\mathbf{U})_{ij}&\approx
		          \begin{cases}
			         U_{ij}&;\quad i<j\\[3pt]
			         -U_{\{ij\}}-\sum_{k\neq i,j} (-1)^{H_{i-k}+H_{j-k}}\,U_{\{ik\}}U_{\{jk\}}&;\quad i>j\\[3pt]
			         1-\sum_{k\neq i} U_{\{ik\}}^{2}/2 &;\quad i=j\\
		          \end{cases} \, , \label{eq:perurbative-U}
	        \end{align}
            where curly brackets $\{\}$ order the indices from smaller to larger, and $H$ is the Heaviside function ($H_{x}=1$ for $x\geq 0$ and $H_{x}=0$ for $x< 0$).
            \item Note that the expressions of \eq{eq:mass-angle-approximation} fail if denominators blow up. Assuming non-degenerate eigenvalues, the main mode of failure is when (multiple) diagonal entries $M_{ii}$ are also parametrically small. For rough estimates, \eq{eq:mass-angle-approximation} can be modified to incorporate the possibility of small diagonal entries $M_{ii}$ by replacing them in denominators by the mass eigenvalues $m_{i}$: 
            \begin{align}
	            m_{i}&\sim M_{ii} + \sum_{k\neq i} \frac{M_{ik}^{2}}{m_{i}-m_{k}}
                \, , &
                \quad
	               U_{ij}&\sim 
                \frac{1}{m_{j}-m_{i}}\;\left(
                M_{ij}-\sum_{k\neq i,j} \frac{M_{ik}M_{jk}}{m_{k}-m_{j}}
        \right) \, .
        \label{eq:mass-angle-approximation2}
        \end{align}
        One can solve this system by computing $m_{i}$ iteratively until consistent, starting with $m_{i}=M_{ii}$, and then using these quantities for the mixing $U_{ij}$.
        Finally, any violation of $U_{ij}\ll 1 $ is a sign that the method is not applicable, since parametric smallness of mixing was the starting assumption.
    \end{enumerate}
    \item 
        \textit{The SM Higgs and the direction of EW VEVs: } \\[8pt]
        We argued in \sect{sec:PSmodel-Yukawa-realistic} that one copy of $\Phi$ and two copies of $\Sigma$ should suffice for a realistic fermion fit. This conclusion was reached by careful parameter counting and comparison with observables. We relied on the fact, however, that EW VEVs, listed in e.g.~the second line of \eq{eq:fit-parameters}, can be taken with arbitrary ratios. This detail is what the discussion below focuses on.
        \par
        The number of weak scalar doublets transforming as $(1,2,\pm\tfrac{1}{2})$ under the SM group is best tracked via the EW VEVs these representations carry. Each irrep $\Phi$ or 
        $\Sigma$ carries $3$ up-type and $3$ down-type EW VEVs,  hence our case $(N_{\Phi},N_{\Sigma})=(1,2)$ corresponds to having $9$ up- and $9$ down-type doublets.  
        Using VEV notation, we can write the basis of $(1,2,+\tfrac{1}{2})$ irreps as
        \begin{align}
            \{
                \vevv{u\Phi}{A},
                \vevv{u\Sigma}{A},
                \vevv{u\Sigma'}{A},
                \vevvc{d\Phi}{A},
                \vevvc{d\Sigma}{A},
                \vevvc{d\Sigma'}{A}
            \},
        \end{align}
        where $A$ is the flavor index. The $18\times 18$ doublet mass matrix can then be written using the above basis for rows and the conjugate basis for columns. To facilitate further exploration of its subparts, we schematically write it in block form as
            \begin{align}
            \begin{pmatrix}
                \Phi_{u}\Phi^{*}_{u}&
                    \Phi_{u}\Sigma^{*}_{u}&
                    \Phi_{u}\Phi_{d}&
                    \Phi_{u}\Sigma_{d}\\
                \Sigma_{u}\Phi^{*}_{u}&
                    \Sigma_{u}\Sigma^{*}_{u}&
                    \Sigma_{u}\Phi_{d}&
                    \Sigma_{u}\Sigma_{d}\\
                \Phi^*_{d}\Phi^{*}_{u}&
                    \Phi^*_{d}\Sigma^{*}_{u}&
                    \Phi^*_{d}\Phi_{d}&
                    \Phi^*_{d}\Sigma_{d}\\
                \Sigma^*_{d}\Phi^{*}_{u}&
                    \Sigma^*_{d}\Sigma^{*}_{u}&
                    \Sigma^*_{d}\Phi_{d}&
                    \Sigma^*_{d}\Sigma_{d}\\
            \end{pmatrix}, \label{eq:doublet-mass-scheme}
            \end{align}
        where block dimension is $3$ if referring to $\Phi$ and $6$ if referring to $\Sigma$, since the latter case takes both $\Sigma$ and $\Sigma'$ into account due to the two irreps being interchangeable in the formation of relevant invariants. Notice that the diagonal $9\times 9$ parts of the full matrix (or $2\times 2$ in block form) are of type $uu$ and $dd$, while the off-diagonal parts are of mixed type, referred to as $ud$.
        \par
        The generic expectation is that most entries in this matrix are of size $V^{2}$. After performing a number of fine-tunings, one doublet mass can be tuned down to the EW scale (the SM Higgs), while the others' masses depend on the number of tunings and the pattern of $V^{2}$-sized entries in the mass matrix, cf.~e.g.~\cite{Berezhiani:1982pj}. Our main goal is to elucidate the pattern for our model and infer the features of the doublet spectrum and required fine-tunings from there.   
        \par
        Consider the renormalizable scalar potential of \eqs{eq:V2}{eq:VC}. The $uu$ and $dd$ parts of \eq{eq:doublet-mass-scheme} are easily populated by a number of terms of size $V^{2}$: 
        \begin{itemize}
            \item 
                The 1-1 and 3-3 entries in \eq{eq:doublet-mass-scheme}, i.e. the $|\Phi|^2$-blocks $\vevv{u\Phi}{A}\vevvc{u\Phi}{B}$ and $\vevv{d\Phi}{A}\vevvc{d\Phi}{B}$, are for example easily populated by terms in $\mathcal{V}_{2,4}$, some of which are
                \begin{align}
                    \left\langle
                        \Phi^{ii'}{}_{A}\;\Phi^*{}_{ii'}{}^{A}
                    \right\rangle
                        &= \vevv{u\Phi}{A}\vevvc{u\Phi}{A}+\vevv{d\Phi}{A}\vevvc{d\Phi}{A}, 
                            \label{eq:doublet-term-1}\\
                    \left\langle
                        \Phi^{ii'}{}_{A}\;\Phi^*{}_{ii'}{}^{B}\;
                        \Delta^{abk'l'AC}\;\Delta^{*}{}_{abk'l'BC}
                    \right\rangle
                        &= \left(\vevv{u\Phi}{A}\vevvc{u\Phi}{B}+\vevv{d\Phi}{A}\vevvc{d\Phi}{B}\right) 
                            \;\vevZ{AC}\vevZc{CB},
                            \label{eq:doublet-term-2}\\
                    \left\langle
                        \Phi^{ii'}{}_{A}\; \Phi^*{}_{ii'}{}^{A}\;
                        \chi^{bj'}{}_{B}\; \chi^*{}_{bj'}{}^{B}
                    \right\rangle
                        &= \left(
                            \vevv{u\Phi}{A}\vevvc{u\Phi}{A}+\vevv{d\Phi}{A}\vevvc{d\Phi}{A}
                            \right)\,
                            \vevV{B}\,\vevVc{B}, 
                            \label{eq:doublet-term-3}\\
                     \left\langle
                        \Phi^{ii'}{}_{A}\; \Phi^*{}_{ii'}{}^{B}\;
                        \chi^{bj'}{}_{B}\; \chi^*{}_{bj'}{}^{A}
                    \right\rangle
                        &= \left(
                            \vevv{u\Phi}{A}\vevvc{u\Phi}{B}+\vevv{d\Phi}{A}\vevvc{d\Phi}{B}
                            \right)\,
                            \vevV{B}\,\vevVc{A},
                            \label{eq:doublet-term-4}
                \end{align}
                where the evaluation of invariants on the EW-broken vacuum is used as a proxy for obtaining the doublet mass terms --- one simply replaces the EW VEVs by the underlying fields. We see that the $3\times 3$ blocks are populated flavor-universally on the diagonal, cf.~\eq{eq:doublet-term-1} and \eqref{eq:doublet-term-3}, while flavor non-universal contributions arise in both diagonal and off-diagonal entries, cf.~\eq{eq:doublet-term-2} and \eqref{eq:doublet-term-4}.
            \item The 2-2 and 4-4 entries in \eq{eq:doublet-mass-scheme}, i.e.~the $|\Sigma|^2$-blocks, are populated in an analogous way to the $|\Phi|^{2}$ blocks by invariants in $\mathcal{V}_{2,4}$, e.g.~by the invariants $|\Sigma|^{2}$, $|\Sigma|^2 |\Delta|^2$ and $|\Sigma|^2 |\chi|^2$ (and all $\Sigma\mapsto \Sigma'$ replacements). The $|\Sigma|^2$-block has, however, an additional contribution from $\mathcal{V}_{3}$ that is conceptually important:
            \begin{align}
                \left\langle
                    \Phi^{ii'}{}_{A}\;\Sigma^*{}_{ij'}{}^{A}\;\xi^{a}{}_{b}{}^{j'k'}\;\epsilon_{i'k'}
                \right\rangle
                    &= \tfrac{\vevW}{\sqrt{6}}\,
                        \left(
                            \vevv{u\Sigma}{A} \vevvc{u\Sigma}{A}
                            - \vevv{d\Sigma}{A} \vevvc{d\Sigma}{A}
                        \right).
                    \label{eq:doublet-term-mix}
            \end{align}
            Notice that it gives a different contribution to the $uu$ and $dd$ part.
            \item The 1-2 and 3-4 blocks in \eq{eq:doublet-mass-scheme} (along with their conjugates 2-1 and 4-3), i.e.~the mixed $\Phi\Sigma^*$-terms (label up to conjugation and transposition), can only be generated from $\mathcal{V}_{C}$. Some examples include
            \begin{align}
                \left\langle
                    \Phi^{ii'}{}_{A}\; \Sigma^*{}_{a}{}^{b}{}_{ij'}{}^{A}
                    \;\xi^{a}{}_{b}{}^{j'k'}
                    \;\epsilon_{i'k'}
                \right\rangle
                    &= -\tfrac{\vevW}{\sqrt{2}}\,
                        \left(
                            \vevv{u\Phi}{A} \vevvc{u\Sigma}{A}
                            - \vevv{d\Phi}{A} \vevvc{d\Sigma}{A}
                        \right), 
                        \label{eq:doublet-term-5} \\
                \left\langle
                    \Phi^{ii'}{}_{A}\; \Sigma^*{}_{a}{}^{b}{}_{ii'}{}^{B} \;
                    \Delta^{acj'k'AC} \; \Delta^*{}_{bcj'k'BC}
                \right\rangle
                    &= -\tfrac{\sqrt{3}}{2} \left(
                        \vevv{u\Phi}{A} \vevvc{u\Sigma}{B} + \vevv{d\Phi}{A} \vevvc{d\Sigma}{B}
                    \right)\,
                    \vevZ{AC}\,\vevZc{BC},
                    \label{eq:doublet-term-6} \\
                \left\langle
                    \Phi^{ii'}{}_{A}\;\Sigma^*{}_{a}{}^{b}{}_{ii'}{}^{A}\;
                    \chi^{aj'}{}_{B} \;\chi^*{}_{bj'}{}^{B}
                \right\rangle
                    &= -\tfrac{\sqrt{3}}{2} \left(
                        \vevv{u\Phi}{A} \vevvc{u\Sigma}{A} + \vevv{d\Phi}{A} \vevvc{d\Sigma}{A}
                    \right)\,
                        \vevV{B}\,\vevVc{B},
                    \label{eq:doublet-term-7} \\   
                \left\langle
                    \Phi^{ii'}{}_{A}\;\Sigma^*{}_{a}{}^{b}{}_{ii'}{}^{B}\;
                    \chi^{aj'}{}_{B} \;\chi^*{}_{bj'}{}^{A}
                \right\rangle
                    &= -\tfrac{\sqrt{3}}{2} \left(
                        \vevv{u\Phi}{A} \vevvc{u\Sigma}{B} + \vevv{d\Phi}{A} \vevvc{d\Sigma}{B}
                    \right)\,
                        \vevV{B}\,\vevVc{A},
                        \label{eq:doublet-term-8}
            \end{align} 
            again consisting of a mix of flavor universal (\eqref{eq:doublet-term-5} and \eqref{eq:doublet-term-7}) and non-universal (\eqref{eq:doublet-term-6} and \eqref{eq:doublet-term-8}) terms. Furthermore, \eq{eq:doublet-term-5} introduces a difference between $uu$ and $dd$ blocks.
        \end{itemize}
        The listed contributions fill up the $uu$ and $dd$ parts of the mass matrix in \eq{eq:doublet-mass-scheme}, with some contributions flavor non-universal and some distinguishing between $uu$ and $dd$ blocks (those with $\xi$), leading to a rich pattern of entry values. The question arises, however, how to obtain parametrically $V^2$-sized contributions in mixed $ud$ entries. For those, the invariant must contain one of the factors $\Phi^2$,
         $\Phi\Sigma$, $\Sigma^2$ or their conjugates, with all other fields necessarily acquiring PS-scale VEVs (order $V$).
        The only candidate in the renormalizable potential would be $\Sigma^2 \Delta^{*2}$ in $\mathcal{V}_{\mathbb{C}}$, but we know this invariant doesn't contribute (to the vacuum, and hence also to the doublet mass matrix), cf.~\app{app:number-of-U1s}.
        \par
        The tree-level mass matrix thus has $V^2$-level entries on diagonal $9\times 9$ blocks, while the off-diagonal $ud$ blocks would only get contributions from $v^{2}$-level terms such as $|\Phi|^4$,  $|\Phi|^2 |\Sigma|^{2}$ and $|\Sigma|^4$. Suppose now this was the final result. Since the SM Higgs, which is effectively the null mode of the full doublet mass matrix, requires an admixture both in the $u$ and the $d$ sector, one is forced to perform a fine-tuning from the PS to the EW scale in both the $uu$ and $dd$ diagonal $9\times 9$ blocks. This requires two fine-tunings, which results in a second weak doublet very close to the EW scale --- not ideal.
        \par
        The situation changes, however, at loop level, as we already analyzed in \app{app:number-of-U1s}. In particular, we obtain a mixed $ud$ contributions of order $V^2$ from quantum corrections. In particular, we see exactly such a $\vevv{u\Phi}{A}\vevv{d\Phi}{B}$ contribution from the operator $\Phi^{2}\Delta^* \chi^2$ in \eq{eq:two-loop-operator-result}, which populates the 1-3 block in \eqref{eq:doublet-mass-scheme} (and the conjugate populates 3-1). Similarly, the effective operators $\Phi\Sigma\Delta^{*}\chi^2$ and $\Sigma^2\Delta^{*}\chi^2$ can be constructed at $2$-loop level from renormalizable couplings. The latter leads to a contribution $\vevv{u\Sigma}{A}\vevv{d\Sigma}{B}$ and populates the 2-4 block entry, while the former leads to contributions $\vevv{u\Phi}{A}\vevv{d\Sigma}{B}$ and $\vevv{u\Sigma}{A}\vevv{d\Phi}{B}$ and populates the blocks 1-4 and 2-3. This accounts for all the mixed $ud$ blocks.
        \par 
        The loop corrected result for the doublet mass matrix thus populates all blocks in \eq{eq:doublet-mass-scheme} by $V^2$-level entries, with $9\times 9$ off-diagonal block entries, i.e.~the mixed $ud$ blocks, having a $2$-loop suppression $1/(16\pi^2)^2$ (for mass-squares).
        This structure requires only a small tuning in the two diagonal $9\times 9$ blocks down to the $2$-loop suppression, and then only one big fine-tuning from there to the EW scale. Altogether, the price of a fermion fit is thus expected to be only one large fine-tuning, with all doublets except for the SM Higgs generically at the PS scale (modulo one of them being $1/(16\pi^2)$ below in mass).
        \par
        The above analysis ultimately shows no major structural hurdle in the doublet mass matrix to align the EW VEVs in the second line of \eq{eq:fit-parameters} in the desired direction, and hence no obvious impediment to a successful fermion fit. Finally, note that flavor off-diagonal couplings in the doublet mass matrix might help to transfer the flavor hierarchy in PS-breaking VEVs $\vevV{A}$ and $\vevZ{AB}$ to the EW VEVs.  
\end{enumerate}

%%%%%%%%%%%%%%%%%%%%%%%%%%%%%%
%%%%%%%%%%%%%%%%%%%%%%%%%%%%%%
\subsection{Perturbativity}
\label{app:PSpert}

With the introduction of a horizontal gauge symmetry, i.e.~the gauged flavor group $\mathrm{SU}(3)_{f_R}$, the number of field degrees of freedom in the scalar sector increases, since many will transform non-trivially under it. For our particular model, we can see from \Table{tab:PSirrep} that all scalar fields transform non-trivially
under the flavor group.

The proliferation of degrees of freedom brings concerns regarding the perturbativity of the model. One aspect of this is the behavior of gauge couplings $g_i$ above the Pati-Salam scale. 

Writing the renormalization group equations (RGE) in terms of $\alpha_{i}^{-1}\equiv 4\pi/g_{i}^{2}$, 
the two loop RGE take the well-known form (see e.g.~\cite{Ellis:2015jwa,Bertolini:2009qj})
\begin{align}
    \dfrac{d}{dt}\,\alpha^{-1}_{i}&= -\dfrac{1}{2\pi} \big(a_{i}+\dfrac{1}{4\pi} \sum_{j} b_{ij} \alpha_{j}\big) \, ,
    \label{eq:RGE}
\end{align}
with $t=\log(\mu/\mu_{0})$ and $\mu_{0}$ being a reference scale. We retained in the beta function only the gauge contributions, since they dominate over the Yukawa contributions.

We use the following conventions: we label the SM gauge couplings $\alpha$ by
$(\alpha_{3},\alpha_{2},\alpha_{1})$ referring to each factor of $\SU(3)_C\times\SU(2)_L\times\mathrm{U}(1)_{Y}$ from left to right,\footnote{We use the GUT-normalization for $\alpha_{1}$, i.e.~$\alpha_{1}=\tfrac{5}{3}\alpha_{Y/2}$, where the hypercharge is normalized so that $Y(e^{c})/2=+1$ for the electron.} and we use $(\alpha_{4C},\alpha_{2L},\alpha_{2R},\alpha_{3F})$ in the Pati-Salam model $\SU(4)_C\times\SU(2)_L\times\SU(2)_R\times\SU(3)_{f_R}$.

The coefficients $a_{i}$ and $b_{ij}$ can be computed in the standard way~\cite{Jones:1981we}. 
Referring with the index $i$ to group factors from left to right using the conventions above, the SM coefficients are
\begin{align}
    a^{\text{(SM)}}_{i}&= (-7,-\tfrac{19}{6},\tfrac{41}{10}) \, , &
    b^{\text{(SM})}_{ij}&= 
        \begin{pmatrix}
            -26 & \tfrac{9}{2} & \frac{11}{10} \\
            12 & \tfrac{35}{6} & \frac{9}{10} \\
            \tfrac{44}{5} & \tfrac{27}{10} & \tfrac{199}{50} \\
        \end{pmatrix} \, , 
        \label{eq:ab-coefficients-SM}
\end{align}
while the Pati-Salam model with the fermion and scalar content of \Table{tab:PSirrep}, and taking $N_\Phi=1$ and $N_\Sigma=2$, yields the the 1- and 2-loop beta coefficients
\begin{align}
    a^{\text{(PS)}}_{i}&= (\tfrac{127}{3},\tfrac{83}{3},\tfrac{224}{3},\tfrac{124}{3}) \, ,&
    b^{\text{(PS)}}_{ij}&= \begin{pmatrix}
        \tfrac{8915}{3} & \tfrac{585}{2} & \tfrac{1563}{2} & 1256 \\[3pt]
        \tfrac{2925}{2} & \tfrac{1220}{3} & 279 & 496 \\[3pt]
        \tfrac{7815}{2} & 279 & \tfrac{5078}{3} & 2144 \\[3pt]
        2355 & 186 & 804 & \tfrac{4900}{3} \\[3pt]
    \end{pmatrix} \, .
    \label{eq:ab-coefficients-PS}
\end{align}
The hypercharge is embedded into the Pati-Salam group via
\begin{align}
    \tfrac{1}{2}\,Y=\tfrac{1}{2}(B-L)+t^{3}_R=\sqrt{\tfrac{2}{3}}\, t^{15}_C+t^{3}_{R} \, ,
\end{align}
where $t^{15}_C=\tfrac{1}{\sqrt{24}}\,\diag(1,1,1,-3)$ and $t^{3}_R=\tfrac{1}{2}\,\diag(1,-1)$ are diagonal generators in $\SU(4)_C$ and $\SU(2)_R$, respectively, and $B$ and $L$ are baryon and lepton number. This connection yields the matching condition at the Pati-Salam scale to be 
\begin{align}
    (\alpha^{-1}_{4C},\;
    \alpha^{-1}_{2L},\;
    \alpha^{-1}_{2R})&= 
    (\alpha^{-1}_{3},  \;
    \alpha^{-1}_{2},\;
    \tfrac{5}{3}\alpha^{-1}_{1}-\tfrac{2}{3}\alpha^{-1}_{3}). \label{eq:RGE-matching}
\end{align}
The flavor gauge coupling $\alpha_F$ is not determined through these, since no remnant of $\SU(3)_{f_{R}}$ is embedded into the SM group.

Eqs.~\eqref{eq:RGE}--\eqref{eq:ab-coefficients-PS} and \eqref{eq:RGE-matching} enable a full bottom-up RGE evolution of gauge couplings, starting with SM values at the $Z$-scale~\cite{Marciano:1991ix,Kuhn:1998ze,Martens:2010nm,Sturm:2013uka,ParticleDataGroup:2020ssz}:
    \begin{align}
        \big(
            \alpha^{-1}_{3},\;
            \alpha^{-1}_{2},\;
            \alpha^{-1}_{1}
        \big)(M_Z)
        &=
        (8.550 \pm 0.065,\;
        29.6261 \pm 0.0051,\;
        59.1054 \pm 0.0031
        ).\label{eq:SM-couplings-at-MZ}
	  \end{align}
The only unknowns are the matching scale $\mu_{PS}$ (where we assume to be all the non-SM fields of the model, neglecting threshold effects) and the flavor gauge coupling $\alpha_{3F}^{-1}(\mu_{PS})$ at that scale, where $\mu_{PS}\sim V\sim 10^{9}\,\mathrm{GeV}$.

\begin{figure}
    \centering
    \includegraphics[width=0.46\linewidth]{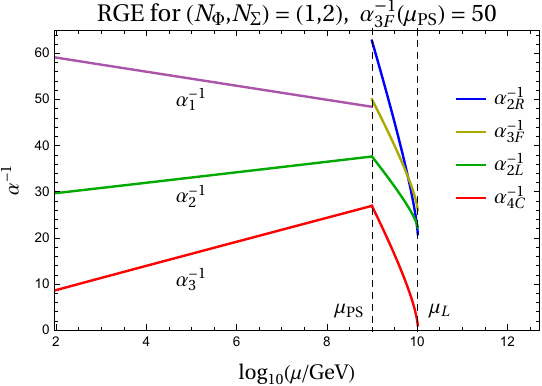}
    \hspace{0.5cm}
    \includegraphics[width=0.46\linewidth]{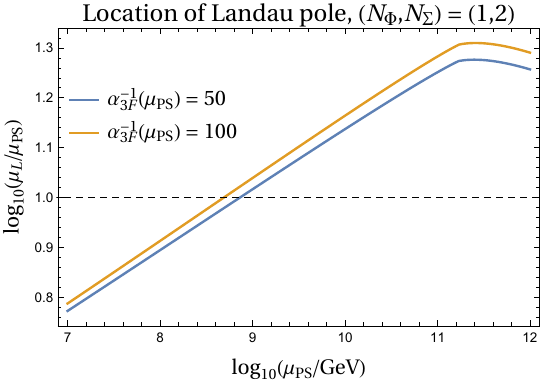}
    \caption{Gauge coupling running for a benchmark point with $\alpha_{3F}^{-1}=50$ at the scale $\mu_{PS}$ (left panel), and the effect of the matching scale $\mu_{PS}$ and the value of the flavor gauge coupling $\alpha_{3F}^{-1}(\mu_{PS})$ on the position of the Landau pole at scale $\mu_{L}$ (right panel).}
    \label{fig:RGE-gauge-couplings}
\end{figure}

A benchmark example with $\alpha^{-1}_{3F}(\mu_{PS})=50$ is shown in the left panel in Fig.~\ref{fig:RGE-gauge-couplings}. It captures the salient feature that the coupling $\alpha_{4C}$ is the one driven to a Landau pole at $\mu_{L}$, i.e., $\alpha_{4C}^{-1}(\mu_L)=0$, unless the coupling $\alpha_{3F}$ is large. Changing the coupling values for $\alpha_{3F}$ has only a slight effect on $\mu_L$, since it drives the other couplings only through the two-loop cross-term. 

With a rising matching scale $\mu_{PS}$, the value $\alpha_{4C}^{-1}$ at that scale rises, while $\alpha_{2R}^{-1}$ decreases, as it can be deduced from the left panel of \fig{fig:RGE-gauge-couplings}. A higher value of $\alpha_{4C}^{-1}$ then implies more running needed before reaching a Landau pole, as seen in the right panel. Beyond around $\mu_{PS}\sim 10^{11.2}\,\mathrm{GeV}$, however, the change of trend is due to $\alpha^{-1}_{2R}$ reaching the Landau pole before $\alpha_{4C}^{-1}$ does.

We thus see that the gauge couplings have a Landau pole around one order of magnitude above the matching scale, i.e.~$\log_{10}(\mu_{L}/\mu_{PS})\simeq 1$. Some non-renormalizable operators are thus expected to be suppressed only by a factor $10^{-1}$, which we still deem acceptable for the theory to be treated  perturbatively at the scale $\mu_{PS}$.

On the other hand, the existence of a Landau pole before the Planck scale raises a more conceptual issue. Unless the theory is tempered by non-perturbative dynamics arising before $\mu_{L}$, the latter scale should be associated to some new physics 
which ensures a proper UV behaviour. 
We then need to require that such new physics does not 
break the PQ symmetry, since our working hypothesis is that the scale responsible for the PQ-breaking operators is $\LUV \sim M_{\rm Pl}$.

Finally, we observe that the issue of perturbativity and the Landau pole arising before the Planck scale should generically arise in models with gauged horizontal symmetries, and it is thus not exclusive for the models under investigation in this paper. 

%%%%%%%%%%%%%%%%%%%%%%%%%%%%%%%%
%%%%%%%%%%%%%%%%%%%%%%%%%%%%%%%%
\section{Reflections on the $\SO(10) \times \SU(3)_f$ 
model of Ref.~\cite{DiLuzio:2020qio}}
\label{sec:SO10model}

In this Appendix we reflect on the $\SO(10)\times\SU(3)_{f}$ model from \cite{DiLuzio:2020qio}, which served as the original inspiration for the Pati-Salam model of this paper, 
provide some corrections to the analysis in Ref.~\cite{DiLuzio:2020qio}
and compare the two models. 

%%%%%%%%%%%%%%%%%%%%%%%%%%%%%%%%
\subsection{Comparing the $\SO(10)$ and Pati-Salam models}

The $\SO(10)\times\SU(3)_{f}$ model of Ref.~\cite{DiLuzio:2020qio} features the field content listed for convenience in \Table{tab:SO10irreps}. The fermions are generically labeled by $\psi$ and the scalars by $\phi$. This compares to the Pati-Salam case of \Table{tab:PSirrep} as follows:
\begin{itemize}[itemsep=0cm,leftmargin=0.5cm]
    \item The fermions $Q_{L}$ and $Q_{R}$ now join into a single representation $\phi_{16}$. The flavor group now simultaneously rotates both $Q_L$ and $Q_R$, so we label it simply as $\SU(3)_f$.
    \item The exotic fermions in light gray ensure $\SU(3)_{f}$ anomaly cancellation, in analogy with $\Psi_{R}$ canceling $\SU(3)_{f_{R}}$ anomalies.
    \item Ignoring the flavor assignment and considering only the inclusion $\SU(4)_{\rm PS}\times\SU(2)_L\times\SU(2)_{R}\subset\SO(10)$, we have 
    \begin{align}
        \Phi&\subset\phi_{10} \, ,&
        \Sigma,\Delta&\subset \phi_{\overline{126}} \, ,&
        \chi^\ast & \subset \phi_{16} \, .
        \label{eq:inclusion-into-SO10}
    \end{align}
    The representations $\phi_{10}$ and $\phi_{\overline{126}}$ function as containers for the SM Higgs, the same as $\Phi$ and $\Sigma$. The representations $\phi_{\overline{126}}$ is also crucial for neutrino masses, analogously to $\Delta$. 
    \item More loosely, one can draw an analogy between $\chi$ and $\phi_{16}$, as well as between the real representations $\xi$ and $\phi_{45}$. Although the details slightly differ between the models, both pairs are involved in breaking the gauge symmetry to the SM,\footnote{While $\xi$ is not strictly necessary for the breaking to the SM group, $\phi_{45}$ is necessary to break the $\SU(5)$ subgroup of $\SO(10)$. For issues related to the breaking of SO(10) with $\phi_{45}$ at the quantum level, see \cite{Bertolini:2009es,Bertolini:2012im,Bertolini:2013vta,Graf:2016znk,Jarkovska:2021jvw,Jarkovska:2023zwv}.} and only in the presence of both representations there is an accidental $\mathrm{U}(1)_{\rm PQ}$ not aligned with $B-L$.
    \item The action under the center $\mathbb{Z}_{4}$ of $\SO(10)$ does not directly correspond to the action of the $\mathbb{Z}_{4}$ center of $\SU(4)_{\rm PS}$. In particular, the PS center acts differently on $\Phi$ and $\Sigma$ 
    compared to how the $\SO(10)$-center acts on the their parent irreps $\phi_{10}$ and $\phi_{\overline{126}}$; for the Peccei-Quinn symmetry and its quality to behave similarly, the transformation of the PS irreps under flavor 
    was thus modified compared to the $\SO(10)$ case.
\end{itemize}

\begin{table}[t!]
$$\begin{array}{c|c|cc|cc|c|c}
\rowcolor[HTML]{C0C0C0} 
\hbox{Field} & \hbox{Lorentz} &  
\SO(10)
& 
\mathbb{Z}_4 
& \SU(3)_{f}  & 
\mathbb{Z}_3 
& \text{Generations} 
& \U(1)_{\rm PQ} \\\hline
\psi_{16} & (1/2,0) & 16 & +i & \3 & e^{i2\pi/3} & 1 & +1 \\ 
\rowcolor{CGray} 
\psi_{1} & (1/2,0) & 1 & +1 & \overline{\3} & e^{i 4\pi/3} & 16 & \phantom{+}0 \\
\hline
\phi_{10} & (0,0) & 10 & -1 & \overline{\6} & e^{i 2\pi/3} & 1 & -2 \\ 
\phi_{16} & (0,0) & 16 & i & \overline{\3} & e^{i 4\pi/3} & 1 & -1 \\ 
\phi_{\overline{126}} & (0,0) & \overline{126} & -1 & \overline{\6} & e^{i 2\pi/3} & 1 & -2 \\ 
\phi_{45} & (0,0) & 45 & 1 & \1 & 1 & 1 & 0 \\ 
\end{array}$$
\caption{Field content of the 
$\SO(10)$ model with gauged flavor from~\cite{DiLuzio:2020qio}. 
For each irrep we state its transformation properties under $\SO(10)$ and $\SU(3)_f$, their respective centers $\mathbb{Z}_{4}$ and $\mathbb{Z}_{3}$, and under the accidental $\U(1)_{\rm PQ}$. 
}
\label{tab:SO10irreps}
\end{table}%

%%%%%%%%%%%%%%%%%%%%%%%%%%%%%%%%
\subsection{The scalar potential of the $\SO(10)$ model}

We provide here some errata regarding the operators in the scalar potential of the $\SO(10)$ model from~\cite{DiLuzio:2020qio}, 
although the main conclusion regarding the emergence of the accidental $\U(1)_{\rm PQ}$ symmetry remains unchanged.
The renormalizable potential of the $\SO(10)$ model, as determined by the methods of \app{app:procedure-invariant-enumeration}, consists of the following invariants:
\begin{align}
    \tilde{\mathcal{V}}&=
        \tilde{\mathcal{V}}_{2}
        +\tilde{\mathcal{V}}_{3}
        +\tilde{\mathcal{V}}_{4}
        +\tilde{\mathcal{V}}_{\mathbb{C}} \, ,\\
    \tilde{\mathcal{V}}_{2}&=|\phi_{10}|^2+|\phi_{\overline{126}}|^2+|\phi_{16}|^2+\phi_{45}^2 \, ,\\
    \tilde{\mathcal{V}}_{3}&= \(|\phi_{10}|^2+|\phi_{\overline{126}}|^2+|\phi_{16}|^2\)\phi_{45} \, ,\\
    \tilde{\mathcal{V}}_{4}&= \(|\phi_{10}|^2+|\phi_{\overline{126}}|^2+|\phi_{16}|^2+\phi_{45}^2\)^{2} \, ,\\
    \tilde{\mathcal{V}}_{\mathbb{C}}&= 
    \phi_{10}\phi_{16}^{\ast 2}\,(1+\phi_{45}) + \phi_{10}\phi_{\overline{126}}^\ast \(\phi_{10}\phi_{\overline{126}}^\ast+|\phi_{\overline{126}}|^{2} + |\phi_{16}|^{2} + \phi_{45}^2 \) + \text{h.c.} \, .
\end{align}
The symbols $\mathcal{V}$ carry a tilde on top to distinguish them from the scalar potential of the PS model from \sect{sec:PS-model}.

Note that two more types of invariants are in principle possible from the given irreps: $\phi_{45}^3$ and $\phi_{16}^{2}\phi_{\overline{126}}^{\ast}\phi_{45}$. These, however, are anti-symmetric with respect to the same-irrep factors, and thus identically vanish given that the multiplicity of all scalar irreps in \Table{tab:SO10irreps} is $1$, i.e., one needs to consider the criteria for invariants all the way up to condition~\ref{item:layers-3} in \app{app:procedure-invariant-enumeration}. While \cite{DiLuzio:2020qio} does not list the invariants of $\tilde{\mathcal{V}}_{3}$ and the last two terms of $\tilde{\mathcal{V}}_{4}$, these clearly have no impact on the preservation of $\mathrm{U}(1)_{\rm PQ}$ from \Table{tab:SO10irreps}, so the associated conclusions are unchanged.

Using the methods of \app{app:procedure-invariant-enumeration} up to condition~\ref{item:layers-2}, we also list the invariants that break PQ symmetry and potentially contribute to the vacuum, analogously to our analysis in \sect{sec:PQquality}. The potentially dangerous invariants for PQ quality in the $\SO(10)$ model are listed in \Table{tab:operators-for-axion-potential-SO10}. Although the full list is much longer than what is given in \cite{DiLuzio:2020qio}, the conclusions on PQ quality (at tree level) remain unchanged. We shall make a more thorough comparison below.
\begin{table}[t!]
    \begin{center}
\begin{minipage}[t]{0.45\linewidth}
    \vspace{0pt}
%%%%%%%%% d=6
\begin{tabular}{lll}
    \toprule
    $\mathcal{O}$ ($d=6$)& $\langle\mathcal{O}\rangle$&$\#$\\
    \midrule
    $\phi_{10}^{6-k}\phi_{\overline{126}}^{k}$ & $v^6$&$7$\\[\SKIP]
    \bottomrule
%\end{tabular}
%%%%%%%%% d=7
%\vskip 0.2cm
%\begin{tabular}{lll}
\addlinespace[2pt]    
    \toprule 
    $\mathcal{O}$ ($d=7$)& $\langle\mathcal{O}\rangle$&$\#$\\
    \midrule
    $\phi_{10}^{4-k}\phi_{\overline{126}}
    ^{k}\phi_{\overline{126}}\phi_{16}^{2}$ & $v^4 V^{3}$&$5$\\[\SKIP]
    $\phi_{10}^{6-k}\phi_{\overline{126}}^{k}\phi_{45}$ & $v^6 V'$&$7$\\[\SKIP]
    $\phi_{10}^{5}\phi_{16}^{2}$ & $v^{6} V$ & $1$\\[\SKIP]
    \bottomrule
%\end{tabular}
%%%%%%%%% d=8
%\vskip 0.2cm
%\begin{tabular}{lll}
\addlinespace[2pt]    
\toprule
    $\mathcal{O}$ ($d=8$)& $\langle\mathcal{O}\rangle$&$\#$\\
    \midrule
    $\phi_{10}^{2-k}\phi_{\overline{126}}^{k}\phi_{\overline{126}}^{2}\phi_{16}^{4}$ & $v^2 V^{6}$&$3$\\[\SKIP]
    $\phi_{10}^{4-k}\phi_{\overline{126}}^{k}\phi_{\overline{126}}\phi_{16}^{2}\phi_{45}$ & $v^{4} V^{3}V'$&$5$\\[\SKIP]
    $\phi_{10}^{3}\phi_{\overline{126}}\phi_{16}^{4}$ & $v^4 V^{4}$&$1$\\[\SKIP]
    $\phi_{10}^{6-k}\phi_{\overline{126}}^{k}\phi_{45}^{2}$ & $v^6 V'^{2}$&$7$\\[\SKIP]
    $\phi_{10}^{5}\phi_{16}^{2}\phi_{45}$ & $v^6 VV'$&$1$\\[\SKIP]
    $\phi_{10}^{6-k}\phi_{\overline{126}}^{k}\phi_{16}\phi_{16}^{\ast}$ & $v^6 V^{2}$&$7$\\[\SKIP]
    $\phi_{10}^{6-k}\phi_{\overline{126}}^{k}\phi_{\overline{126}}\phi_{\overline{126}}^{\ast}$ & $v^6 V^{2}$&$7$\\[\SKIP]
    $\phi_{10}^{4}\phi_{16}^{4}$ & $v^6 V^{2}$&$1$\\[\SKIP]
    $\phi_{10}^{6-k}\phi_{\overline{126}}^{k}\phi_{10}\phi_{10}^{\ast}$ & $v^8$&$7$\\[\SKIP]
    $\phi_{10}^{7}\phi_{\overline{126}}^{\ast}$ & $v^8$&$1$\\[\SKIP]
    $\phi_{10}\phi_{\overline{126}}^{\ast 7}$ & $v^8$&$1$\\[\SKIP]
    \bottomrule
\end{tabular}
\end{minipage}
%%%%%% d=9
\begin{minipage}[t]{0.45\linewidth}
    \vspace{0pt}
\begin{tabular}{lll}
    \toprule
    $\mathcal{O}$ ($d=9$)& $\langle\mathcal{O}\rangle$&$\#$\\
    \midrule
    $\phi_{\overline{126}}^{3}\phi_{16}^{6}$ & $V^{9}\NO$ &$1$\\[\SKIP]
    %% v2
    $\phi_{10}^{2-k}\phi_{\overline{126}}^{k}\phi_{\overline{126}}^{2}\phi_{16}^{4}\phi_{45}$ & $v^2 V^{6} V'$&$3$\\[\SKIP]
    $\phi_{10}\phi_{\overline{126}}^{2}\phi_{16}^{6}$ & $v^2 V^{7}$&$1$\\[\SKIP]
    %% v4
    $\phi_{10}^{4-k}\phi_{\overline{126}}^{k}\phi_{\overline{126}}\phi_{16}^{2}\phi_{45}^{2}$ & $v^4 V^{3} V'^{2}$&$5$\\[\SKIP]
    $\phi_{10}^{3}\phi_{\overline{126}}\phi_{16}^{4}\phi_{45}$ & $v^4 V^{4} V'$&$1$\\[\SKIP]
    $\phi_{10}^{4-k}\phi_{\overline{126}}^{k}\phi_{\overline{126}}^{2}\phi_{16}^{2}\phi_{\overline{126}}^{\ast}$ & $v^4 V^{5}$&$5$\\[\SKIP]
    $\phi_{10}^{4-k}\phi_{\overline{126}}^{k}\phi_{\overline{126}}\phi_{16}^{3}\phi_{16}^{\ast}$ & $v^4 V^{5}$&$5$\\[\SKIP]
    $\phi_{10}^{2}\phi_{\overline{126}}\phi_{16}^{6}$ & $v^4 V^{5}$&$1$\\[\SKIP]
    %% v6
    $\phi_{10}^{6-k}\phi_{\overline{126}}^{k}\phi_{45}^{3}$ & $v^6 V'^{3}$&$7$\\[\SKIP]
    $\phi_{10}^{5}\phi_{16}^{2}\phi_{45}^{2}$ & $v^6 V V'^{2}$&$1$\\[\SKIP]
    $\phi_{10}^{4}\phi_{16}^{4}\phi_{45}$ & $v^6 V^{2} V'$&$1$\\[\SKIP]
    $\phi_{10}^{6-k}\phi_{\overline{126}}^{k}\phi_{\overline{126}}\phi_{\overline{126}}^{\ast}\phi_{45}$ & $v^6 V^{2} V'$&$7$\\[\SKIP]
    $\phi_{10}^{6-k}\phi_{\overline{126}}^{k}\phi_{16}\phi_{16}^{\ast}\phi_{45}$ & $v^6 V^{2} V'$&$7$\\[\SKIP]
    $\phi_{10}^{4-k}\phi_{\overline{126}}^{k}\phi_{10}\phi_{10}^{\ast}\phi_{\overline{126}}\phi_{16}^{2}$ & $v^6 V^{3}$&$5$\\[\SKIP]
    $\phi_{10}\phi_{16}^{\ast 2}\phi_{\overline{126}}^{\ast 6}$ & $v^6 V^{3}$&$1$\\[\SKIP]
    $\phi_{10}^{5}\phi_{16}^{3}\phi_{16}^{\ast}$ & $v^6 V^{3}$&$1$\\[\SKIP]
    $\phi_{10}^{3}\phi_{16}^{6}$ & $v^6 V^{3}$&$1$\\[\SKIP]
    $\phi_{10}^{5}\phi_{\overline{126}}\phi_{\overline{126}}^{\ast}\phi_{16}^{2}$ & $v^6 V^{3}$&$1$\\[\SKIP]
    %% v8
    $\phi_{10}^{6-k}\phi_{\overline{126}}^{k}\phi_{10}\phi_{10}^{\ast}\phi_{45}$ & $v^8 V'$&$7$\\[\SKIP]
    $\phi_{10}^{7}\phi_{\overline{126}}^{\ast}\phi_{45}$ & $v^8 V'$&$1$\\[\SKIP]
    $\phi_{10}\phi_{\overline{126}}^{\ast 7}\phi_{45}$ & $v^8 V'$&$1$\\[\SKIP]
    $\phi_{10}^{6}\phi_{16}^{2}\phi_{\overline{126}}^{\ast}$ & $v^8 V$&$1$\\[\SKIP]
    $\phi_{10}^{6}\phi_{10}^{\ast}\phi_{16}^{2}$ & $v^8 V$&$1$\\[\SKIP]
    $\phi_{10}^{6-k}\phi_{\overline{126}}^{k}\phi_{10}\phi_{16}^{\ast 2}$ & $v^8 V$&$7$\\[\SKIP]
    $\phi_{16}^{2}\phi_{\overline{126}}^{\ast 7}$ & $v^8 V$&$1$\\[\SKIP]
%%%
    \bottomrule
\end{tabular}
\end{minipage}
\end{center}
    \caption{Non-renormalizable operators $\mathcal{O}$ of dimensions $d\leq 9$ in the $\SO(10)$ model from~\cite{DiLuzio:2020qio}. Only these may (but not necessarily) contribute to the axion potential: they violate PQ and have a VEV product with vanishing $B-L$, see main text for further details. Their potential vacuum contribution is denoted by $\langle\mathcal{O}\rangle$, while $\#$ specifies the number of operators given the running index $k$. Note that the index contraction for the $V^{9}$ contribution makes it vanish, see main text, as indicated by using the notation $\NO$.
        \label{tab:operators-for-axion-potential-SO10}
    }
\end{table}

To compile \Table{tab:operators-for-axion-potential-SO10}, certain subtleties need to be considered. We gather the relevant comments below:
\begin{itemize}
    \item In the $\SO(10)$ model, the breaking pattern is 
        \begin{align}
            \label{eq:SO10breakpatt}
            & \ \SO(10) \times \SU(3)_{f} \times \U(1)_{\rm PQ} \nonumber \\
            & \xrightarrow[]{\langle \phi_{45}\rangle}  \SU(3)_c \times \SU(2)_L \times \SU(2)_{R} \times \U(1)_{B-L} \times \SU(3)_{f} \times \U(1)_{\rm PQ} \nonumber \\
            & \xrightarrow[]{\langle \phi_{\overline{126}},\phi_{16}\rangle}  \SU(3)_c \times \SU(2)_L \times \U(1)_{Y} \nonumber \\
            & \xrightarrow[]{\langle \phi_{10},\phi_{\overline{126}},\phi_{16} \rangle} \SU(3)_c \times \U(1)_{\rm EM} \, , 
        \end{align}
        where we assumed the flavor group to be broken in one step for simplicity. The three steps of symmetry breaking happen at the GUT scale $V'$, the PQ scale $V$ and the EW scale $v$, respectively. These labels are used in the contributions $\langle\mathcal{O}\rangle$ in \Table{tab:operators-for-axion-potential-SO10}. 
        The intermediate symmetry between the GUT group and SM is the left-right symmetry group. Note that flavor and PQ breaking ($\SU(3)_{f}$ and $\U(1)_{\rm PQ}$) occurs at the scale $V$, the same as in our PS model, while the GUT scale $V'$ is not present there. We have the hierarchy $v\ll V\ll V'$, and we can assume the values of \eq{eq:vev-scaling} and $V'\sim 10^{16}\,\mathrm{GeV}$.
     
     \begin{table}[t!]
    \begin{center}
        \vskip 0.2cm
        \begin{tabular}{lc@{$\qquad$}cc@{$\qquad$}cc@{$\qquad$}c}
         \toprule
        irrep& 
            $\phi_{10}$ &
            \multicolumn{2}{c@{\qquad}}{$\phi_{16}$}&
            \multicolumn{2}{c@{\qquad}}{$\phi_{\overline{126}}$}&
            $\phi_{45}$\\
        \midrule
        VEV type &
            $v$&$V$&$v$&$V$&$v$&$V'$ \\
        $B-L$ of VEV &$0$&$+1$&$-1$&$-2$&$0$&$0$\\ 
        \bottomrule
    \end{tabular}
    \end{center}
        \caption{The $B-L$ numbers carried by various VEVs of the irreps in the $\SO(10)$ model. An irrep can carry multiple types of VEVs. \label{tab:SO10-BL}}   
        \end{table}
    
    \item Some irreps carry multiple VEVs, in particular the fields $\phi_{\overline{126}}$ and $\phi_{16}$ can acquire both a PQ-breaking VEV $V$ and a EW VEV $v$ (in different components). 
    The $B-L$ numbers of these VEVs are given in \Table{tab:SO10-BL}. A contribution $\langle\mathcal{O}\rangle$ to the vacuum is possible only if the sum of $B-L$ numbers in the VEV directions of each field vanishes. With multiple VEVs present in some irrep, we assume all possible VEV alignment combinations, and we specify for $\langle\mathcal{O}\rangle$ the largest consistent contribution: total $B-L$ vanishes, and the number of EW VEVs is even. Note that the EW VEV in $\phi_{16}$ carries a non-vanishing $B-L$ charge; this VEV has no analog in our Pati-Salam model of \Table{tab:PSirrep}.
    \item Comparing the $\SO(10)$ contributions in \Table{tab:operators-for-axion-potential-SO10} with the Pati-Salam contributions in \Table{tab:operators-for-axion-potential}, the dominant candidate for each fixed power of $v$ are the same: $v^{6}$, $v^4 V^{3}$, $v^{2} V^{6}$, $V^{9}$. The $v^{6}$ and $V^{9}$ candidates have also been identified in \cite{DiLuzio:2020qio}, while the other two have been revealed only from the extended list in this appendix. 
    \par
    Furthermore, the $V^{9}$ contribution in fact identically vanishes 
    in $\SO(10)$, indicated by an arrow in \Table{tab:operators-for-axion-potential-SO10}). This was also the case in the Pati-Salam model, see 
    \app{app:PSmodel-PQ-at-tree-level}, and the same argument from there
    carries over to the $\SO(10)$ case. Indeed, in the $\SO(10)$ model the VEVs $V$ come from the same type of Pati-Salam irreps $\Delta$ and $\chi$, which are now  part of larger $\SO(10)$ irreps (and with the same flavor assignment), cf.~\eq{eq:inclusion-into-SO10}. The $V^{9}$ contribution would thus come from the Pati-Salam part $\Delta^{3} \chi^{\ast 6}\subset \phi_{\overline{126}}^{3}\phi_{16}^{6}$ that was shown to vanish. The crucial argument given in \app{app:PSmodel-PQ-at-tree-level} was the anti-symmetry of the flavor contraction, and it relied on having only one copy of irreps $\Delta$ and $\chi$, which is now still the case for $\phi_{\overline{126}}$ and $\phi_{16}$. 
    \par
    To summarize, the size of the dominant PQ contributions in the $\SO(10)$ model are the same as in the Pati-Salam model, so the conclusions on 
    the PQ quality are analogous. 
    Similar considerations apply as well to the 
    associated axion phenomenology, as 
    represented by \fig{fig:plotfa}. 
    \item 
    In the $\SO(10)$ case, a potential danger is represented by the presence of the larger GUT-scale VEV $V'$ associated with $\phi_{45}$. 
    Structurally, however, $\phi_{45}$ acts like a neutral element not crucial for the construction of an invariant, similar to how the irrep $\xi$ behaves in the Pati-Salam model. An invariant with contribution $v^{m} V^{n} V'^{k}$ thus has an analog with omitted $\phi_{45}$ and a contribution $v^{m} V^{n}$. The original thus has an extra suppression factor $(V'/\Lambda)^{k}$ compared to the lower dimensional analog, and hence  invariants with $V'$ can be omitted when searching for the most dangerous contributions.   
\end{itemize}

%%%%%%%%%%%%%%%%%%%%%%%%%%%%%%%%
\subsection{Perturbativity in the $\SO(10)$ model}

The issue of perturbativity of the $\SO(10)$ model was not addressed in~\cite{DiLuzio:2020qio}. Since it has even more degrees of freedom than our PS model, the expectation is that a Landau pole in the gauge coupling develops even sooner. 

We do not address the details of gauge coupling unification here, but merely 
perform an analysis in the UV theory analogous to the one in \sect{app:PSpert}. Above the GUT scale, the field content is that of \Table{tab:SO10irreps}, which yields the beta coefficients for the gauge group $\SO(10)\times\SU(3)_{f}$ to be
\begin{align}
    a^{\text{(GUT)}}_{i}&= 
       (50,\tfrac{347}{3}) \, , &
    b^{\text{(GUT)}}_{ij}&= 
        \begin{pmatrix}
        \tfrac{23157}{2} & 2928 \\[3pt]
        16470 & \tfrac{16118}{3} \\
        \end{pmatrix} \, ,
    \label{eq:ab-coefficients-SO10}
\end{align}
where the order of beta function coefficients corresponds to the order of factors in the gauge group: $(\alpha^{-1}_{\rm GUT},\alpha^{-1}_{3F})$.

\begin{figure}[t!]
    \begin{center}
    \includegraphics[width=0.5\linewidth]{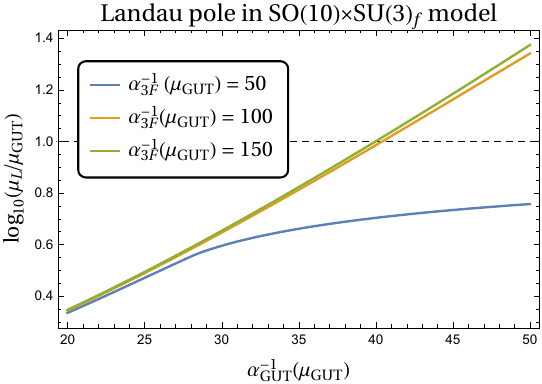}
    \caption{The scale $\mu_L$, measured in powers of $10$ above $\mu_{\rm GUT}$, when the Landau pole is reached. It is computed as a function of the value of the unified gauge coupling at the GUT scale $\alpha^{-1}_{\rm GUT}(\mu_{\rm GUT})$. The various lines show the cases with different starting values of the flavor gauge coupling 
    $\alpha^{-1}_{3F}(\mu_{\rm GUT})$. \label{fig:Landau-pole-SO10}}
    \end{center}
\end{figure}

These coefficients lead to a Landau pole soon above the GUT scale, as shown in \fig{fig:Landau-pole-SO10}. For a typical value $\alpha^{-1}_{\rm GUT}(\mu_{\rm GUT})=40$ and a small flavor coupling (large $\alpha^{-1}_{3F}$), the Landau pole is reached within an order of magnitude from the GUT scale. The $\SO(10)$ scenario is thus borderline perturbative in this particular regime, but becomes non-perturbative as soon as unification happens with a larger value of the unified coupling (a smaller value of $\alpha^{-1}_{\rm GUT}(\mu_{\rm GUT})$). 

Furthermore, notice that the one-loop beta coefficient for $\alpha^{-1}_{3F}$ is larger than for $\alpha^{-1}_{\rm GUT}$. 
Small values of the flavor coupling are now $\alpha^{-1}_{3F}(\mu_{\rm GUT})\geq 100$. At such values, the flavor coupling has little effect on the Landau pole, namely only through the two-loop off-diagonal coefficient. A value $\alpha^{-1}_{3F}(\mu_{\rm GUT})=50$ (blue curve), on the other hand, already corresponds to a large flavor coupling. The Landau pole is then reached much sooner and in the flavor coupling, again corresponding to a non-perturbative scenario.  

\begin{small}

\bibliographystyle{utphys.bst}
\setlength{\bibsep}{5pt plus 0.3ex}
%\bibliography{bibliography}
\providecommand{\href}[2]{#2}\begingroup\raggedright\endgroup

\end{small}

\clearpage 

\end{document}